\documentclass[twoside,12pt]{article}
\usepackage{epsfig}
\usepackage{amssymb}
\usepackage{amsthm}
\usepackage{amsmath, bm}
\usepackage{color}
\numberwithin{equation}{section}
\newcommand{\be}{\begin{equation}}
\newcommand{\ee}{\end{equation}}
\newcommand{\bea}{\begin{eqnarray}}
\newcommand{\eea}{\end{eqnarray}}

\begin{document}

\title{ \vspace{1cm} Quarkonium in Quark-Gluon Plasma: \\ Open Quantum System Approaches Re-examined}
\author{Y. Akamatsu\\
\\
Department of Physics, Osaka University, Japan}
\maketitle
\begin{abstract}
Dissociation of quarkonium in quark-gluon plasma (QGP) is a long standing topic in relativistic heavy-ion collisions because it has been believed to signal one of the fundamental natures of the QGP -- Debye screening due to the liberation of color degrees of freedom.
Among recent new theoretical developments is the application of open quantum system framework to quarkonium in the QGP.
Open system approach enables us to describe how dynamical as well as static properties of QGP influences the time evolution of quarkonium in a coherent way.

Currently, there are several master equations for quarkonium corresponding to various scale assumptions, each derived in different theoretical frameworks.
In this review, all of the existing master equations are systematically rederived as Lindblad equations in a unified framework.
Also, as one of the most relevant descriptions in relativistic heavy-ion collisions, quantum Brownian motion of heavy quark pair in the QGP is studied in detail.
The quantum Brownian motion is parametrized by a few fundamental quantities of QGP such as real and imaginary parts of heavy quark potential (complex potential), heavy quark momentum diffusion constant, and thermal dipole self-energy constant, which constitute in-medium self-energy of a static quarkonium.
This indicates that the yields of quarkonia such as $J/\psi$ and $\Upsilon$ in the relativistic heavy-ion collisions have the potential to determine these fundamental quantities.
\end{abstract}

\eject
\tableofcontents

\newpage
\section{Introduction}
Recently, theory of open quantum systems \cite{BRE02, gardiner2004quantum, weiss2012quantum} is widely applied to subatomic physics and cosmology as well as to the fields of condensed matter physics and atomic, molecular, and optical (AMO) physics.
For instance, the effects of quantum friction on nuclear collective dynamics are studied in nuclear fusion and fission processes \cite{tokieda2017quantum, bulgac2019unitary}, 
entanglement of valence partons with the rest of degrees of freedom in high-energy hadronic wave function is studied by analyzing the rapidity evolution of their density matrix \cite{armesto2019color, li2020jimwlk},
and dissipation of inflaton due to coupling to environmental fields is found to suppress the power spectrum of temperature anisotropy in the cosmic microwave background at large scales \cite{boyanovsky2015effective, boyanovsky2016effective}.
Also, in the condensed matter and AMO physics, non-Hermitian physics, which naturally arises in certain limit of open system set up, is now developing very rapidly \cite{Ashida:2020dkc}.
As is common to these examples, the open quantum system refers to a quantum system of our interest that couples to the environmental degrees of freedom, which is a ubiquitous situation in physics.

In this review paper, I re-examine recent applications of the open quantum system approach to a particular subfield of high energy nuclear collisions -- in-medium properties of quarkonia in quark-gluon plasma (QGP).
In particular, I show that all of the microscopic derivations of open system dynamics of quarkonia to date \cite{akamatsu2013real, Akamatsu:2014qsa, Blaizot:2015hya, Blaizot:2017ypk, Blaizot:2018oev, Brambilla:2016wgg, Brambilla:2017zei, Yao:2018nmy} can be reformulated in a unified framework on the basis of a simple quantum mechanical picture of quarkonium coupled with QGP particles, which is somewhat close to the method adopted in \cite{Blaizot:2017ypk, Blaizot:2018oev}.
I believe that putting the previous efforts in a unified framework must provide a useful guide to this field because the descriptions used in the original derivations differ from one another: influence functional in the path integral formalism \cite{akamatsu2013real, Akamatsu:2014qsa, Blaizot:2015hya}, dynamical evolution of density matrix in the operator formalism \cite{Blaizot:2017ypk, Blaizot:2018oev, Yao:2018nmy}, and Dyson-Schwinger equation of correlation functions on the closed-time path \cite{Brambilla:2016wgg, Brambilla:2017zei}.

In what follows, I will first briefly review the history of studies of quarkonium in the QGP.
In 1980s, it was not clear whether the QGP, an extremely hot matter in which quarks and gluons are liberated from inside the hadrons, can be experimentally produced and identified if produced by colliding two large nuclei at ultrarelativistic energies.
Proposing signals for the QGP production is one of the important theoretical works in the community of relativistic heavy-ion collisions.
Among other proposals, it was predicted by Matsui and Satz \cite{Matsui:1986dk} that the yield of $J/\psi$, the ground state of charm quark pair $c\bar c$ with $J^{PC} = 1^{--}$, is suppressed if the QGP is created\footnote{
More precisely, $J/\psi$ yield is suppressed compared to that in proton-proton collisions multiplied by a proper scaling factor $N_{\rm coll}$ that estimates the number of binary nucleon-nucleon collisions in a heavy-ion collision.
The latter is a theoretical estimate for the case without QGP production.
}.
This scenario of $J/\psi$ suppression is based on a very simple but essential physics of the deconfined plasma -- Debye screening.
In the quark-gluon plasma, liberated color degrees of freedom screen the local color charges and the color interaction gets short ranged.
The qualitative difference between the confining force in the vacuum and the screened force in the deconfined phase results in a drastic change of the spectrum of $J/\psi$ as the temperature increases and eventually leads to its disappearance at high temperature.
Extending this reasoning to other charmonium states such as  $\psi(2{\rm S})$ and bottomonia $\Upsilon(n{\rm S})$ (bound states of bottom quark pair $b\bar b$ with $J^{PC}=1^{--}$) is straightforward and different melting temperatures for each quarkonium will be obtained.
One may even conclude that the different melting temperatures can be used as a thermometer of the QGP.
Theoretically, however, there is no unambiguous definition of melting temperature and thus the resolution of thermometer is not high at best.

There were two important developments that demonstrated and supplemented the idea of Matsui and Satz in terms of real-time quantities with clear field theory definitions -- quarkonium spectral function and real-time heavy quark potential, which are closely related to each other.
In short, the $J/\psi$ suppression scenario was based only on static nature of QGP.
These two developments shed light on dynamical nature of QGP that influences the in-medium quarkonium properties.

Spectral function of quarkonium at finite temperature provides us with a useful information on the nature of a quarkonium bound state in the QGP and its interaction with the QGP constituents.
It was first calculated for $J/\psi$ and $\eta_c$ by Asakawa and Hatsuda \cite{asakawa2004j} using the maximum entropy method to reconstruct the spectral image from Euclidean lattice QCD data \cite{asakawa2001maximum}.
They found that the peak structures of $J/\psi$ and $\eta_c$ survive up to $T\simeq 1.6T_c$ and disappear below $1.9T_c$, which signals the melting of $J/\psi$.
From this example, one can see that the definition of melting temperature inevitably contains ambiguity because it must specify the conditions for existence and disappearance of the spectral peak.
The lattice QCD simulation of quarkonium spectral functions at finite temperatures is still developing.
Recent studies of charmonium and bottomonium spectral functions are found in \cite{ohno2011charmonium, ding2012charmonium, borsanyi2014charmonium, ikeda2017medium} and \cite{aarts2014bottomonium, kim2018quarkonium, larsen2019thermal, larsen2020excited} respectively.

The other important quantity, real-time heavy quark potential, gives a precise definition of potential for static heavy quarks at finite temperature.
It is defined in terms of spectral decomposition of thermal Wilson loop and was first calculated perturbatively by Laine et al. \cite{Laine:2006ns} and by others \cite{Beraudo:2007ky, Brambilla:2008cx}.
It was also computed from the Euclidean lattice QCD data and has been continually updated and improved \cite{Rothkopf:2011db, Burnier:2014ssa, Burnier:2015tda, Burnier:2016mxc, petreczky2019realistic, bala2020nonperturbative}.
The real-time heavy quark potential at finite temperatures is complex valued and thus often called complex potential.
In the potential model, the complex potential is the potential that must be substituted in the time-dependent Schr\"odinger equation for in-medium wave function.
The Schr\"odinger equation with complex potential (and with finite heavy quark mass) is a useful tool to calculate the quarkonium spectral function within the potential model \cite{laine2007resummed, burnier2008heavy}, relating these two quantities.
Indeed, the imaginary part of the potential causes the spectral broadening even if the real part of the potential admits bound states, as observed in the lattice QCD simulations.

Although these two developments clarified to a large extent the real-time aspects of the $J/\psi$ suppression scenario, the notion of in-medium wave function limits the predictive power of the formalism\footnote{
If $J/\psi$ decays in the hot medium and is in kinetic equilibrium in the relativistic heavy-ion collisions, the spectral function may be observed through dilepton spectrum.
$J/\psi$ mass shift \cite{hashimoto1986mass} was proposed as a signal for the QGP formation around the same time with the $J/\psi$ suppression \cite{Matsui:1986dk}.
In reality, $J/\psi$ decays mostly in the vacuum outside the hot medium after out-of-equilibrium evolution.
}.
As is the case in non-equilibrium physics, it is the time evolution of the expectation values of physical quantities that one is eager to know.
For such purposes, the evolution of in-medium wave function is not enough; one needs to know the evolution of density matrix, the central object for the theory of open quantum systems.
One of the first applications of open quantum system approach to quarkonium physics in the QGP was the interpretation of the complex potential as statistical average of stochastic evolutions with fluctuating real potential, namely the stochastic potential \cite{Akamatsu:2011se, Rothkopf:2013kya, Kajimoto:2017rel}.
The stochastic potential explains the physical origin of imaginary potential in terms of thermal fluctuations and thus succeeds for the first time in describing both the static and dynamical effects of thermal fluctuations in one potential picture.
Moreover, it generates ensemble of wave functions from which one can reconstruct the mixed state density matrix.

After the introduction of stochastic potential and a few earlier applications of open quantum system approach \cite{Young:2010jq, Borghini:2011yq, Borghini:2011ms}, the development of this field is remarkable
\cite{akamatsu2013real, Akamatsu:2014qsa, Blaizot:2015hya, Blaizot:2017ypk, Blaizot:2018oev, Brambilla:2016wgg, Brambilla:2017zei, Yao:2018nmy, Akamatsu:2015kaa, Katz:2015qja, DeBoni:2017ocl, Akamatsu:2018xim, Brambilla:2019tpt, Miura:2019ssi, sharma2020quantum, Alund:2020ctu, Brambilla:2020qwo, Brambilla:2021wkt, Omar:2021kra, Akamatsu:2021vsh, Akamatsu:2021dot}.
In particular, it has reached to a point where several Lindblad equations for quarkonium in the QGP have been derived \cite{Akamatsu:2014qsa, Blaizot:2017ypk, Blaizot:2018oev, Brambilla:2016wgg, Brambilla:2017zei, Yao:2018nmy}.
On the other hand, there are several differences in the obtained equations, such as the regimes of applicabilities, whether or not the dissipation is included, and the treatments of singlet-octet transitions of quarkonium color states, in addition to the formalisms used in the derivations.
It is thus desirable at this moment that there exists a review paper that summarizes all of the existing Lindblad equations for quarkonium in the QGP in a unified theoretical framework, which is the main topic of this review.
There are also recent review papers \cite{Sharma:2021vvu, Yao:2021lus} that contain the same topic from different perspectives.

Finally, I list below the related topics, which however are not included in this review.
It is partly because I do not have enough knowledge and partly because there already exist nice review papers on these topics.
Interested readers should consult the references.
\begin{itemize}
\item Quarkonium in the vacuum and its production in proton-proton collisions \cite{brambilla2011heavy}
\item Experimental data and phenomenological studies of quarkonium in relativistic heavy-ion collisions.
See \cite{andronic2016heavy} for a general review and \cite{Rothkopf:2019ipj} for recent theoretical developments.
Popular descriptions in the phenomenological studies are:
\begin{itemize}
\item Schr\"odinger equation with complex potential \cite{Strickland:2011mw, Strickland:2011aa, Krouppa:2015yoa, Krouppa:2016jcl, Krouppa:2017jlg, Islam:2020gdv}
\item Kinetic equation with chemical reactions between quarkonium and heavy quarks \cite{Rapp:2008tf, Zhao:2010nk, Zhao:2011cv, Emerick:2011xu, Du:2015wha, Du:2017qkv, Du:2019tjf, Liu:2010ej, Zhou:2014kka, Zhou:2016wbo, Song:2011xi, Song:2011nu, Yao:2017fuc, Yao:2020xzw}
\end{itemize}
\item Quarkonium in other environments such as QGP with viscous corrections \cite{Dumitru:2007hy, Dumitru:2009ni, Burnier:2009yu, Dumitru:2009fy, Margotta:2011ta, Thakur:2020ifi}, QGP with finite velocity \cite{chu1989dynamic, Escobedo:2013tca, Sharma:2012dy}, and QGP in an external magnetic field \cite{Bonati:2016kxj, Hasan:2017fmf, Singh:2017nfa}
\end{itemize}

This review paper is organized as follows.
In Section~\ref{sec:Basics}, basics of the open quantum system is reviewed.
After the introduction of the Lindblad equation as a general form of Markovian master equation, weak coupling methods are explained for two different regimes of open quantum systems -- the quantum optical regime and the quantum Brownian motion. 
Formulas to obtain the Lindblad equation from microscopic Hamiltonian are given when the system-environment coupling is weak.
In Section~\ref{sec:NRQCD}, Lindblad equations are obtained for single heavy quark and quarkonium in the QGP that are described by non-relativistic QCD, an effective theory for heavy quarks.
In Section~\ref{sec:pNRQCD}, Lindblad equations are obtained for quarkonium in the QGP that is described by potential non-relativistic QCD, an effective theory for quarkonium as a color dipole.
Section~\ref{sec:summary} is devoted for summary.

\newpage
\section{Basics of open quantum systems}
\label{sec:Basics}
When a system of our interest is in contact with its surrounding {\it environment}, the former is called an open system, or simply a {\it system}.
Quantum mechanically, the total Hilbert space $\mathcal H_{\rm tot} = \mathcal H_{S} \otimes \mathcal H_{E}$ is a direct product of the system Hilbert space $\mathcal H_{S}$ and the environment Hilbert space $\mathcal H_{E}$.
The information of the total system is encoded as the total density matrix:
\begin{align}
\rho_{\rm tot}(t) \equiv \sum_{\alpha} w_{\alpha}|\alpha (t)\rangle\langle\alpha (t)|, \quad
|\alpha\rangle \in \mathcal H_{\rm tot}, \quad
\langle\alpha|\alpha\rangle = 1,
\end{align}
where $w_{\alpha}\geq 0$ is the probability to find a state $\alpha$ and satisfies $\sum_{\alpha} w_{\alpha} = 1$.
In the Schr\"odinger picture, the time evolution of $\rho_{\rm tot}(t)$ is governed by the von-Neumann equation:
\begin{align}
\label{eq:vonNeumann}
\frac{d}{dt}\rho_{\rm tot}(t) = \frac{1}{i\hbar}\left[H_{\rm tot}, \rho_{\rm tot}\right], \quad
H_{\rm tot} = H_S \otimes I_E + I_S\otimes H_E + \sum_i V^{(i)}_S \otimes V^{(i)}_E,
\end{align}
where  $H_{\rm tot}$ is the total Hamiltonian, $I_{E(S)}$ is the identity operator in the environment (system) Hilbert space and the last term is the interaction Hamiltonian between the system and the environment.
Since we are interested in the system part, we often need to evaluate the expectation value of a system operator $\mathcal O_S\otimes I_E$ that leaves the environment state unchanged:
\begin{align}
\langle\mathcal O_S(t)\rangle
= {\rm Tr}_{\rm tot}\left[\rho_{\rm tot}(t)(\mathcal O_S\otimes I_E)\right]
={\rm Tr}_S\left[\mathcal O_S\left({\rm Tr}_E\rho_{\rm tot}(t)\right)\right].
\end{align}
Note that ${\rm Tr}_{\rm tot} = {\rm Tr}_S{\rm Tr}_E =  {\rm Tr}_E{\rm Tr}_S$.
If we know the evolution of $\rho_S(t)\equiv {\rm Tr}_E\rho_{\rm tot}(t)$, we can calculate the expectation value as if $\rho_S$ is the density matrix of the system:
\begin{align}
\langle\mathcal O_S(t)\rangle={\rm Tr}_S\left[\rho_S(t)\mathcal O_S\right].
\end{align}
Theory of open quantum system \cite{BRE02, gardiner2004quantum, weiss2012quantum} describes the time evolution of {\it reduced density matrix} $\rho_S(t)$ by a differential equation, namely the {\it master equation}.
In the section \ref{sec:Basics_Lindblad}, I introduce a particular form of the master equation, the {\it Gorini-Kossakowski-Sudarshan-Lindblad equation} \cite{gorini1976completely, lindblad1976generators} or the {\it Lindblad equation} in short, which ensures that the evolved reduced density matrix fulfills desired physical properties.
In the section \ref{sec:Basics_Approx}, I summarize how and when the master equation can be obtained in the Lindblad form.
In particular, I will emphasize the necessity of time scale hierarchies and give their intuitive physical interpretation.
For advanced mathematics of open quantum systems, I will recommend an excellent textbook \cite{rivas2012open}.

\subsection{Lindblad equation}
\label{sec:Basics_Lindblad}
\subsubsection{Dynamical map}
\label{sec:Basics_Lindblad_map}
Evolution of the reduced density matrix $\rho_S(t)$ is quite involved.
Let us first observe basic properties of $\rho_S(t)$.
We assume that the system and the environment are initially decoupled $\rho_{\rm tot}(0) = \rho_S(0)\otimes \rho_E(0)$.
Then the reduced density matrix at later time $t$ is obtained as
\begin{align}
\rho_S(t) ={\rm Tr}_E \left\{U(t,0)\left[\rho_S(0)\otimes \rho_E(0)\right]U^{\dagger}(t,0)\right\}, \quad
U(t_1,t_2) \equiv e^{-iH_{\rm tot}(t_1-t_2)}.
\end{align}
Clearly, the evolution from $\rho_S(0)$ to $\rho_S(t)$ is linear in $\rho_S(0)$, which is called {\it dynamical map} and denoted as
\begin{align}
\rho_S(t) = \mathcal V_{\rm dyn}(t)\rho_S(0)
\end{align}
with a superoperator $\mathcal V_{\rm dyn}(t)$.
The environment density matrix $\rho_E(0)$ is Hermitian, positive semi-definite, and trace-normalized ${\rm Tr}_E\rho_E(0)=1$, so that it can be decomposed by its orthonormal eigenvectors $|\varphi_{\alpha}\rangle$ and non-negative eigenvalues $\lambda_{\alpha}$ as
\begin{align}
\label{eq:rhoE_diag}
\rho_E(0) = \sum_{\alpha}\lambda_{\alpha}|\varphi_{\alpha}\rangle\langle\varphi_{\alpha}|, \quad
\sum_\alpha \lambda_{\alpha}=1.
\end{align}
With this basis, the superoperator $\mathcal V_{\rm dyn}(t)$ is explicitly given as
\begin{align}
\label{eq:dyn_map}
\mathcal V_{\rm dyn}(t) \rho_S(0) = \sum_{\alpha, \beta}W_{\alpha\beta}(t) \rho_S(0) W_{\alpha\beta}^{\dagger}(t), \quad
W_{\alpha\beta}(t) \equiv \sqrt{\lambda_{\beta}}\langle\varphi_{\alpha}|U(t,0)|\varphi_{\beta}\rangle.
\end{align}
$W_{\alpha\beta}(t)$ labeled by $\alpha$ and $\beta$ is an operator in the system Hilbert space $\mathcal H_S$.
With this form, it is clear that $\mathcal V_{\rm dyn}(t)$ maps a positive operator $\rho_S(0)$ to another positive operator $\rho_S(t)$, namely $\mathcal V_{\rm dyn}(t)$ is a {\it positive} map.
The operator $W_{\alpha\beta}$ also possesses a property $\sum_{\alpha,\beta}W_{\alpha\beta}^{\dagger}(t)W_{\alpha\beta}(t) = \sum_{\beta}\lambda_{\beta}=I_S$, so that the trace of $\rho_S$ is preserved ${\rm Tr}_S\rho_S(t) = {\rm Tr}_S\rho_S(0)=1$.
Therefore, the dynamical map $\mathcal V_{\rm dyn}(t)$ is a linear map that preserves the positivity and trace of the reduced density matrix $\rho_S$.

To be precise, $\mathcal V_{\rm dyn}(t)$ in Eq.~\eqref{eq:dyn_map} is not only positive but also {\it completely positive}.
The latter requires that in addition to the positivity of $\mathcal V_{\rm dyn}(t)$, the superoperator $\mathcal V_{\rm dyn}(t)\otimes \mathcal I_{A}$ in an arbitrarily enlarged Hilbert space $\mathcal H_S\otimes \mathcal H_{A}$ is also a positive map in the composite system.
Here, $\mathcal I_{A}$ is the identity superoperator acting on the operators in $\mathcal H_A$.
Here is the proof of complete positivity of $\mathcal V_{\rm dyn}(t)$.
Prepare an enlarged system $\mathcal H_{S'} = \mathcal H_S\otimes \mathcal H_{A}$ such that the subsystem $A$ does not interact with either $S$ or $E$:
\begin{subequations}
\begin{align}
H_{\rm tot'} &= H_{S'} \otimes I_E + I_{S'}\otimes H_E + \sum_i V^{(i)}_{S'} \otimes V^{(i)}_E, \\
H_{S'} &= H_S\otimes I_A + I_S\otimes H_A, \quad
I_{S'} = I_S\otimes I_A, \quad
V^{(i)}_{S'} =V^{(i)}_S\otimes I_A.
\end{align}
\end{subequations}
Then, the dynamical map of the composite system $S'$ is
\begin{subequations}
\begin{align}
\mathcal V'_{\rm dyn}(t) \rho_{S'}(0) &= \sum_{\alpha, \beta}\left[W_{\alpha\beta}(t)\otimes e^{-iH_At}\right] \rho_{S'}(0) \left[W_{\alpha\beta}(t)\otimes e^{-iH_At} \right]^{\dagger}
=\left[\mathcal V_{\rm dyn}(t)\otimes \mathcal U_{A}\right]\rho_{S'}(0), \\
\mathcal U_{A}(t)\rho_A(0) &= e^{-iH_At}\rho_A(0)e^{iH_At},
\end{align}
\end{subequations}
and thus the superoperator is given by
\begin{align}
\mathcal V'_{\rm dyn}(t) = \mathcal V_{\rm dyn}(t)\otimes \mathcal U_{A}(t)
= \left[ \mathcal V_{\rm dyn}(t)\otimes \mathcal I_{A}\right]\cdot \left[\mathcal I_{S}\otimes \mathcal U_{A}(t)\right].
\end{align}
By repeating the same argument that lead to the positivity of $\mathcal V_{\rm dyn}(t)$, we can show that $\mathcal V'_{\rm dyn}(t)$ is positive as well.
Since $\mathcal I_{S}\otimes \mathcal U^{-1}_{A}(t)$ is a (reversed) unitary evolution and is a positive map, it follows that $\mathcal V_{\rm dyn}(t)\otimes \mathcal I_{A} = \mathcal V'_{\rm dyn}(t)\cdot\left[\mathcal I_{S}\otimes \mathcal U^{-1}_{A}(t)\right]$ is positive for any subsystem $A$ and thus $\mathcal V_{\rm dyn}(t)$ is completely positive.

Mathematically, it is not the positivity but the complete positivity that fully characterizes the dynamical map.
For example, transposition $\rho_S\to \rho_S^T$ is positive but not completely positive, therefore positivity alone cannot exclude such class of operations.
Indeed, it is proved by Kraus \cite{kraus1983states} that any linear map which is completely positive can be written with a set of operators $K_{\alpha}$ as
\begin{align}
\label{eq:Kraus_decomp}
\rho_S(0) \to \rho_S(t) = \sum_{\alpha}K_{\alpha}(t,0) \rho_S(0) K_{\alpha}^{\dagger}(t,0),
\end{align}
known as Kraus decomposition.
Conversely, it is straightforward (as above) to show that any map given by a Kraus decomposition is completely positive.
Therefore, mathematically precise classification of the dynamical map is a linear map which is Completely Positive and Trace-Preserving, or a CPTP map.

So far, we have assumed that the initial condition takes the product form $\rho_{\rm tot}(0) = \rho_S(0)\otimes \rho_E(0)$.
Here, let us discuss the effect of initial correlation between the system and the environment:
\begin{align}
\label{eq:init_corr}
\rho_{\rm tot}(0) = \rho_S(0)\otimes \rho_E(0) + \rho_{\rm corr}(0), \quad
{\rm Tr}_S\rho_{\rm corr}(0) = {\rm Tr}_E\rho_{\rm corr}(0) = 0.
\end{align}
At time $t$, the reduced density matrix is
\begin{align}
\label{eq:dyn_map_init_corr}
\rho_S(t) = \sum_{\alpha,\beta}W_{\alpha\beta}(t)\rho_S(0)W_{\alpha\beta}^{\dagger}(t)
+ {\rm Tr}_E \left[U(t,0)\rho_{\rm corr}(0)U^{\dagger}(t,0)\right],
\end{align}
where $W_{\alpha\beta}(t)$ is defined in Eq.~\eqref{eq:dyn_map}.
It is clear that the initial correlation does play a role.
The first term is in the form of the Kraus decomposition, which depends on $\rho_E(0)$ (through $W_{\alpha\beta}(t)$) and linearly on $\rho_S(0)$.
The second term does not directly depend on $\rho_S(0)$.
Therefore, when $\rho_E(0)$ and $\rho_{\rm corr}(0)$ are fixed, the evolution of $\rho_S(t)$ is inhomogeneous but is linear in $\rho_S(0)$, i.e. if $\rho_S(0) = w \rho_S^{(1)}(0)+ (1-w)\rho_S^{(2)}(0)$ with $0\leq w \leq 1$, it evolves to $\rho_S(t) = w \rho_S^{(1)}(t)+ (1-w)\rho_S^{(2)}(t)$.
However, due to the presence of the initial correlation $\rho_{\rm corr}(0)$, not all the initial $\rho_S(0)$s can give a positive $\rho_{\rm tot}(0)$ by \eqref{eq:init_corr}.
In this sense, the map \eqref{eq:dyn_map_init_corr} is not a positive map and thus cannot be written as the Kraus decomposition \cite{vstelmachovivc2001dynamics}.

From the point of view of constructing a dynamical map from $\rho_S(0)$ to $\rho_S(t)$, the first step is to assign an initial total density matrix
\begin{align}
\label{eq:assignment_map}
\rho_{\rm tot}(0) = \Phi[\rho_S(0)],
\end{align}
where $\Phi$ is called {\it assignment map}.
It seems natural to require the following three conditions on $\Phi$:
(a) linear map of $\rho_S(0)$,
(b) consistent $\rho_S(0) = {\rm Tr}_E\Phi [\rho_S(0)]$, and
(c) $\Phi[\rho_S(0)]$ is positive for any positive $\rho_S(0)$.
It is proved in \cite{pechukas1994reduced} that if we demand all of these, the only possible assignment map is in the form
\begin{align}
\Phi[\rho_S(0)] = \rho_S(0)\otimes\rho_E(0),
\end{align}
with $\rho_E(0)$ being a positive density matrix independent of $\rho_S(0)$.
When the system-environment coupling is strong, the initial correlation is often non-negligible and $\Phi$ fails to meet at least one of the conditions (a)-(c) \cite{pechukas1994reduced, alicki1995comment, pechukas1995pechukas}.
The initial condition \eqref{eq:init_corr} with $\rho_E(0)$ and $\rho_{\rm corr}(0)$ independent of $\rho_S(0)$ corresponds to a $\Phi$ which satisfies (a) and (b) but fails to satisfy (c) as we see above.
In some cases, $\Phi$ satisfies (c) but fails (a) or (b).
For example, the total initial state $\rho_{\rm tot}(0)$ is prepared from the total equilibrium $\rho_{\rm tot}^{\rm eq}$ by performing projective measurements on the system:
\begin{align}
\rho_{\rm tot}(0) = \sum_n w_n P_n 
\otimes {\rm Tr}_S\left[P_n \rho_{\rm tot}^{\rm eq}\right]/{\rm Tr}_{\rm tot}\left[P_n \rho_{\rm tot}^{\rm eq}\right], \quad
0<w_n\leq 1, \quad \sum_n w_n = 1,
\end{align}
where $P_n = |n\rangle\langle n|$ denotes a projective measurement and $w_n$ is an arbitrary weight factor to make a mixed state $\rho_S(0)=\sum_n w_n P_n$.
In this case, $\Phi$ is a nonlinear map of $\rho_S(0)$ and fails to satisfy (a) unless $\rho_{\rm tot}^{\rm eq}$ is a product state \cite{alicki1995comment}.

\subsubsection{Markovian limit}
\label{sec:Basics_Lindblad_Markov}
If the environment is static and loses its correlation in a finite short time, there is expected to be a time scale in which the system dynamics becomes Markovian and the dynamical map $\mathcal V_{\rm dyn}(t)$ satisfies a semigroup property $\mathcal V_{\rm dyn}(t_1)\mathcal V_{\rm dyn}(t_2)=\mathcal V_{\rm dyn}(t_1+t_2)$ for $t_1, t_2 \geq 0$.
The semigroup property is slightly puzzling because the system and the environment are not any more decoupled in $\rho_{\rm tot}(t_2)$, but it can be explained as follows.
If the system evolution is Markovian, $\rho_S(t_1+t_2)$ can be expressed in two ways, $\rho_S(t_1+t_2) = \tilde {\mathcal V}(t_1)\rho_S(t_2) = \tilde {\mathcal V}(t_1)\mathcal V_{\rm dyn}(t_2)\rho_S(0) =\mathcal V_{\rm dyn}(t_1+t_2)\rho_S(0)$ for any $\rho_S(0)$, and thus $\tilde {\mathcal V}(t_1)\mathcal V_{\rm dyn}(t_2) = \mathcal V_{\rm dyn}(t_1+t_2)$ holds.
Here the Markovian evolution operator $\tilde {\mathcal V}(t_1)$ depends only on the time difference $t_1$ because we assume a static environment\footnote{
One might think that the update from $t_2$ to $t_1+t_2$ still depends on both $t_2$ and $t_1+t_2$ even in the static case because initial time is 0, i.e. $\tilde {\mathcal V}(t_1+t_2-t_0, t_2-t_0)$ if the initial time is $t_0$.
However the dependence on $t_0$ contradicts the Markovian assumption and we get the form $\tilde {\mathcal V}(t_1)$.}.
Choosing $t_2=0$, we get $\tilde {\mathcal V}(t_1) = \mathcal V_{\rm dyn}(t_1)$ and the semigroup property follows.
In this time scale, $\mathcal V_{\rm dyn}(t)$ describes a quantum Markov process and we can define a generator $\mathcal L$ of the dynamical map $\mathcal V_{\rm dyn}(t) = \exp[\mathcal L t]$.
The generator must be written in the Lindblad form \cite{gorini1976completely, lindblad1976generators}
\begin{align}
\label{eq:Lindblad}
\frac{d}{dt}\rho_S
=\mathcal L \rho_S =  -i\left[H, \rho_S\right] 
+ \sum_k \left(L_k\rho_S L_k^{\dagger} - \frac{1}{2}L_k^{\dagger}L_k\rho_S - \frac{1}{2}\rho_S L_k^{\dagger}L_k\right),
\end{align}
where the second term in the right hand side is called the {\it dissipator}.

Here we provide a short derivation of this form in a finite dimensional Hilbert space $\mathcal H_S$ following \cite{BRE02}.
For a finite dimensional Hilbert space (${\rm dim} \ \mathcal H_S=N$), there is an operator basis $F_i \ (i=1,2,\cdots, N^2)$ by which any operator can be expanded.
It is convenient to define an inner product of the operators by $(A, B) \equiv {\rm Tr}_S(A^{\dagger}B)$.
Since $(F_i, F_j) = (F_j, F_i)^*$ is Hermitian, one can take a proper linear combination to make an orthonormal basis $(F_i, F_j) = \delta_{ij}$.
We can choose, for the later purpose, to define $F_{N^2} = \frac{1}{\sqrt{N}}I_S$ and then all the other basis operators are traceless ${\rm Tr}_S F_i = 0 \ (i=1,2,\cdots, N^2-1)$.
The dynamical map $\mathcal V_{\rm dyn}(t)$ can be expanded by this operator basis
\begin{subequations}
\begin{align}
\rho_S(t)&=\mathcal V_{\rm dyn}(t) \rho_S(0)
=\sum_{\alpha, \beta}W_{\alpha\beta}(t) \rho_S(0) W_{\alpha\beta}^{\dagger}(t)
=\sum_{i,j} c_{ij}(t) F_i \rho_S(0) F_j^{\dagger}, \\
c_{ij}(t)&\equiv \sum_{\alpha, \beta}(F_i, W_{\alpha\beta}(t))(F_j, W_{\alpha\beta}(t))^*,
\end{align}
\end{subequations}
where the coefficient matrix $c_{ij}(t)$ is positive semi-definite.
The action of $\mathcal L$ is obtained by
\begin{align}
\label{eq:dyn_generator}
\mathcal L\rho_S &= \lim_{\epsilon\to 0}\frac{1}{\epsilon}\left[\mathcal V_{\rm dyn}(\epsilon)\rho_S - \rho_S\right] \\
&=\lim_{\epsilon\to 0}\left[
\frac{1}{\epsilon}\left(\frac{c_{N^2N^2}(\epsilon)}{N} -1\right)\rho_S
+\frac{1}{\sqrt{N}}\sum_{i=1}^{N^2-1}\left(\frac{c_{iN^2}(\epsilon)}{\epsilon} F_i\rho_S
+ \frac{c_{N^2i}(\epsilon)}{\epsilon} \rho_SF_i^{\dagger}\right)
+\sum_{i,j=1}^{N^2-1}\frac{c_{ij}(\epsilon)}{\epsilon} F_i \rho_S F_j^{\dagger}
\right]. \nonumber
\end{align}
If the Markovian description is available, the following limit for the coefficients must exist after coarse graining in time
\begin{align}
a_{N^2N^2} \equiv \lim_{\epsilon\to 0} \frac{c_{N^2N^2}(\epsilon)-N}{\epsilon}, \quad
a_{iN^2} = a_{N^2i}^* \equiv \lim_{\epsilon\to 0}\frac{c_{iN^2}(\epsilon)}{\epsilon}, \quad
a_{ij}\equiv \lim_{\epsilon\to 0}\frac{c_{ij}(\epsilon)}{\epsilon}, \quad
(i,j = 1,2,\cdots,N^2-1),
\end{align}
by which we also define new operators and rewrite \eqref{eq:dyn_generator} as
\begin{subequations}
\begin{align}
F&\equiv \frac{1}{\sqrt{N}}\sum_{i=1}^{N^2-1}a_{iN^2} F_i, \quad
H\equiv\frac{1}{2i} (F^{\dagger} - F), \quad
G\equiv\frac{1}{2N}a_{N^2N^2}I_S + \frac{1}{2}(F^{\dagger} + F), \\
\mathcal L\rho_S&=
-i[H,\rho_S] + \left\{G,\rho_S \right\}
+\sum_{i,j=1}^{N^2-1} a_{ij}F_i\rho_SF_j^{\dagger}.
\end{align}
\end{subequations}
Furthermore, the dynamical map $\mathcal V_{\rm dyn}(t)$ preserves the trace of $\rho_S$, so that $\mathcal L$ must satisfy
\begin{subequations}
\begin{align}
0&={\rm Tr}_S\left(\mathcal L\rho_S\right) = {\rm Tr}_S\left[\left(2G + \sum_{i,j=1}^{N^2-1}a_{ij}F_j^{\dagger}F_i\right)\rho_S\right], \quad
G=-\frac{1}{2}\sum_{i,j=1}^{N^2-1}a_{ij}F_j^{\dagger}F_i, \\
\label{eq:Lindblad_generator}
\mathcal L\rho_S &= -i[H,\rho_S] 
+\sum_{i,j=1}^{N^2-1} a_{ij} \left( F_i\rho_SF_j^{\dagger}
-\frac{1}{2}\left\{F_j^{\dagger}F_i,\rho_S \right\}\right).
\end{align}
\end{subequations}
Since $a_{ij}$ $(i,j=1,2,\cdots, N^2-1)$ is also positive semi-definite, we can diagonalize the quadratic forms $a_{ij}F_i\rho_SF_j^{\dagger}$ and $a_{ij}F_j^{\dagger}F_i$ by a unitary transformation of the operator basis and then rescale the operators to get
\begin{align}
\mathcal L\rho_S &= -i[H,\rho_S] 
+\sum_{k} \left(L_k\rho_SL_k^{\dagger}
-\frac{1}{2}L_k^{\dagger}L_k\rho_S - \frac{1}{2}\rho_SL_k^{\dagger}L_k\right).
\end{align}
The operators $L_k$ are called Lindblad operators and their number can be smaller than $N^2-1$ because $a_{ij}$ may have zero eigenvalues.
By this construction, ${\rm Tr}_S H={\rm Tr}_S L_k=0$ and ${\rm Tr}_S(L_k^{\dagger}L_l)=(L_k, L_l)\propto \delta_{kl}$.

A few remarks are in order here.
First, note that $H$ is not necessarily the system Hamiltonian $H_S$.
There can be a new contribution by the coupling to the environment.
Second, note also that the Lindblad operators $L_k$ are not unique.
The Lindblad equation is invariant under transformations
(i) $L_k \to L_k' = \sum_l u_{kl}L_l$ with a unitary matrix $\sum_l u_{kl}u_{ml}^*=\delta_{km}$ and
(ii) $L_k \to L_k' = L_k + a_k$, $H\to H'=H+\frac{1}{2i}\sum_k(a_k^*L_k-a_kL_k^{\dagger}) + b$ with $a_k\in \bf C$ and $b\in \bf R$.
With these transformations taken into consideration, the relations ${\rm Tr}_S H={\rm Tr}_S L_k=0$ and ${\rm Tr}_S(L_k^{\dagger}L_l)=(L_k, L_l)\propto \delta_{kl}$ do not hold in general.

Original derivation of Eq.~\eqref{eq:Lindblad} by Gorini, Kossakowski, and Sudarshan \cite{gorini1976completely} took a different approach.
Here we briefly sketch their statement because it illustrates the importance of complete positivity and the connection to classical Markov processes.
They started from a theorem by Kossakowski \cite{kossakowski1972necessary} (proof can also be found in \cite{rivas2012open}), which states that, for $\mathcal L$ to be a generator of a positive dynamical semigroup $\mathcal V_{\rm dyn}(t)=\exp\left[\mathcal L t\right]$, the necessary and sufficient condition is to satisfy
\begin{subequations}
\label{eq:Kossakowski_conditions}
\begin{align}
{\rm Tr}_S P_r \mathcal L (P_s) &\geq 0 \quad (r\neq s = 1,2,\cdots, N={\rm dim} \ \mathcal H_S), \\
\sum_{r=1}^{N} {\rm Tr}_S P_r \mathcal L (P_s) &= 0 \quad (s= 1,2,\cdots, N),
\end{align}
\end{subequations}
where $P_r \ (r=1,2,\cdots, N)$ is a set of complete and mutually orthogonal one-dimensional projection operator of $\mathcal H_S$.
These conditions are known as Kossakowski conditions.
Furthermore, a superoperator $\mathcal L$ is a generator of a completely positive dynamical semigroup if and only if $\mathcal L\otimes \mathcal I_A$ satisfies the Kossakowski conditions \eqref{eq:Kossakowski_conditions} for the extended Hilbert space $\mathcal H_S\otimes\mathcal H_A$ with ${\rm dim} \ \mathcal H_A=N$ so that ${\rm dim} \ (\mathcal H_S\otimes\mathcal H_A)=N^2$, which is a consequence of a theorem by Choi \cite{choi1972positive}\footnote{
It states that a linear map $\Gamma$ of the operators in $\mathcal H_S$ is completely positive if a map $\Gamma\otimes \mathcal I_A$ of the operators in $\mathcal H_S\otimes \mathcal H_A$ with ${\rm dim} \ \mathcal H_S = {\rm dim} \ \mathcal H_A$ is positive.
The generator of $\mathcal V_{\rm dyn}(t)\otimes  \mathcal I_A$ is $\mathcal L\otimes \mathcal I_A$.
}.
Using these conditions, they proved that $\mathcal L$ is a generator of a complete positive dynamical semigroup if and only if it is given by \eqref{eq:Lindblad_generator} with a positive coefficient matrix $a_{rs}$.

With the probability distribution $p_r(t) \equiv {\rm Tr}_SP_r\rho_S(t)$, the connection to the classical Markov process is obtained by an ansatz $\rho_S(t) = \sum_r p_r(t) P_r$, where off-diagonal parts are neglected.
In general, classical Markov process is described by
\begin{align}
\frac{dp_r}{dt} = \sum_{s=1}^N\Gamma_{rs}p_s, \quad
\Gamma_{rs} \geq 0 \quad (r\neq s=1,2,\cdots, N), \quad \sum_{r=1}^N\Gamma_{rs} = 0 \quad (s=1,2,\cdots, N).
\end{align}
The Kossakowski conditions are interpreted as quantum extension of the conditions on the transition rates $\Gamma_{rs}$, by replacing $\Gamma_{rs}$ with ${\rm Tr}_S P_r\mathcal L(P_s)$.

The above derivations are for finite dimensional Hilbert space $\mathcal H_S$ while in general ${\rm dim} \ \mathcal H_S$ can be infinite.
Mathematically, it is proven in a general Hilbert space that if the Lindblad operators form a countable set, any bounded generator $\mathcal L$ for a trace-preserving completely positive map $\mathcal V_{\rm dyn}(t)$ can be written in the form of Eq.~\eqref{eq:Lindblad} \cite{lindblad1976generators}. 
Although many physical examples of $\mathcal L$ are unbounded (e.g. Hamiltonian) or the label $k$ of $L_k$ runs continuous variables (e.g. momentum), they are written in the form \eqref{eq:Lindblad} or can be slightly modified to take the form \eqref{eq:Lindblad}.

\subsubsection{Steady states}
\label{sec:Basics_Lindblad_SS}
Given a Lindblad equation \eqref{eq:Lindblad}, what is its steady state solution?
Although its mathematical definition is simple $\mathcal L\rho_{ss}=0$ and can be formulated as a concrete problem in linear algebra in finite dimensional case, not much is known about the steady states.
If there is a unique steady state and every initial reduced density matrix evolves into that steady state
\begin{align}
\forall \rho_S(0): \quad \lim_{t\to\infty}e^{\mathcal Lt}\rho_S(0) = \rho_{ss},
\end{align}
such dynamical semigroup is called {\it relaxing}.
Below, after giving general discussions, we list at the end some of the known facts about the steady states when ${\rm dim} \ \mathcal H_S=N$ is finite.

As defined in Sec.~\ref{sec:Basics_Lindblad_Markov}, the inner product of the operators $(A,B)\equiv {\rm Tr}_S(A^{\dagger}B)$ is useful.
Here again we adopt the orthonormal operator basis $F_i \ (i=1,2,\cdots, N^2)$ satisfying $(F_i,F_j)=\delta_{ij}$.
Then, the condition for the steady state $\mathcal L\rho_{ss}=0$ can be written as the following linear algebraic problem:
\begin{align}
[\mathcal L]_{ij} \equiv (F_i,\mathcal L F_j), \quad
[\rho_{ss}]_i \equiv (F_i, \rho_{ss}), \quad
\sum_{j=1}^{N^2}[\mathcal L]_{ij}[\rho_{ss}]_j = 0 \quad (i=1,2,\cdots, N^2).
\end{align}
In this section, we choose Hermitian operator basis $F_i=F_i^{\dagger}$, by which $[\mathcal L]_{ij}$ is a real matrix and $[\rho_{ss}]_i$ is a real vector.
Note that real vectors correspond to Hermitian matrices but they are not necessarily positive.
A steady state is a zero mode of $[\mathcal L]_{ij}$ corresponding to a positive Hermitian matrix\footnote{
``State" and ``density matrix" imply a positive Hermitian matrix with unit trace.
}.

In general, $[\mathcal L]_{ij}$ is not diagonalizable and is only similar to the Jordan normal form.
In the language of (super-)operators, it means that we can divide the $N^2$-dimensional space of matrices into invariant subspaces with respect to the action of $\mathcal L$.
Each invariant subspace, labeled by $i$, has an eigenvalue $\lambda_i$ and a dimension $k_i$, and consists of operator basis satisfying 
\begin{align}
(\mathcal L-\lambda_i\mathcal I_S) \sigma^{(n)}_i=\sigma^{(n-1)}_i \quad (n=2,3,\cdots, k_i), \quad
(\mathcal L-\lambda_i\mathcal I_S) \sigma^{(1)}_i=0.
\end{align}
Here the matrix $\sigma^{(n)}_i$ is not necessarily Hermitian.
For a real eigenvalue $\lambda_i=\lambda_i^*$, one can take $\sigma^{(n)}_i (n=1,2,\cdots, k_i)$ to be Hermitian.
For a complex eigenvalue $\lambda_j$, its complex conjugate is also an eigenvalue $\lambda_{j'}=\lambda_j^*$ and the basis of these subspaces are related by $\sigma^{(n)}_j = \sigma^{(n)\dagger}_{j'}  (n=1,2,\cdots, k_j=k_{j'})$. 
The real part of the eigenvalues never exceeds zero ${\rm Re}\lambda_i \leq 0$ because $e^{\mathcal Lt}$ is a {\it contraction} map\footnote{
We adopt the trace norm defined as $\|A\| \equiv {\rm Tr} \sqrt{A^{\dagger}A}\geq 0$.
In the context here, we consider a linear operation that maps a Hermitian operator $\sigma\neq 0$ to another Hermitian operator.
If the trace norm does not increase from the original one for all the initial Hermitian operators, such map is a contraction.
See \cite{rivas2012open} for the proof that a dynamical map is a contraction.
}.
Otherwise, we can easily construct counterexamples.
For a real eigenvalue $\lambda_i>0$, norm of the eigenmode $\| e^{\mathcal Lt}\sigma_i^{(1)}\|$ increases. 
For a pair of complex eigenvalues $\lambda_{j}=\lambda_{j'}^*=\lambda_{Rj} + i\lambda_{Ij}$, a Hermitian operator $\sigma_j^{(1)}+\sigma_{j'}^{(1)}=\sigma_j^{(1)}+\sigma_{j}^{(1)\dagger}$ evolves as
\begin{align}
e^{\mathcal Lt}\left(\sigma_j^{(1)}+\sigma_{j}^{(1)\dagger}\right)
=e^{\lambda_{Rj}t}\left(
\sigma_j^{(1)} e^{i\lambda_{Ij}t} + \sigma_j^{(1)\dagger} e^{-i\lambda_{Ij}t}
\right),
\end{align}
whose norm at $t=2\pi/\lambda_{Ij}$ is larger than at $t=0$.
Furthermore, the subspace of a zero mode $\lambda_o = 0$ is one dimensional $k_o=1$.
If $k_o\geq2$ for a zero mode, it follows that 
\begin{align}
\mathcal L\sigma_o^{(2)} = \sigma_o^{(1)}, \quad \mathcal L\sigma_o^{(1)}=0, \quad
e^{\mathcal Lt} \sigma_o^{(2)} = \sigma_o^{(2)} + t\sigma_o^{(1)}.
\end{align}
Since $e^{\mathcal Lt}$ is a trace preserving map, $\sigma_o^{(1)}$ must be traceless.
Even if $\sigma_o^{(1)}$ is traceless, $t\sigma_o^{(1)}$ dominates in $e^{\mathcal Lt} \sigma_o^{(2)}$ at a large enough time $t$ and the norm increases from the initial state and contradicts that $e^{\mathcal Lt}$ is a contraction map.

Finally, let us quote two known facts about the steady states for a finite-dimensional case, with only a few comments added.
\begin{itemize}
\item There is at least one steady state.
\begin{itemize}
\item One can construct it by taking the long-time average
\begin{align}
\bar\rho_S \equiv \lim_{T\to\infty}\frac{1}{T}\int_0^T dt e^{\mathcal Lt}\rho_S(0), \quad
\mathcal L(\bar\rho_S)=0.
\end{align}
The long-time average is shown to converge because $e^{\mathcal Lt}$ is a contraction \cite{rivas2012open}.
\item For an infinite-dimensional case, the existence of a steady state is not necessarily ensured.
\end{itemize}
\item The dynamical semigroup is relaxing if and only if the spectrum of $[\mathcal L]_{ij}$ contains a non-degenerate zero eigenvalue and no pure imaginary eigenvalues.
\begin{itemize}
\item If there are pure imaginary eigenvalues, oscillatory behavior is allowed and the dynamics is not relaxing.
Therefore, their non-existence is explicitly mentioned in the condition for the spectrum.
In \cite{schirmer2010stabilizing}, a stronger statement is proved without mentioning the absence of pure imaginary eigenvalues, so one can omit ``and no pure imaginary eigenvalues" in the above.
\item The ``if part" is trivial.
Indeed, if the spectrum contains only one zero mode, it must be a density matrix because its existence is ensured as above.
\item The ``only if part" is nontrivial.
One needs to exclude the possibility that there exists another zero mode $\sigma_{o,\text{non-pos.}}^{(1)}$ corresponding to a non-positive Hermitian matrix.
If this is the case, one can construct two steady states by choosing an initial density matrix $\rho_S(0)$ such that $C\rho_S(0) + \sigma_{o,\text{non-pos.}}^{(1)} \ (C>0)$ is positive.
Then, there exist two different long-time averages $\bar\rho_S$ and $C\bar\rho_S + \sigma_{o,\text{non-pos.}}^{(1)}$. 
After normalization, both of them are steady states and the dynamics turns out not relaxing.
\end{itemize}
\end{itemize}

\subsection{Approximations for master equation}
\label{sec:Basics_Approx}
The Lindblad equation \eqref{eq:Lindblad} is the general form of the Markovian master equation which preserves trace and positivity of the reduced density matrix $\rho_S$.
In this section, we derive from a microscopic theory the Lindblad equation when the coupling between the system and the environment is small.
As is mentioned in the previous section, it is essential that the correlation time of the environment is short to get a coarse grained description for the dynamical map $\mathcal V_{\rm dyn}(t)=\exp[\mathcal L t]$.

Let us first introduce the following three time scales: environment correlation time $\tau_E$, system intrinsic time scale $\tau_S$, and system relaxation time $\tau_R$.
The environment correlation time $\tau_E$ is the time scale by which the environment correlation function decays.
The system intrinsic time scale $\tau_S$ is estimated by the system Hamiltonian through the energy-time uncertainty relation $\Delta \epsilon\cdot\tau_S\sim 1$, where $\Delta \epsilon$ is typical energy gap between the system eigenstates.
The system relaxation time $\tau_R$ is the resulting time scale of the master equation.
Corresponding to different time scale hierarchies in the Table~\ref{tab:oqs_regimes}, there are two regimes in the open quantum systems; one is the {\it quantum optical limit} and the other is the {\it quantum Brownian motion}.
Below I will explicitly show in each regime how to coarse grain $\mathcal V_{\rm dyn}(t)$ using proper approximation schemes; namely rotating wave approximation for the former, gradient expansion for the latter, and the {\it Born-Markov approximation} for the both.

\begin{table}[t]
\centering
\caption{Two regimes of open quantum systems}
\label{tab:oqs_regimes}
\vspace{3mm}
\begin{tabular}{c||c|c}
Regimes & Time scales & Approximation schemes\\
\hline 
Quantum optical limit & $\tau_R\gg\tau_E$, $\tau_R\gg\tau_S$  & Born-Markov \& Rotating wave approximations  \\
Quantum Brownian motion & $\tau_R\gg\tau_E$, $\tau_S\gg\tau_E$ & Born-Markov approximation \& Gradient expansion
\end{tabular}
\end{table}

\subsubsection{Born-Markov approximation}
\label{sec:Basics_Approx_BM}
First we derive the master equation in the Born-Markov approximation.
Depending on the time scale hierarchies, a further approximation, the rotating wave approximation or the gradient expansion, will be later applied to this master equation in Sec.~\ref{sec:Basics_Approx_QOL} and Secs.~\ref{sec:Basics_Approx_QBM}, \ref{sec:Basics_Approx_exQBM} respectively.

Our starting point is the von-Neumann equation in the interaction picture:
\begin{subequations}
\begin{align}
\frac{d}{dt}\rho_{\rm tot}(t) &= -i\left[V(t), \rho_{\rm tot}(t)\right], \\
V(t) &=\sum_i\left(e^{iH_St}V_S^{(i)}e^{-iH_St}\right)\otimes \left(e^{iH_Et}V_E^{(i)}e^{-iH_Et}\right)
=\sum_i V_{S}^{(i)}(t)\otimes V_{E}^{(i)}(t).
\end{align}
\end{subequations}
For simplicity, we omit $I$ for $\rho_{\rm tot}^{(I)}$ or $V_I$ which denotes that the interaction picture is adopted.
By substituting a formal solution of the von-Neumann equation
\begin{align}
\rho_{\rm tot}(t) = \rho_{\rm tot}(0) - i\int_0^t ds\left[V(s), \rho_{\rm tot}(s)\right],
\end{align}
into its right hand side, we get
\begin{align}
\frac{d}{dt}\rho_{\rm tot}(t) =-i\left[V(t), \rho_{\rm tot}(0)\right]
- \int_0^t ds\left[V(t), \left[V(s), \rho_{\rm tot}(s)\right]\right].
\end{align}
Taking the trace in the environment, the reduced density matrix obeys
\begin{align}
\label{eq:master_iterative}
\frac{d}{dt}\rho_S(t) = -\int_0^t ds {\rm Tr}_E\left[V(t),\left[V(s), \rho_{\rm tot}(s)\right]\right].
\end{align}
Here we assume ${\rm Tr}_E \left[V(t), \rho_{\rm tot}(0)\right] = 0$.
When the initial total density matrix factorizes $\rho_{\rm tot}(0)=\rho_S(0)\otimes \rho_{\rm E}(0)$, this assumption is made to hold true by subtracting the one-point functions from the interaction $V_E^{(i)}(t) \to V_E^{(i)}(t) - {\rm Tr}_E\left(\rho_E(0)V_E^{(i)}(t)\right)$ and adding the corresponding terms to the system Hamiltonian $H_S$.
In particular, when $\rho_{\rm E}(0)$ is a function of $H_E$ such as the Boltzmann distribution in the equilibrium, ${\rm Tr}_E\left(\rho_E(0)V_E^{(i)}(t)\right)$ is independent of time.

Equation \eqref{eq:master_iterative} is exact but not yet obtained in a closed form of $\rho_S$.
To obtain a master equation, we need to make two approximations: the Born approximation and the Markov approximation.
In the Born approximation, we use an ansatz $\rho_{\rm tot}(s)\approx \rho_S(s)\otimes\rho_E(0)$.
This is justified when the environment is so large that it is not much affected by its weak coupling to the system.
In the Markov approximation, we can replace $\rho_S(s)$ with $\rho_S(t)$ because the error introduced in the evolution equation is of higher order in the weak coupling expansion.
Furthermore, since there is a time scale hierarchy $\tau_R\gg\tau_E$, at typical time $t\sim \tau_R$ the environmental correlation to the initial time $(t=0)$ is lost, so that we can extend the domain of time integration to $(-\infty, t)$.
After changing the integration variable, Eq.~\eqref{eq:master_iterative} becomes
\begin{align}
\label{eq:master_BM}
\frac{d}{dt}\rho_S(t) = -\int_0^{\infty} ds {\rm Tr}_E\left[V(t),\left[V(t-s), \rho_S(t)\otimes\rho_E(0)\right]\right],
\end{align}
which is Markovian and is in the closed form of $\rho_S(t)$.

\begin{figure}
\centering
\includegraphics[width=0.8\textwidth]{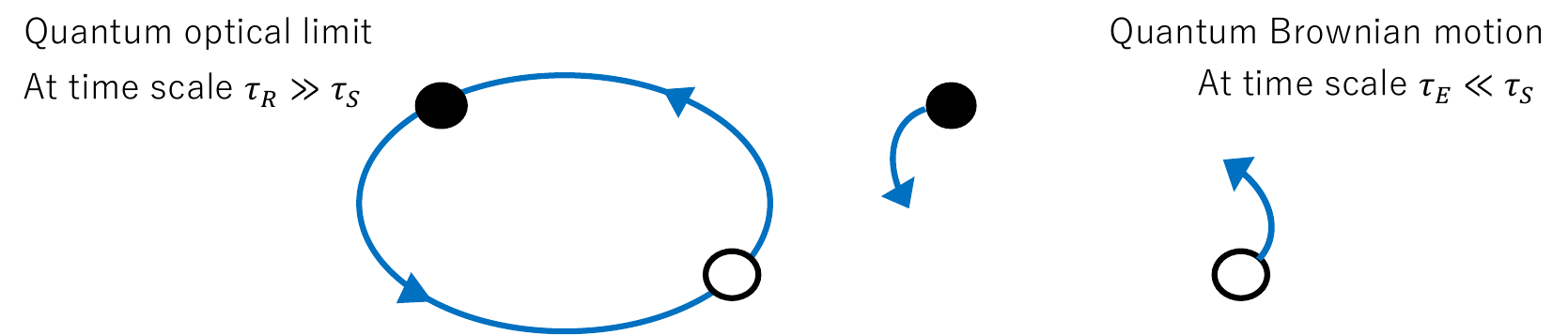}
\caption{Physical picture of two regimes of open quantum systems.
In the quantum optical limit, the system eigenstates are the good basis.
In the quantum Brownian motion, phase space variables provide good descriptions.
}
\label{fig:oqs_regimes}
\end{figure}

\subsubsection{Quantum optical limit}
\label{sec:Basics_Approx_QOL}
The master equation \eqref{eq:master_BM} in the Born-Markov approximation only assumes one of the time scale hierarchies $\tau_R\gg\tau_E$.
Another relation $\tau_R\gg\tau_S$ will be used to make a further approximation to Eq.~\eqref{eq:master_BM}, the rotating wave approximation, to get a Lindblad equation \cite{davies1974markovian,dumcke1979proper}.
Physical condition corresponding to $\tau_R\gg\tau_S$ is shown in Fig.~\ref{fig:oqs_regimes}.
In Sec.~\ref{sec:pNRQCD_deep}, the quantum optical master equation is applied to deeply bound quarkonia in quark-gluon plasma.

Let us define a projector to a system eigenstate $\Pi(\epsilon)\equiv |\epsilon\rangle\langle\epsilon|$ with an eigenenergy $\epsilon$.
It will turn out useful to decompose the system operator $V_S^{(i)}$ by transition energies
\begin{align}
\label{eq:transitions}
V_S^{(i)}(\omega) \equiv \sum_{\epsilon'-\epsilon=\omega}\Pi(\epsilon) V_S^{(i)}\Pi(\epsilon'), \quad
V_S^{(i)} = \sum_{\omega} V_S^{(i)}(\omega), \quad
V_S^{(i)}(t) = \sum_{\omega} e^{-i\omega t}V_S^{(i)}(\omega).
\end{align}
The time scale of $V_S^{(i)}(t)$ is governed by $\tau_S\sim 1/\Delta \epsilon$, where $\Delta \epsilon$ is a typical energy gap $\omega$.
The interaction Hamiltonian $V(t)$ in the interaction picture is then expanded as
\begin{align}
V(t) = \sum_{i,\omega} e^{-i\omega t}V_S^{(i)}(\omega) \otimes V_E^{(i)}(t)
=\sum_{i,\omega} e^{i\omega t}V_S^{(i)\dagger}(\omega) \otimes V_E^{(i)}(t).
\end{align}
Substituting this form into Eq.~\eqref{eq:master_BM}, we get
\begin{align}
\frac{d}{dt}\rho_S(t) &= 
\sum_{\omega,\omega'}\sum_{i,j}e^{i(\omega'-\omega)t}
\int_0^{\infty} ds e^{i\omega s}{\rm Tr}_E\left(\rho_E(0)V_E^{(i)}(t)V_E^{(j)}(t-s)\right) \nonumber \\
&\quad \times \left[V_S^{(j)}(\omega) \rho_S(t)V_S^{(i)\dagger}(\omega') -  V_S^{(i)\dagger}(\omega')V_S^{(j)}(\omega) \rho_S(t)\right] \quad + \quad h.c. ,
\end{align}
where $\it h.c.$ denotes the Hermitian conjugate of the former part.
The phase factor $e^{i(\omega'-\omega)t}$ is oscillating with time scale $\tau_S$, which is rapid compared to the system relaxation time $\tau_R$.
Therefore, in the time scale of $\tau_R$, the phase factor for $\omega'\neq\omega$ averages to 0 and only $\omega=\omega'$ remains in the summation.
This is the rotating wave approximation and the master equation is now
\begin{subequations}
\begin{align}
\frac{d}{dt}\rho_S(t) &= \sum_{\omega}\sum_{i,j}
\Gamma_{ij}(\omega)
\left[V_S^{(j)}(\omega) \rho_S(t)V_S^{(i)\dagger}(\omega) -  V_S^{(i)\dagger}(\omega)V_S^{(j)}(\omega) \rho_S(t)\right] \quad + \quad h.c. , \\
\label{eq:env_corr}
\Gamma_{ij}(\omega)
&\equiv \int_0^{\infty} ds e^{i\omega s}{\rm Tr}_E\left(\rho_E(0)V_E^{(i)}(s)V_E^{(j)}(0)\right).
\end{align}
\end{subequations}
Here we assume that the environment is invariant under the time translation.
Although the system time scale is previously estimated as $\tau_S\sim 1/\Delta \epsilon$ for simplicity, the derivation here makes it clear that $\tau_S$ is the inverse of the typical gap of transition energies (not the energy gap).

Physical picture of rotating wave approximation is as follows.
Consider a pure state $\rho_S(0)=|\epsilon\rangle\langle\epsilon|$ as an initial state.
The interaction with the environment induces a transition to another state $|\epsilon'\rangle$ as long as $\langle\epsilon'|V_S^{(i)}|\epsilon\rangle\neq 0$.
In general, simultaneous transition to yet another state $|\epsilon''\rangle$ is also allowed if $\langle\epsilon''|V_S^{(i)}|\epsilon\rangle\neq 0$.
Therefore, the system wave function is a superposition of many different system eigenstates.
The energy-time uncertainty principle tells that it takes at least $\tau\gtrsim1/|\epsilon'-\epsilon''|$ before one can specify which energy eigenstate is realized.
In other words, it takes $\tau\gtrsim1/|\epsilon'-\epsilon''|$ before a quantum mechanical superposition state turns into a mixed state with classical probability.
This situation is typical of the quantum optics, where a few discrete atomic levels compose the system of interest and the photon gas constitutes the environment.
This is why open systems of this type are called the {\it quantum optical regime}.

The coefficient matrix $\Gamma_{ij}(\omega)$ is given by the environment correlation functions and is decomposed into Hermitian and anti-Hermitian parts:
\begin{align}
\Gamma_{ij}(\omega) = \frac{1}{2}\gamma_{ij}(\omega) + iS_{ij}(\omega), \quad
\gamma_{ij}^*(\omega)=\gamma_{ji}(\omega), \quad
S_{ij}^*(\omega)=S_{ji}(\omega).
\end{align}
Using this expression, the master equation (in the interaction picture) is obtained as
\begin{subequations}
\label{eq:Lindblad_QOL}
\begin{align}
\frac{d}{dt}\rho_S(t) &= -i\left[\Delta H_S, \rho_S(t) \right] + \sum_{\omega}\sum_{i,j}
\gamma_{ij}(\omega)\left[V_S^{(j)}(\omega) \rho_S(t)V_S^{(i)\dagger}(\omega)
- \frac{1}{2} \left\{V_S^{(i)\dagger}(\omega)V_S^{(j)}(\omega), \rho_S(t)\right\}\right] , \\
\Delta H_S &\equiv \sum_{\omega}\sum_{i,j}S_{ij}(\omega)V_S^{(i)\dagger}(\omega)V_S^{(j)}(\omega).
\end{align}
\end{subequations}
Here $\Delta H_S$ is a modification to the system Hamiltonian $H_S$ due to the coupling to the environment.
They commute with each other $[\Delta H_S, H_S]=0$ and $\Delta H_S$ introduces renormalization of the energy eigenvalues of the system and thus is called Lamb shift term.
Getting back to the Schr\"odinger picture, the master equation has a Hamiltonian term $-i\left[H_S + \Delta H_S, \rho_S(t) \right]$ and the dissipator is unchanged because the extra phases cancel between $V_S^{(j)}(\omega)$ and $V_S^{(j)\dagger}(\omega)$:
\begin{align}
e^{-iH_S t} V_S^{(j)}(\omega)e^{iH_S t} = e^{i\omega t}V_S^{(j)}(\omega), \quad
e^{-iH_S t} V_S^{(i)\dagger}(\omega)e^{iH_S t} = e^{-i\omega t}V_S^{(i)\dagger}(\omega).
\end{align}
If $\gamma_{ij}(\omega)$ is positive semi-definite\footnote{
Without mathematical rigor, we can show that $\gamma_{ij}(\omega)$ is positive semi-definite when $\rho_E(0)$ is a function only of $H_E$ such as the Boltzmann weight.
Using the eigenstates of $H_E$ as the basis for the decomposition \eqref{eq:rhoE_diag}, we get for any $c_i$
\begin{align}
\sum_{i,j}c_i^*\gamma_{ij}(\omega)c_j
=\sum_{E,E'}\lambda_{E}\Bigl|\langle E'|\sum_i c_iV_E^{(i)}(0) |E\rangle\Bigr|^2 2\pi\delta(\omega + E - E') \geq 0.
\end{align}
}, one can obtain the Lindblad equation \eqref{eq:Lindblad} by taking an appropriate linear combination of $V_S^{(i)}(\omega)$ that diagonalizes and rescales $\gamma_{ij}(\omega)$ to $\delta_{ij}$.

Lastly, let us analyze the steady state solution of the Lindblad equation in the quantum optical regime when the environment is a thermal bath with temperature $T=1/\beta$. 
In this case, the Kubo-Martin-Schwinger (KMS) relation $\gamma_{ij}(\omega) = e^{\beta\omega}\gamma_{ji}(-\omega)$ is satisfied by the environment correlation functions (see Appendix \ref{app:Thermal_Corr} for details).
It is then reasonable to guess that $\rho_S\propto e^{-\beta H_S}$ is a steady state solution.
Indeed, by using similar properties as above
\begin{align}
e^{\beta H_S}V_S^{(j)}(\omega)e^{-\beta H_S} = e^{-\beta\omega}V_S^{(j)}(\omega), \quad
e^{\beta H_S}V_S^{(i)\dagger}(\omega)e^{-\beta H_S} = e^{\beta\omega}V_S^{(i)\dagger}(\omega),
\end{align}
and $V_S^{(i)\dagger}(\omega) = V_S^{(i)}(-\omega)$, we can show
\begin{subequations}
\begin{align}
V_S^{(j)}(\omega)e^{-\beta H_S}V_S^{(i)\dagger}(\omega)
&= e^{-\beta\omega}e^{-\beta H_S}V_S^{(j)}(\omega)V_S^{(i)\dagger}(\omega) 
= e^{-\beta\omega}e^{-\beta H_S}V_S^{(j)\dagger}(-\omega)V_S^{(i)}(-\omega), \\
V_S^{(i)\dagger}(\omega)V_S^{(j)}(\omega)e^{-\beta H_S}
&= e^{-\beta H_S}V_S^{(i)\dagger}(\omega)V_S^{(j)}(\omega).
\end{align}
\end{subequations}
Putting all these together, the right hand side of the Lindblad equation is shown to vanish because
\begin{align}
\sum_{\omega}\sum_{i,j}\gamma_{ij}(\omega)V_S^{(j)}(\omega)e^{-\beta H_S}V_S^{(i)\dagger}(\omega)
&=\sum_{\omega}\sum_{i,j}\gamma_{ij}(\omega)e^{-\beta\omega}e^{-\beta H_S}V_S^{(j)\dagger}(-\omega)V_S^{(i)}(-\omega) \nonumber\\
&=\sum_{\omega}\sum_{i,j}\gamma_{ji}(-\omega)e^{-\beta H_S}V_S^{(j)\dagger}(-\omega)V_S^{(i)}(-\omega)\nonumber\\
&=\sum_{\omega}\sum_{i,j}\gamma_{ij}(\omega)e^{-\beta H_S}V_S^{(i)\dagger}(\omega)V_S^{(j)}(\omega),
\end{align}
and $\Delta H_S$ and $H_S$ commute.
As is naturally expected, it is proven that the Boltzmann distribution $\rho_S\propto e^{-\beta H_S}$ is a steady state solution of the Lindblad equation in the quantum optical regime.

\subsubsection{Quantum Brownian motion}
\label{sec:Basics_Approx_QBM}
There is another interesting regime of the open quantum system, namely the quantum Brownian motion.
Most of the examples in Sections \ref{sec:NRQCD} and \ref{sec:pNRQCD} are in this regime.
The master equation in the Born-Markov approximation \eqref{eq:master_BM} can be further approximated by using the fact that the system's intrinsic time scale is long compared to the environment correlation time $\tau_S\gg \tau_E$.
See Fig.~\ref{fig:oqs_regimes} for a physical picture corresponding to this condition.
The explicit form of \eqref{eq:master_BM} is
\begin{align}
\label{eq:master_BM_explicit}
\frac{d}{dt}\rho_S(t) &=
\int_0^{\infty} ds \sum_{i,j} {\rm Tr}_E\left(\rho_E(0)V_E^{(i)}(t)V_E^{(j)}(t-s)\right) \nonumber \\
&\quad \times \left[
V_S^{(j)}(t-s)\rho_S(t) V_S^{(i)}(t) 
-V_S^{(i)}(t)  V_S^{(j)}(t-s)\rho_S(t)
\right] \quad + \quad h.c..
\end{align}
The environment correlation function takes finite values only for short time $\tau_E \ll \tau_S$, so that the system operators can be approximated by the gradient expansion:
\begin{align}
V_S^{(i)}(t-s) &\simeq V_S^{(i)}(t) - s \dot V_S^{(i)}(t) + \cdots 
=V_S^{(i)}(t) - is \left[H_S, V_S^{(i)}(t)\right] + \cdots.
\end{align}
Since this time scale hierarchy is typically realized in the Brownian motion, this regime of the open quantum system is called the {\it quantum Brownian motion}.
The master equation for the quantum Brownian motion was first derived in the well-known paper by Caldeira and Leggett \cite{caldeira1983path} using the influence functional method \cite{feynman1963theory}, which is the path-integral formulation of the open quantum systems.
In the Caldeira-Leggett master equation, the gradient expansion terminates at the first order.
As will be clear in a specific example of quantum Brownian motion, the first-order gradient term represents quantum dissipation by recoil of the Brownian particle during a collision.
Therefore, truncating the gradient expansion at leading order is called the {\it recoilless limit}.

Introducing the environment correlation functions with zero frequency:
\begin{subequations}
\begin{align}
\Gamma_{ij}(\omega=0)
&=\int_0^{\infty} ds{\rm Tr}_E\left(\rho_E(0)V_E^{(i)}(s)V_E^{(j)}(0)\right)
\equiv \frac{1}{2}\gamma_{ij} + iS_{ij}, \quad 
\gamma_{ij}^*=\gamma_{ji}, \quad
S_{ij}^*=S_{ji},\\
-i\frac{d}{d\omega}\Gamma_{ij}(\omega)\Bigr|_{\omega=0}
&=\int_0^{\infty} ds \
s \ {\rm Tr}_E\left(\rho_E(0)V_E^{(i)}(s)V_E^{(j)}(0)\right) \equiv \eta_{ij},
\end{align}
\end{subequations}
the master equation is
\begin{align}
\label{eq:master_QBM}
\frac{d}{dt}\rho_S(t) =
\sum_{i,j}\left[
\begin{aligned}
&\gamma_{ij}\left(
V_S^{(j)}(t)\rho_S(t) V_S^{(i)}(t)
- \frac{1}{2}\left\{V_S^{(i)}(t)  V_S^{(j)}(t),\rho_S(t)\right\}
\right) \\
&-iS_{ij}\left(
V_S^{(i)}(t) V_S^{(j)}(t)\rho_S(t)
- \rho_S(t) V_S^{(i)}(t) V_S^{(j)}(t)
\right) \\
&-\eta_{ij}\left(
\dot V_S^{(j)}(t)\rho_S(t) V_S^{(i)}(t) - V_S^{(i)}(t)\dot V_S^{(j)}(t)\rho_S(t)
\right) \\
&-\eta_{ij}^*\left(
V_S^{(i)}(t)\rho_S(t) \dot V_S^{(j)}(t) - \rho_S(t)\dot V_S^{(j)}(t)V_S^{(i)}(t)
\right)
\end{aligned}
\right].
\end{align}
Here again it is assumed that the environment is invariant under the time translation.

Let us assume $\gamma_{ij}$ is positive definite (not just semi-definite), so that it has an inverse matrix $\gamma^{-1}_{ij}$.
Then, up to the first-order gradient expansion, the master equation is equivalent to the following Lindblad equation:
\begin{subequations}
\label{eq:Lindblad_QBM}
\begin{align}
\frac{d}{dt}\rho_S(t) &= -i\left[\Delta H_S(t), \rho_S(t) \right]
+\sum_{i,j}\gamma_{ij}\left(
\tilde V_S^{(j)}(t)\rho_S(t)\tilde V_S^{(i)\dagger}(t)
-\frac{1}{2}\left\{\tilde V_S^{(i)\dagger}(t)\tilde V_S^{(j)}(t),\rho_S(t)\right\}
\right),\\
\Delta H_S(t)&\equiv\sum_{i,j} S_{ij}V_S^{(i)}(t)V_S^{(j)}(t)
+\frac{i}{2}\sum_{i,j}\left(
\eta_{ij}V_S^{(i)}(t)\dot V_S^{(j)}(t)
-\eta_{ij}^*\dot V_S^{(j)}(t)V_S^{(i)}(t)
\right),\\
\tilde V_S^{(i)}(t)&\equiv V_S^{(i)}(t)- \sum_{jk}\gamma^{-1}_{ij}\eta_{jk}\dot V_S^{(k)}(t), \quad
\tilde V_S^{(i)\dagger}(t)= V_S^{(i)}(t)-\sum_{jk}\left(\eta_{jk}^*\dot V_S^{(k)}(t)\right) \gamma^{-1}_{ji}.
\end{align}
\end{subequations}
In the Schr\"odinger picture, all the operators in the above Lindblad equation becomes independent of time and the Hamiltonian term receives modification $-i\left[H_S + \Delta H_S, \rho_S(t)\right]$ due to the coupling to the environment.
Note that $\dot V_S^{(i)} \equiv i \left[H_S, V_S^{(i)}\right]$ is still well-defined and independent of time in the Schr\"odinger picture.
In the Lindblad equation \eqref{eq:Lindblad_QBM}, the gradient expansion is not strict in the sense that only some of the second order terms are included in the master equation.
The Caldeira-Leggett master equation, which is obtained in the strict first-order gradient expansion, is not in the Lindblad form as is clear from this argument.
The Lindblad extension of the Caldeira-Leggett master equation was obtained similarly by adding some of the second order terms \cite{diosi1993high, diosi1993calderia, gao1997dissipative,vacchini2000completely}.

When the environment is a thermal bath $\rho_E(0)=\rho_E^{\rm th}$ and the $V_E^{(i)}$s have the same sign under the time-reversal transformation, the spectral density
\begin{align}
\sigma_{ij}(\omega)&\equiv
\int_{-\infty}^{\infty}dt e^{i\omega t}{\rm Tr}_E
\left(\rho_E^{\rm th} \left[V_E^{(i)}(t), V_E^{(j)}(0)\right]\right)
\end{align}
is shown to be real, odd in $\omega$, and symmetric in the indices.
Then, the coefficient matrices $\gamma_{ij}$, $S_{ij}$, and $\eta_{ij}$ are related to the spectral density as follows (see Appendix \ref{app:Thermal_Corr} for details):
\begin{subequations}
\begin{align}
\gamma_{ij} &= T\frac{d\sigma_{ij}}{d\omega}\Bigr|_0, \quad
S_{ij} = -\frac{1}{2}\int_{-\infty}^{\infty}\frac{d\omega}{2\pi} \frac{\sigma_{ij}(\omega)}{\omega}, \\
\eta_{ij} &=-\frac{1}{2}\int_{-\infty}^{\infty}\frac{d\omega}{2\pi}\frac{1}{\omega}
\frac{d}{d\omega}\left[
\coth\left(\frac{\beta\omega}{2}\right)\sigma_{ij}(\omega) \right]
-i\frac{1}{4}\frac{d\sigma_{ij}}{d\omega}\Bigr|_0.
\end{align}
\end{subequations}
Let us emphasize here that $\gamma_{ij}$ and ${\rm Im} \eta_{ij}$ are given by a common transport coefficient $d\sigma_{ij}/d\omega|_0$.

Finally, let us make a remark on an approximation ${\rm Re}\eta_{ij}=0$ which is often used in the literature.
By the assumption that the environment correlation time is $\tau_E$, the spectral density $\sigma_{ij}(\omega)$ is cut off at some frequency $\Omega\sim \tau_E^{-1}$, for example by the Lorentz-Drude cutoff
\begin{align}
\sigma_{ij}(\omega) \simeq \omega\frac{d\sigma_{ij}}{d\omega}\Bigr|_0\frac{\Omega^2}{\omega^2 + \Omega^2}.
\end{align}
Then, ${\rm Re}\eta_{ij}$ is roughly evaluated as
\begin{align}
-\frac{1}{2}\int_{-\infty}^{\infty}\frac{d\omega}{2\pi}\frac{1}{\omega}
\frac{d}{d\omega}\left[
\coth\left(\frac{\beta\omega}{2}\right)\sigma_{ij}(\omega) \right]
&=-\frac{1}{2}\frac{d\sigma_{ij}}{d\omega}\Bigr|_0\int_{-\infty}^{\infty}\frac{d\nu}{2\pi}\frac{1}{\nu}
\frac{d}{d\nu}\left[
\coth\left(\frac{\beta\Omega\nu}{2}\right)\frac{\nu}{\nu^2+1} \right] \nonumber \\
&\sim \frac{\gamma_{ij}}{T}\times \left\{\begin{aligned}
&(\beta\Omega)^{-1} &(\beta\Omega\ll 1)\\
&\quad 1 &(\beta\Omega\sim 1)
\end{aligned}\right\} \sim \gamma_{ij} \tau_E
\end{align}
by noting that the integrand is cutoff at $|\nu|\sim 1$.
In Eq.~\eqref{eq:master_QBM}, the terms containing ${\rm Re}\eta_{ij}$ appear only in the combination of
\begin{align}
\sum_j\left(\gamma_{ij} V_S^{(j)}(t) - 2{\rm Re}\eta_{ij} \dot V_S^{(j)}(t)\right)
\sim \sum_j \gamma_{ij} \left(1 - \frac{\tau_E}{\tau_S}\right) V_S^{(j)}(t).
\end{align}
Since $\tau_E/\tau_S\ll 1$ by the assumption of quantum Brownian motion, ${\rm Re}\eta_{ij}$ only gives rise to Hermitian correction to the system-environment coupling.
Therefore we can neglect ${\rm Re}\eta_{ij}$ and substitute $\eta_{ij}\approx -\frac{i}{4T}\gamma_{ij}$ in the master equation \eqref{eq:master_QBM} and the Lindblad equation \eqref{eq:Lindblad_QBM}.
The Lindblad operator in this approximation is
\begin{align}
\label{eq:Lindblad_op_qbm}
\tilde V_S^{(i)}(t)&\simeq V_S^{(i)}(t) + \frac{i}{4T}\dot V_S^{(i)}(t), \quad
\tilde V_S^{(i)\dagger}(t)\simeq V_S^{(i)}(t)-\frac{i}{4T}\dot V_S^{(i)}(t),
\end{align}
where the gradient expansion introduces anti-Hermitian correction to the system-environment coupling, which is Hermitian.
Notice here that the ratio of the first two terms in \eqref{eq:Lindblad_op_qbm} is $\dot V_S^{(i)}/TV_S^{(i)} \sim 1/T\tau_S$ and is small because $\tau_S\gg 1/T$ holds typically.
However, this ratio derives from the KMS relation and does not reflect the smallness of $\tau_E/\tau_S$.
Therefore, although it might sound slightly confusing, this is the leading order result in the limit $\tau_E/\tau_S\ll 1$.
There are corrections of order $\tau_E/\tau_S$ to both Hermitian and anti-Hermitian parts of $\tilde V_S^{(i)}$, which may come from $\Gamma_{ij}(\omega)$ at finite $\omega$ and from higher order gradient expansion of $V^{(i)}_S(t-s)$.

\subsubsection{Exceptional case for quantum Brownian motion}
\label{sec:Basics_Approx_exQBM}
In the section \ref{sec:pNRQCD_weak}, we encounter an exceptional example of spectral density with $d\sigma_{ij}/d\omega|_0=0$, which is the spectral density of free gluons ($V_E^{(i)}=\vec E^{a}$) at high temperatures.
Since $\gamma_{ij}=Td\sigma_{ij}/d\omega|_0=0$ in this case, the derivation of the Lindblad equation in the previous section \ref{sec:Basics_Approx} does not apply here.

We provide a general framework by which to derive the Lindblad equation in this class of exceptional cases.
Therefore, we assume that the spectral density is real, odd in $\omega$, and symmetric in the indices (because $\vec E^a$s are Hermitian and time reversal even).
By the assumption, the spectral density is approximated by
\begin{align}
\sigma_{ij}(\omega)\simeq 
\frac{1}{3!}\frac{d^3\sigma_{ij}}{d\omega^3}\Bigr|_0\omega^3 + \frac{1}{5!}\frac{d^5\sigma_{ij}}{d\omega^5}\Bigr|_0\omega^5 + \cdots .
\end{align}
for small $\omega$.
Then the real symmetric matrices $\gamma_{ij}(\omega)$ and $S_{ij}(\omega)$ are also expanded as
\begin{subequations}
\begin{align}
\gamma_{ij}(\omega) 
&\simeq \frac{1}{2!}\gamma_{ij}^{(2)}\omega^2 + \frac{1}{3!}\gamma_{ij}^{(3)}\omega^3 + \cdots
\simeq \frac{T}{3!}\frac{d^3\sigma_{ij}}{d\omega^3}\Bigr|_0 \omega^2\left(
1 + \frac{\omega}{2T}
\right) + \mathcal O(\omega^4),\\
S_{ij}(\omega)
&\simeq S_{ij}^{(0)} + S_{ij}^{(1)}\omega + \frac{1}{2!}S_{ij}^{(2)}\omega^2 + \cdots.
\end{align}
\end{subequations}
In the following analysis, we keep the expansion up to the second order in $\omega$ in order to accomodate the leading term of $\gamma_{ij}(\omega)$.

Starting point is again the master equation in the Born-Markov approximation \eqref{eq:master_BM_explicit}.
Up to the first order gradient expansion, the right hand side of \eqref{eq:master_BM_explicit}  is obtained by setting $\gamma_{ij}=0$, $S_{ij}=S_{ij}^{(0)}$, and $\eta_{ij}=S_{ij}^{(1)}$ in \eqref{eq:master_QBM}
\begin{align}
\frac{d}{dt}\rho_S(t)\Bigr|_{\partial^0, \partial^1}=
\sum_{i,j}\left[
\begin{aligned}
&-iS^{(0)}_{ij}\left(
V_S^{(i)}(t) V_S^{(j)}(t)\rho_S(t)
- \rho_S(t) V_S^{(i)}(t) V_S^{(j)}(t)
\right) \\
&-S^{(1)}_{ij}\left(
\begin{aligned}
&\dot V_S^{(j)}(t)\rho_S(t) V_S^{(i)}(t) - V_S^{(i)}(t)\dot V_S^{(j)}(t)\rho_S(t)\\
&+V_S^{(i)}(t)\rho_S(t) \dot V_S^{(j)}(t) - \rho_S(t)\dot V_S^{(j)}(t)V_S^{(i)}(t)
\end{aligned}
\right)
\end{aligned}
\right].
\end{align}
The second order expansion contains several new terms from $\gamma_{ij}^{(2)}$ and $S_{ij}^{(2)}$
\begin{align}
\frac{d}{dt}\rho_S(t)\Bigr|_{\partial^2}=
\sum_{i,j}\left[
\begin{aligned}
&-\frac{1}{2}\left(
\frac{1}{2}\gamma_{ij}^{(2)}
+iS_{ij}^{(2)}
\right)
\left(
\ddot V_S^{(j)}(t)\rho_S(t) V_S^{(i)}(t) - V_S^{(i)}(t)\ddot V_S^{(j)}(t)\rho_S(t)
\right)\\
&-\frac{1}{2}\left(
\frac{1}{2}\gamma_{ij}^{(2)}-iS^{(2)}_{ij}
\right)
\left(
V_S^{(i)}(t)\rho_S(t) \ddot V_S^{(j)}(t) - \rho_S(t)\ddot V_S^{(j)}(t)V_S^{(i)}(t)
\right)
\end{aligned}
\right].
\end{align}

It is not at all obvious that the above master equation is in the Lindblad form.
Actually, we make it a Lindblad equation by the following technique of {\it integration by parts}, which involves a redefinition of the density matrix \footnote{
In the influence functional, this procedure literally means the integration by parts and is used for example in \cite{diosi1993high, Akamatsu:2014qsa}.
The density matrix needs to be redefined because of the surface term in the partial integration.
}.
Let us suppose that a master equation is derived perturbatively (in the interaction picture) and contains a term such as
\begin{align}
\frac{d}{dt}\rho_S(t) = A(t)\rho_S(t) \dot B(t) + \cdots + \mathcal O(V_S^3), \quad
A(t)\sim B(t)\sim\mathcal O(V_S).
\end{align}
We then make a redefinition of the density matrix
\begin{align}
\bar\rho_S(t) \equiv \rho_S(t) - A(t)\rho_S(t) B(t), \quad
\rho_S(t)\simeq \bar\rho_S(t) + \mathcal O(V_S^2).
\end{align}
The master equation for the new density matrix $\bar\rho_S(t)$ is obtained up to $\mathcal O(V_S^2)$ accuracy,
\begin{align}
\frac{d}{dt}\bar\rho_S(t) = -\dot A(t)\bar\rho_S(t)B(t) + \cdots + \mathcal O(V_S^3).
\end{align}
In the same way, we can freely move the time derivatives on the right hand side of the master equation to get $A(t)\dot B(t)\rho_S(t) \to -\dot A(t) B(t)\bar\rho_S(t)$ and so on.
Using this transformation and renaming $\bar\rho_S(t)$ as $\rho_S(t)$, the master equation (in the interaction picture) is rewritten in the Lindblad form
if $\gamma_{ij}^{(2)}$ is positive semi-definite:
\begin{subequations}
\label{eq:Lindblad_exQBM}
\begin{align}
\frac{d}{dt}\rho_S(t) &= -i\left[\Delta H_S(t), \rho_S(t) \right]
+\sum_{i,j}\frac{1}{2}\gamma_{ij}^{(2)}\left(
\dot V_S^{(j)}(t)\rho_S(t)\dot V_S^{(i)}(t)
-\frac{1}{2}\left\{\dot V_S^{(i)}(t)\dot V_S^{(j)}(t),\rho_S(t)\right\}
\right),\\
\Delta H_S(t)&\equiv\sum_{i,j}\left(
S_{ij}^{(0)}V_S^{(i)}(t)V_S^{(j)}(t)
+\frac{i}{2}
S_{ij}^{(1)}\left[V_S^{(i)}(t), \dot V_S^{(j)}(t)\right]
+\frac{1}{2}S_{ij}^{(2)}\dot V_S^{(i)}(t)\dot V_S^{(j)}(t)
\right).
\end{align}
\end{subequations}

\subsection{Concluding remarks of Section \ref{sec:Basics}}
In this section, the basics of open quantum system is first reviewed in Sec.~\ref{sec:Basics_Lindblad}, such as completely positive and trace preserving (CPTP) map, effects of initial system-environment correlation, Lindblad equation in the Markov limit, and the steady state solutions.
Then, using the approximation methods, the Lindblad equations are derived from microscopic theory depending on the regimes of time scale hierarchy.
Let us summarize here the obtained equations when the system-environment coupling is weak and is given by $\sum_i V_S^{(i)}\otimes V_E^{(i)}$.
Below, the Schr\"odinger picture is adopted for the time dependence.
\begin{itemize}
\item Lindblad equation in the quantum optical limit ($\tau_R\gg \tau_E, \ \tau_R\gg\tau_S$)
\begin{align}
\frac{d}{dt}\rho_S(t) &= -i\left[H_S + \Delta H_S, \rho_S(t) \right] + \sum_{\omega}\sum_{i,j}
\gamma_{ij}(\omega)\left[V_S^{(j)}(\omega) \rho_S(t)V_S^{(i)\dagger}(\omega)
- \frac{1}{2} \left\{V_S^{(i)\dagger}(\omega)V_S^{(j)}(\omega), \rho_S(t)\right\}\right] , \nonumber\\
\Delta H_S &\equiv \sum_{\omega}\sum_{i,j}S_{ij}(\omega)V_S^{(i)\dagger}(\omega)V_S^{(j)}(\omega), \quad
V_S^{(i)}(\omega) \equiv \sum_{\epsilon'-\epsilon=\omega}\Pi(\epsilon) V_S^{(i)}\Pi(\epsilon'), \nonumber \\
\Gamma_{ij}(\omega)
&= \int_0^{\infty} ds e^{i\omega s}{\rm Tr}_E\left(\rho_E(0)V_E^{(i)}(s)V_E^{(j)}(0)\right)
\equiv\frac{1}{2}\gamma_{ij}(\omega) + iS_{ij}(\omega). \nonumber
\end{align}
\begin{itemize}
\item $\gamma_{ij}(\omega)$ and $S_{ij}(\omega)$ are Hermitian, $\gamma_{ij}(\omega)$ is positive semi-definite.
\end{itemize}
\item Lindblad equation in the quantum Brownian regime ($\tau_R\gg \tau_E, \ \tau_S\gg\tau_E$)
\begin{align}
\frac{d}{dt}\rho_S(t) &= -i\left[H_S + \Delta H_S, \rho_S(t) \right]
+\sum_{i,j}\gamma_{ij}\left(
\tilde V_S^{(j)}\rho_S(t)\tilde V_S^{(i)\dagger}
-\frac{1}{2}\left\{\tilde V_S^{(i)\dagger}\tilde V_S^{(j)},\rho_S(t)\right\}
\right),\nonumber \\
\Delta H_S&\equiv\sum_{i,j} S_{ij}V_S^{(i)}V_S^{(j)}
+\frac{1}{8T}\sum_{i,j}\gamma_{ij}\left\{V_S^{(i)}, \dot V_S^{(j)}\right\}, \quad \tilde V_S^{(i)}\equiv V_S^{(i)} + \frac{i}{4T}\dot V_S^{(i)}, \quad
\dot V_S^{(i)}\equiv i\left[H_S, V_S^{(i)}\right], \nonumber \\
\Gamma_{ij}(0) &=\int_0^{\infty} ds{\rm Tr}_E\left(\rho_E(0)V_E^{(i)}(s)V_E^{(j)}(0)\right)
\equiv \frac{1}{2}\gamma_{ij} + iS_{ij}. \nonumber
\end{align}
\begin{itemize}
\item Assumption: $V_E^{(i)}$s have the same sign under time reversal.
\item $\gamma_{ij}$ and $S_{ij}$ are real and symmetric, $\gamma_{ij}$ is positive semi-definite.
\end{itemize}
\item Lindblad equation in the quantum Brownian regime with $\gamma_{ij}=0$
\begin{align}
\frac{d}{dt}\rho_S(t) &= -i\left[H_S + \Delta H_S, \rho_S(t) \right]
+\sum_{i,j}\frac{1}{2}\gamma_{ij}^{(2)}\left(
\dot V_S^{(j)}\rho_S(t)\dot V_S^{(i)}
-\frac{1}{2}\left\{\dot V_S^{(i)}\dot V_S^{(j)},\rho_S(t)\right\}
\right),\nonumber \\
\Delta H_S&\equiv\sum_{i,j}\left(
S_{ij}^{(0)}V_S^{(i)}V_S^{(j)}
+\frac{i}{2}
S_{ij}^{(1)}\left[V_S^{(i)}, \dot V_S^{(j)}\right]
+\frac{1}{2}S_{ij}^{(2)}\dot V_S^{(i)}\dot V_S^{(j)}
\right), \nonumber \\
\gamma_{ij} (\omega) &\simeq \frac{1}{2!}\gamma_{ij}^{(2)}\omega^2 + \cdots, \quad
S_{ij}(\omega)\simeq S_{ij}^{(0)} + S_{ij}^{(1)}\omega + \frac{1}{2!}S_{ij}^{(2)}\omega^2 + \cdots. \nonumber
\end{align}
\begin{itemize}
\item Assumption: $V_E^{(i)}$s have the same sign under time reversal.
\item $\gamma_{ij}^{(2)}$ and $S_{ij}^{(0,1,2)}$ are real and symmetric, $\gamma_{ij}^{(2)}$ is assumed to be positive semi-definite.
\end{itemize}
\end{itemize}

\newpage
\section{Lindblad equations from non-relativistic QCD (NRQCD)}
\label{sec:NRQCD}
In this section, I first review in Sec.~\ref{sec:NRQCD_Review} the basics of an effective field theory (EFT) for non-relativistic heavy quarks: non-relativistic QCD (NRQCD).
Then I will introduce how to derive Lindblad equations from NRQCD in the general framework of Sec.~\ref{sec:Basics}, in particular that of the quantum Brownian motion.
Before discussing the quantum Brownian motion of quarkonium, I first give introductory remarks on the quantum Brownian motion in general in Sec.~\ref{sec:NRQCD_HQ} by using a simpler example of quantum Brownian motion of single heavy quark.
In Sec.~\ref{sec:NRQCD_QQbar}, I will illustrate various aspects of the quantum Brownian motion of quarkonium by taking several interesting limits such as recoilless limit (Sec.~\ref{sec:NRQCD_QQbar_Stochastic}), static limit (Sec.~\ref{sec:NRQCD_QQbar_Static}), classical limit (Sec.~\ref{sec:NRQCD_QQbar_CL}), and small dipole limit (Sec.~\ref{sec:NRQCD_QQbar_Dipole}).

\subsection{A short review of non-relativistic QCD (NRQCD)}
\label{sec:NRQCD_Review}
The NRQCD is an effective description of almost on-shell heavy quarks, typically of heavy quark-antiquark pair, based on the existence of a rest frame where the motion of heavy quarks stays non-relativistic.
Such a rest frame is guaranteed to exist when the heavy quark mass is much larger than the strength of the (quantum and thermal) fluctuations in the environment ($M\gg \Lambda_{\rm QCD}, T$), so that the velocity of the heavy quarks stays small ($v\ll 1$).
The NRQCD Lagrangian is defined in a standard manner of EFT construction:
\begin{itemize}
\item The Lagrangian is systematically expanded by the inverse of heavy quark mass $1/M$, with a power counting scheme in terms of heavy quark velocity $v$, and respecting the symmetries of QCD such as color SU(3) gauge symmetry, up to a desired precision one needs to accomplish.
\item The expansion coefficients, or the {\it Wilson coefficients}, are determined by matching the Green's functions calculated by QCD and NRQCD at the ultraviolet scale ($\mu_{\rm NR}$) of the latter.
\end{itemize}
The degrees of freedom in NRQCD are non-relativistic heavy quarks represented by the two-component Pauli spinors ($\psi$ for heavy quark annihilation and $\chi$ for heavy antiquark creation), light quarks ($q$), and gluons ($A$), and all of these are below the cutoff scale $p^{\mu} < \mu_{\rm NR}$.
Since the typical energy and momentum of the heavy quarks are $E\sim Mv^2\ll M$ and $p\sim Mv\ll M$, we can put a cutoff scale at $Mv, \Lambda_{\rm QCD}, T \ll \mu_{\rm NR}\ll M$ where perturbative matching between QCD and NRQCD is available.
With the above considerations, the NRQCD Lagrangian is
\begin{align}
\mathcal L_{\rm NRQCD}
=\mathcal L_{q+A} 
+\psi^{\dagger} \left[iD_t + \frac{\vec D^2}{2M} + \frac{c_F\vec \sigma\cdot  g\vec B}{2M}\right]\psi
+\chi^{\dagger} \left[iD_t - \frac{\vec D^2}{2M} - \frac{c_F\vec \sigma\cdot  g\vec B}{2M}\right]\chi
+\mathcal O(1/M^2),
\end{align}
where $\mathcal L_{q+A}$ is the QCD Lagrangian for light quarks and gluons, $D_{\mu}\equiv \partial_{\mu} + igA^a_{\mu}t_F^a$ is the covariant derivatives for fundamental quarks, and $c_F(\mu_{\rm NR})=1 + \mathcal O(\alpha)$ is the Wilson coefficient for the spin-color magnetic coupling (the subscript $F$ for $c_F$ comes from Fermi contact interaction).
The $\mathcal O(1/M^2)$ terms consist of the spin-orbit and Darwin terms as well as higher order expansions of the kinetic energy in the heavy quark bilinears, four-fermi contact interactions between the heavy quark pair including their inclusive annihilation, and corrections to the light particle sector due to the heavy quark loops.
Note that the matching is done at $\mu_{\rm NR}\gg T$, so that there is no temperature dependence in the Wilson coefficients.

Establishing a power counting scheme is of fundamental importance because it provides predictive power to an effective field theory with a non-renormalizable Lagrangian.
A beautiful application of NRQCD power counting in terms of heavy quark velocity is the factorization formula for the inclusive decay widths of quarkonia \cite{Bodwin:1994jh}.
There is a technical subtlety in maintaining the velocity power counting at loop orders, which is resolved by the prescription given in \cite{Manohar:1997qy}.
For further details and applications of the NRQCD, see the reviews \cite{brambilla2005effective, petrov2015effective}.

Let us now estimate the relevant scales of heavy quarks and gluons to derive power counting schemes in terms of heavy quark velocity.
Here, we restrict ourselves to considering two extreme cases: (i) quarkonium in a Coulombic bound state and (ii) heavy quark pair far apart and in kinetic equilibrium.
The physical motivation of (ii) is to confirm that the NRQCD can describe the dissociation of heavy quark pair in the quark-gluon plasma.
Since our primary interest is in a quarkonium in the quark-gluon plasma, we implicitly assume that there is no confining force between the heavy quark pair.
We adopt the Coulomb gauge ($\vec\nabla \cdot \vec A = 0$) in which the power counting scheme becomes particularly simple.

For the case (i), let us quote the result \cite{Lepage:1992tx}
\begin{subequations}
\begin{align}
\partial_i &\sim Mv, \quad \partial_t\sim gA_0 \sim Mv^2, \quad v\sim g^2/4\pi \ll 1, \\
\psi&\sim\chi\sim (Mv)^{3/2}, \quad gA_i \sim Mv^3.
\end{align}
\end{subequations}
The first line is easily confirmed by the solution of non-relativistic bound state problem in the Coulomb potential.
The scale of heavy quark spinors is estimated by the normalization condition ($\int_x |\psi|^2=\int_x |\chi|^2 = 1$).
The strength of the transverse gauge field is perturbatively estimated by solving $\Box \vec A = \vec j_T$, where $\vec j_T$ is the transverse color current in the bound states.
With this power counting, the gauge-invariant NRQCD Lagrangian starts from $\sim Mv^2 \cdot (Mv)^3$
\begin{align}
\label{eq:NRQCD_LO}
\mathcal L_{\rm NRQCD}
=\mathcal L_{q+A} 
+\psi^{\dagger} \left[iD_t + \frac{\vec D^2}{2M} \right]\psi
+\chi^{\dagger} \left[iD_t - \frac{\vec D^2}{2M} \right]\chi
+\cdots,
\end{align}
where spin-color magnetic coupling contributes as $\mathcal O(v^2)$ correction to the leading terms and is dropped hereafter.
For the case (ii), heavy quark scales are typically
\begin{align}
\partial_t&\sim Mv^2\sim T, \quad \partial_i \sim Mv, \quad v\sim \sqrt{\frac{T}{M}}\ll 1,
\end{align}
while the thermal fluctuations of gauge fields are simply estimated as $A_{\mu}\sim T,$ $F_{\mu\nu}\sim T^2$.
We do not and need not specify the scale of heavy quark spinors $\psi$ and $\chi$ for an unbound heavy quark pair.
By keeping the leading terms of light particle and heavy quark sectors, the spin-color magnetic coupling does not contribute and we get the same Lagrangian \eqref{eq:NRQCD_LO} in this case.
In both cases (i) and (ii), the transverse gauge field $A_i$ is subdominant to the heavy quark canonical momentum $\partial_i$ and in the analysis below we approximate $D_i\simeq\partial_i$.

In order to apply the formula derived in the Section~\ref{sec:Basics} to NRQCD, the Hamiltonian formalism is more convenient.
The creation or annihilation of heavy quark pair occurs only at the order $\mathcal O(1/M^2)$ and is neglected in the NRQCD Lagrangian at the leading order, which allows us to formulate the heavy quark dynamics by means of non-relativistic quantum mechanics.
The Hamiltonian for the Fock state containing a heavy quark pair is thus
\begin{align}
H=\frac{p_Q^2}{2M} + gA^a_0(\vec x_Q)t^a_Q + \frac{p_{Q_c}^2}{2M} - gA^a_0(\vec x_{Q_c})t^{a*}_{Q_c},
\end{align}
where $(\vec x_Q, \vec p_Q, t_Q^a)$ and $(\vec x_{Q_c}, \vec p_{Q_c}, t_{Q_c}^{a*})$ are operators for the heavy quark and antiquark respectively.
The matrices $t^a_Q$ and $-t^{a*}_{Q_c}$ are the color SU($N_c$) algebras in the fundamental representation and its complex conjugate.
To explicitly distinguish the system (heavy quarks) and the environment (light particles) as in Eq.~\eqref{eq:vonNeumann}, the total Hamiltonian $H_{\rm tot}$ is written as
\begin{align}
\label{eq:NRQCD_Hamiltonian}
H_{\rm tot} = \left(\frac{p_Q^2}{2M} + \frac{p_{Q_c}^2}{2M}\right)\otimes I_E
+ I_S \otimes H_{q+A}
+ \int d^3 x \left[
\delta(\vec x - \vec x_Q) t^a_Q - \delta(\vec x - \vec x_{Q_c}) t^{a*}_{Q_c}
\right] \otimes gA_0^a(\vec x),
\end{align}
where we add the Hamiltonian for the light particle sector $H_{q+A}$.
In the last term, $\vec x$ is just a label, whereas $\vec x_{Q}$ and $\vec x_{Q_c}$ are system operators and $A^a_0$ is an environment operator.

\subsection{Quantum Brownian motion of a heavy quark}
\label{sec:NRQCD_HQ}
In this section, we derive and analyze the Lindblad equation for quantum Brownian motion of a heavy quark in a weakly coupled quark-gluon plasma (QGP).
The scale hierarchy for the quantum Brownian motion ($\tau_R\gg \tau_E$ and $\tau_S\gg \tau_E$) is satisfied as follows.
The system relaxation time is estimated by its kinetic equilibration time $\tau_R\sim M/g^4 T^2$, the system time scale $\tau_S=\infty$, and the environment correlation time is the duration of ($t$-channel) collisions, that is $\tau^{\rm (soft)}_E\sim 1/gT$ for soft collisions with momentum transfer $\sim gT$ and $\tau^{\rm (hard)}_E\sim 1/T$ for hard collisions with momentum transfer $\sim T$.
Since $M\gg T$, this is the regime of the quantum Brownian motion.

\subsubsection{Lindblad equation}
\label{sec:NRQCD_HQ_Lindblad}
We apply the formula for quantum Brownian motion \eqref{eq:Lindblad_QBM} to single non-relativistic heavy quark in the QGP.
The total Hamiltonian in this case is
\begin{align}
H_{\rm tot} =\frac{p_Q^2}{2M} \otimes I_E + I_S \otimes H_{q+A}
+ \int d^3 x \left[\delta(\vec x - \vec x_Q) t^a_Q \right] \otimes gA_0^a(\vec x).
\end{align}
It is easy to see the operator correspondence
\begin{subequations}
\begin{align}
V_S^{(i)} &\leftrightarrow \delta(\vec x - \vec x_Q) t^a_Q \equiv V_S^a(\vec x), \\
\dot V_S^{(i)}=i[H_S, V_S^{(i)}] &\leftrightarrow \left[
-\frac{i}{2M}\nabla_x^2\delta(\vec x - \vec x_Q)
-\frac{1}{M}\vec\nabla_x\delta(\vec x - \vec x_Q)\cdot\vec p_Q
\right] t^a_Q \equiv \dot V_S^a(\vec x),
\end{align}
\end{subequations}
and the coefficients in the Lindblad equation
\begin{subequations}
\label{eq:gluon_corr}
\begin{align}
\gamma_{ab}(\vec x-\vec y) &= T\frac{d}{d\omega}\sigma_{ab}(\omega, \vec x -\vec y)\Bigr|_{\omega=0}, \quad
S_{ab}(\vec x -\vec y) = -\frac{1}{2}\int_{-\infty}^{\infty} \frac{d\omega}{2\pi}\frac{\sigma_{ab}(\omega, \vec x-\vec y)}{\omega}, \\
\sigma_{ab}(\omega, \vec x -\vec y) &\equiv
\int_{-\infty}^{\infty} dt e^{i\omega t} {\rm Tr}_E\left(\rho_E^{\rm th}\left[gA_0^a(t,\vec x), gA_0^b(0,\vec y)\right]\right)
\propto \delta_{ab}.
\end{align}
\end{subequations}
Analytic expressions for $\gamma_{ab}(\vec x)=\gamma(\vec x)\delta_{ab}$ and $S_{ab}(\vec x)=S(\vec x)\delta_{ab}$ at the soft scale $|\vec x|\sim 1/gT$ are given in the Appendix \ref{app:Thermal_Corr_Gluons}.
As explained at the end of Sec.~\ref{sec:Basics_Approx_QBM}, we can approximate using the hierarchy $\tau_E/\tau_S\ll 1$
\begin{align}
\label{eq:gluon_corr_eta}
\eta_{ab}(\vec x-\vec y)&\simeq -\frac{i}{4T}\gamma_{ab}(\vec x -\vec y).
\end{align}
Then, the Lindblad operator reads
\begin{align}
\tilde V_S^{(i)} &\leftrightarrow 
V_S^a(\vec x) + \frac{i}{4T} \dot V_S^a(\vec x) \nonumber \\
& \qquad =
\left[\delta(\vec x - \vec x_Q)
+\frac{1}{8MT}\nabla_x^2\delta(\vec x - \vec x_Q)
-\frac{i}{4MT}\vec\nabla_x\delta(\vec x - \vec x_Q)\cdot\vec p_Q\right]t^a_Q \equiv \tilde V_S^a(\vec x).
\end{align}
One can show that the correction to the Hamiltonian
\begin{align}
\Delta H_S &= \int d^3x\int d^3y \left[
S_{ab}(\vec x-\vec y)V_S^a(\vec x)V_S^b(\vec y)
+\frac{1}{8T}\gamma_{ab}(\vec x-\vec y) \left(V_S^a(\vec x)\dot V_S^b(\vec y) + \dot V_S^b(\vec y)V_S^a(\vec x)\right)
\right] 
\end{align}
is a constant and thus neglected, using a property $\vec\nabla\gamma_{ab}(\vec x=\vec 0)=0$ which follows from the rotational invariance.
Collecting these results, we obtain the Lindblad equation
\begin{align}
\label{eq:Lindblad_NRQCD_HQx}
\frac{d}{dt}\rho_S(t) = -i\left[H_S, \rho_S\right]
+\int_x\int_y \gamma(\vec x-\vec y)\left[
\tilde V_S^a(\vec y)\rho_S \tilde V_S^{a\dagger}(\vec x)
-\frac{1}{2}\left\{
\tilde V_S^{a\dagger}(\vec x)\tilde V_S^a(\vec y), \rho_S
\right\}
\right],
\end{align}
where $\gamma_{ab}(\vec x) = \gamma(\vec x)\delta_{ab}$ and $\int_x\equiv \int d^3x$.
Performing the Fourier transform\footnote{
As mentioned in Sec.~\ref{sec:Basics_Lindblad}, this is an example of the unitary transformation of the Lindblad operator basis.
}, the Lindblad operators are diagonalized:
\begin{subequations}
\label{eq:Lindblad_NRQCD_HQk}
\begin{align}
\frac{d}{dt}\rho_S(t) &= -i\left[H_S, \rho_S\right]
+\int_k \gamma(\vec k)\left[
\tilde V_S^a(\vec k)\rho_S \tilde V_S^{a\dagger}(\vec k)
-\frac{1}{2}\left\{
\tilde V_S^{a\dagger}(\vec k)\tilde V_S^a(\vec k), \rho_S
\right\}
\right],\\
\tilde V_S^{a}(\vec k) &= e^{i\vec k\cdot\vec x_Q/2}\left(1-\frac{\vec k\cdot\vec p_Q}{4MT}\right)e^{i\vec k\cdot\vec x_Q/2} t^a_Q,
\end{align}
\end{subequations}
where $\int_k\equiv\int d^3k/(2\pi)^3$ and we use $f(\vec k)\equiv \int_x e^{i\vec k\cdot\vec x}f(\vec x)$ to denote the Fourier transform of $f(\vec x)$ in general.
In this diagonal form, physical meaning of the Lindblad equation becomes clear as we discuss below.
The Lindblad equation is further simplified by taking the trace in the internal color space for the heavy quark $\bar\rho_S(t)\equiv {\rm Tr}_{\rm color}\rho_S(t)$:\footnote{
We believe that readers are not confused with the same notation for a different meaning in the last part of Sec.~\ref{sec:Basics_Approx_exQBM}
}
\begin{subequations}
\label{eq:Lindblad_NRQCD_HQk_traced}
\begin{align}
\frac{d}{dt}\bar\rho_S(t) &= -i\left[H_S, \bar\rho_S\right]
+\int_k C_F\gamma(\vec k)\left[
\tilde V_S(\vec k)\bar\rho_S \tilde V_S^{\dagger}(\vec k)
-\frac{1}{2}\left\{
\tilde V_S^{\dagger}(\vec k)\tilde V_S(\vec k), \bar\rho_S
\right\}
\right],\\
\tilde V_S(\vec k) &= e^{i\vec k\cdot\vec x_Q/2}\left(1-\frac{\vec k\cdot\vec p_Q}{4MT}\right)e^{i\vec k\cdot\vec x_Q/2},
\end{align}
\end{subequations}
where $C_F = (N_c^2-1)/2N_c$ is the fundamental Casimir of SU($N_c$).

Let us see the physical meaning of the Lindblad equation.
At the leading order in the gradient expansion, the Lindblad operator $e^{i\vec k \cdot\vec x_Q}t^a_Q$ shifts the heavy quark momentum by $\vec k$ and rotates the color by $t^a_Q$.
Corresponding microscopic process is a scattering $Q+g\to Q+g$ or $Q+q (\bar q)\to Q+q(\bar q)$ with momentum transfer $\vec k$.
The rate for this process is $\gamma(\vec k)$ and equals the rate for the inverse process $\gamma(-\vec k)=\gamma(\vec k)$.
Including the next-to-leading order, the Lindblad operator is $e^{i\vec k\cdot\vec x_Q/2}\left(1-\frac{\vec k\cdot\vec p_Q}{4MT}\right)e^{i\vec k\cdot\vec x_Q/2} t^a_Q = e^{i\vec k\cdot\vec x_Q}\left(1-\frac{\vec k\cdot\vec p_Q}{4MT} - \frac{k^2}{8MT}\right) t^a_Q$ and improves on the description of the scattering event.
At this order, the Lindblad operator shifts and rotates the momentum and color of the heavy quark, but now with momentum dependent rates
\begin{subequations}
\begin{align}
\Gamma_{\vec p_Q\to \vec p_Q + \vec k}&=\gamma(\vec k)\left(1-\frac{\vec k\cdot\vec p_Q}{4MT} - \frac{k^2}{8MT}\right)^2
=\gamma(\vec k)\left(
1+\frac{E_{\vec p_Q} - E_{\vec p_Q+\vec k}}{4T}
\right)^2, \\
\Gamma_{\vec p_Q + \vec k \to \vec p_Q}&=\gamma(\vec k)\left(1+\frac{\vec k\cdot\vec p_Q}{4MT} + \frac{k^2}{8MT}\right)^2
=\gamma(\vec k)\left(
1+\frac{E_{\vec p_Q+\vec k} - E_{\vec p_Q}}{4T}
\right)^2,
\end{align}
\end{subequations}
with $E_{\vec p} \equiv p^2/2M$.
Using an approximation for small $x$ (relative error is less than $3\%$ for $|x|<0.25$)
\begin{align}
\left(\frac{1+x}{1-x}\right)^2 = 1+4x+8x^2 + 12 x^3 + \cdots \simeq e^{4x} + \mathcal O(x^3),
\end{align}
the detailed balance holds approximately
\begin{align}
\frac{\Gamma_{\vec p_Q\to \vec p_Q + \vec k}}{\Gamma_{\vec p_Q + \vec k \to \vec p_Q}}
\simeq \exp\left(\frac{E_{\vec p_Q} - E_{\vec p_Q+\vec k}}{T}\right),
\end{align}
so that the equilibrium distribution is very close to the Boltzmann distribution.
The next-to-leading order term includes recoil effect of the heavy quark and thus the leading order approximation is often referred to as the {\it recoilless limit}.
Note that without the next-to-leading order term, the detailed balance does not hold even approximately because the inverse process takes place at the same rate.
In other words, the detailed balance for unphysical temperature $T=\infty$ holds and the heavy quark system continues to heat up.
Therefore the master equation in the leading order gradient expansion, or in the {\it recoilless limit}, can only describe the heavy quark evolution in a shorter time scale than its relaxation.

The Lindblad equation \eqref{eq:Lindblad_NRQCD_HQk} for a heavy quark in the QGP was first derived in \cite{Akamatsu:2014qsa}.
Its Abelian limit, or equivalently \eqref{eq:Lindblad_NRQCD_HQk_traced}, was first numerically simulated in 1 dimension in \cite{Akamatsu:2018xim} using the Quantum State Diffusion method \cite{gisin1992quantum,percival1998quantum}.
The equilibrium distribution is found to be very close to the Boltzmann distribution.
Also, it is confirmed that the heavy quark energy continues to rise without the dissipation term.

\subsubsection{Decoherence vs dissipation}
\label{sec:NRQCD_HQ_DecDiss}
Let us examine how the system density matrix evolves according to \eqref{eq:Lindblad_NRQCD_HQk_traced}.
In the analysis below, it is convenient to choose the position basis $\bar\rho_S(t, \vec x, \vec y) = \langle\vec x|\bar\rho_S(t)|\vec y\rangle$ for the density matrix.
The Lindblad equation at the leading order gradient expansion is
\begin{align}
\label{eq:decoherence_HQ}
\frac{\partial}{\partial t}\bar\rho_S(t, \vec x, \vec y) = i\frac{\nabla_x^2 - \nabla_y^2}{2M} \bar\rho_S(t, \vec x, \vec y)
-C_F\left[\gamma(\vec 0) - \gamma(\vec x-\vec y)\right]\bar\rho_S(t, \vec x, \vec y).
\end{align}
The effect of environment is summarized in the last term, which causes {\it decoherence} to the heavy quark wave function: the off-diagonal element ($\vec x\neq\vec y$) of the density matrix decays in a time scale $1/C_F\left[\gamma(0) - \gamma(\vec x-\vec y)\right]$.
This form of the master equation generally holds true for the description of collisional decoherence \cite{gallis1990environmental,hornberger2003collisional}, which was experimentally measured \cite{hornberger2003matterwave}.
Since $r = |\vec x - \vec y|$ is the spatial extent of the heavy quark wave function and $\gamma(\vec 0) - \gamma(\vec r)$ is an increasing function of $r$ that starts from $0$ and saturates to $\gamma (\vec 0)$ at very large $r$, the decoherence proceeds more quickly for extended states.
If the density matrix is evolved with the master equation long enough, it becomes nearly diagonal.
The process of the diagonalization itself is physical, but the relevant size of the coefficient $C_F\left[\gamma(0) - \gamma(\vec x-\vec y)\right]$ continues to get smaller and the gradient expansion is not justified at the late stage of the diagonalization/decoherence.
Therefore, we need to perform the gradient expansion up to the next-to-leading order and see how the next-to-leading term affects the diagonalization process.

The explicit form of the Lindblad equation \eqref{eq:Lindblad_NRQCD_HQk_traced} is
\begin{align}
\label{eq:Lindblad_NRQCD_HQ_explicit}
\frac{\partial}{\partial t}\bar\rho_S(t,\vec x,\vec y)
&=i\frac{\nabla_x^2-\nabla_y^2}{2M}\bar\rho_S(t,\vec x,\vec y)
-C_F\left[F_1(\vec 0) - F_1(\vec x-\vec y)\right]
\bar\rho_S(t,\vec x,\vec y) \nonumber\\
& \quad + C_F\left[
\vec F_2(\vec x-\vec y)\cdot (\vec \nabla_x -\vec \nabla_y)
+ F^{ij}_3(\vec x-\vec y)\partial_i^{(x)}\partial_j^{(y)} +F^{ii}_3(\vec 0)\frac{\nabla^2_x+\nabla^2_y}{6}
\right]
\bar\rho_S(t,\vec x,\vec y),
\end{align}
which not only contains the full leading and next-to-leading terms but also some of the next-to-next-to-leading terms in the gradient expansion.
The coefficients are defined as 
\begin{align}
F_1(\vec r)&\equiv \gamma(\vec r)+\frac{\nabla^2 \gamma(\vec r)}{4MT}+\frac{\nabla^4 \gamma(\vec r)}{64M^2T^2}, 
\quad \vec F_2(\vec r)\equiv \vec\nabla\left(\frac{\gamma(\vec r)}{4MT}
+\frac{\nabla^2 \gamma(\vec r)}{32M^2T^2}\right),
\quad F^{ij}_3(\vec r)\equiv -\frac{\partial_i\partial_j \gamma(\vec r)}{16M^2T^2}.
\end{align}
To illustrate the fate of density matrix under decoherence at late enough time, we take the small wave packet limit by expanding $\gamma(\vec r)$ to the quadratic order.
The master equation then becomes
\begin{align}
\label{eq:Lindblad_NRQCD_HQ_wp}
\frac{\partial}{\partial t}\bar\rho_S(t,\vec x,\vec y)
&=i\frac{\nabla_x^2-\nabla_y^2}{2M}\bar\rho_S(t,\vec x,\vec y) \\
& \quad + \frac{C_F\nabla^2\gamma(0)}{6}\left[
(\vec x-\vec y)^2 + \frac{\vec x-\vec y}{2MT}\cdot \left(\vec \nabla_x-\vec \nabla_y\right)
- \frac{1}{16M^2T^2}\left(\vec \nabla_x + \vec \nabla_y\right)^2
\right]\bar\rho_S(t,\vec x,\vec y). \nonumber
\end{align}
This master equation is equivalent to the well-known Caldeira-Leggett master equation for the quantum Brownian motion if we ignore the last term in the second line of \eqref{eq:Lindblad_NRQCD_HQ_wp}.
Near equilibrium, the wave packet has typical extension $|\vec x -\vec y|\sim 1/\sqrt{MT}$ and momentum $|\vec \nabla|\sim \sqrt{MT}$.
Therefore, the leading and the next-to-leading terms in the gradient expansion are comparable\footnote{
As we see in Sec.~\ref{sec:NRQCD_HQ_CL}, the next-to-next-to-leading term turns out to be smaller than the leading and the next-to-leading order terms and the lack of these terms in the Caldeira-Leggett master equation is physically justified.
} and the latter cannot be simply ignored.
As will be fully discussed in the next section \ref{sec:NRQCD_HQ_CL}, the physical meaning of this observation becomes clear when we take the classical limit: the leading order term represents random force from the thermal environment and the next-to-leading term describes the friction, and these forces balance in the equilibrium according to the fluctuation-dissipation theorem.

A few technical remarks are in order here.
The master equation \eqref{eq:Lindblad_NRQCD_HQ_wp} is written as the Lindblad equation \eqref{eq:Lindblad} with
\begin{align}
\tilde V_S^{(i)} \leftrightarrow  \left(\vec x_Q + \frac{i \vec p_Q}{4MT}\right)_i, \quad
\gamma_{ij} \leftrightarrow -\frac{C_F}{3}\nabla^2\gamma(0)\delta_{ij} > 0, \quad
\Delta H_S = -\frac{C_F\nabla^2\gamma(0)}{12MT} \left\{\vec x_Q, \vec p_Q \right\},
\end{align}
which is a particular generalization of the Caldeira-Leggett master equation \cite{diosi1993high,diosi1993calderia,gao1997dissipative,vacchini2000completely}.
This is verified by explicit substitution or by expanding the operators of \eqref{eq:Lindblad_NRQCD_HQk_traced} up to the second order in $\vec x_Q$.
The explicit $\vec x_Q$ dependence in the Lindblad operator and the Hamiltonian may seem to indicate the violation of the translational invariance $\vec x_Q \to \vec x_Q + \vec c$, but the Lindblad equation is actually invariant under this transformation thanks to the property mentioned in the last part of Sec.~\ref{sec:Basics_Lindblad}.
In relation to the Caldeira-Leggett master equation, the Lindblad equation \eqref{eq:Lindblad_NRQCD_HQ_explicit} is regarded as its generalization to a Brownian particle with extended wave functions.
Similar master equations are also obtained in \cite{gallis1993models,diosi1995quantum,vacchini2001test,hornberger2006master}.

\subsubsection{Equilibration in the classical limit}
\label{sec:NRQCD_HQ_CL}
The master equation \eqref{eq:Lindblad_NRQCD_HQ_explicit} describes the quantum decoherence and quantum dissipation.
Moreover, it describes  the transition from quantum to classical \cite{zurek306072decoherence} --- after a long enough time, the decoherence and dissipation eventually balance and the heavy quark is described by a localized wave packet that can be approximated by a classical point particle.
For a localized wave packet, the density matrix is nearly diagonal, so that the master equation \eqref{eq:Lindblad_NRQCD_HQ_wp} is a good approximation.
To gain further insight, it is useful to define the Wigner function
\begin{align}
\label{eq:Wigner}
f(t,\vec r, \vec p) \equiv \int d^3 s  e^{-i\vec p\cdot\vec s}\rho_S\left(t,\vec r+\frac{\vec s}{2}, \vec r - \frac{\vec s}{2}\right),
\end{align}
which is a quantum mechanical counterpart of the classical phase space distribution but does not necessarily take positive values.
The evolution of $f(t,\vec r, \vec p)$ is governed by
\begin{align}
\label{eq:Kramerslike}
\left[
\frac{\partial}{\partial t}
+\frac{\vec p}{M}\cdot\vec \nabla_r - \frac{\kappa}{2MT}\frac{\partial}{\partial \vec p}
\cdot\left(
\vec p + MT\frac{\partial}{\partial \vec p}
\right)
-\frac{\kappa}{32M^2T^2} \nabla_r^2
\right]f(t,\vec r, \vec p) = 0,
\end{align}
with $\kappa \equiv -\frac{C_F\nabla^2\gamma(0)}{3}>0$ being the momentum diffusion constant\footnote{
Since the master equation is obtained by weak coupling expansion, $\kappa$ here should be the leading order perturbative one
$\kappa_{\rm LO} = \frac{C_F g^4 T^3}{18\pi}\left[
N_c\left(\ln\frac{2T}{m_D} + \xi \right) + \frac{N_f}{2}\left(\ln\frac{4T}{m_D} + \xi \right)
\right]$
with $\xi\simeq  -0.64718$ \cite{moore2005much}.
See Appendix \ref{app:Thermal_Corr_Gluons} for details.
}.
It is clear that $\partial^2/\partial p^2$, $(\partial/\partial \vec p)\cdot\vec p$, and $\nabla_r^2$ terms derive from the leading, next-to-leading, and next-to-next-to-leading terms respectively.
If we neglect the $\nabla_r^2$ term, the equation is equivalent to the Kramers equation for the classical Brownian motion:
\begin{align}
\label{eq:Langevin_HQ}
\frac{d\vec r}{dt} = \frac{\vec p}{M}, \quad
\frac{d\vec p}{dt} = -\frac{\kappa}{2MT}\vec p + \vec \xi, \quad
E_{\xi}\left[\xi_i(t)\xi_j(t')\right] = \kappa\delta_{ij}\delta(t-t'),
\end{align}
where $E_{\xi}[\mathcal O]$ denotes the expectation value of $\mathcal O(\xi)$ over the stochastic variable $\xi$.
Since the classical Brownian particle diffuses with a diffusion constant $D=2T^2/\kappa\sim 1/T$, the extra $\nabla_r^2$ term just adds to the diffusion constant $D$ by a fraction $T^2/M^2\ll 1$, so that it can be safely neglected.
In any case, it is easy to see that the equilibrium distribution is the classical Boltzmann distribution $\propto \exp\left[-p^2/2MT\right]$.

Here is a technical comment.
Although we get \eqref{eq:Kramerslike} from the Lindblad equation \eqref{eq:Lindblad_NRQCD_HQ_wp} by Wigner transformation, we can directly calculate the evolution equation of $f(t,\vec r, \vec p)$ from the Lindblad equation \eqref{eq:Lindblad_NRQCD_HQ_explicit} by taking the small wave packet limit.
In this case, we need to expand $F_1(\vec s)$, $\vec F_2(\vec s)$, and $F^{ij}_3(\vec s)$ in terms of $\vec s$, which yields extra contribution from higher order derivatives of $\gamma(\vec s)$ other than $\nabla^2\gamma(0)$.
To systematically count the small contributions, it is natural to recover $\hbar$ in Eq.~\eqref{eq:Lindblad_NRQCD_HQ_explicit}
by $\gamma(\vec s) \to \gamma(\vec s)/\hbar$, $\partial_t \to \hbar\partial_t$, and $\nabla\to \hbar\nabla$ and in Eq.~\eqref{eq:Wigner} by $\exp(-i\vec p\cdot\vec s) \to \exp(-i\vec p\cdot\vec s/\hbar)$.
The leading contribution in $\hbar\to 0$ limit results in Eq.~\eqref{eq:Kramerslike} without the last term, which is suppressed by $\hbar^2$, and the Kramers equation is obtained.

\subsection{Quantum Brownian motion of a quarkonium}
\label{sec:NRQCD_QQbar}
In this section, we derive and analyze the Lindblad equation for quantum Brownian motion of a quarkonium in a weakly coupled quark-gluon plasma (QGP).
Let us see when the scale hierarchy for the quantum Brownian motion ($\tau_R\gg \tau_E$ and $\tau_S\gg \tau_E$) is satisfied.
We estimate the system relaxation time $\tau_R\sim M/g^4 T^2$ by its kinetic equilibration time, the system time scale $\tau_S=1/E$ by the inverse of the binding energy $E$, and the environment correlation time $\tau^{\rm (soft,hard)}_E\sim 1/gT, 1/T$ by the duration of the collisions.
The relation $\tau_R\gg \tau_E$ is satisfied because $M\gg T$.
Another relation $\tau_S\gg \tau_E$ is satisfied when $E\ll gT$.
One might wonder whether the latter relation is close to the situation realized in the heavy-ion collisions.
To make the argument explicit, let us take the Coulombic binding energy for bottomonium in the QGP.
Using the textbook results for the hydrogen atom with singlet potential $-C_F\alpha_s/r$
\begin{align}
\frac{1}{\tau_S} = E = \frac{1}{2} (M_b/2) (C_F\alpha_s)^2 \simeq 0.5\cdot (4.7 [{\rm GeV}]/2) \cdot (0.3\text{-}0.4)^2 \simeq (0.11\text{-}0.19) [{\rm GeV}],
\end{align}
where $M_b\simeq 4.7 [{\rm GeV}]$ is the bottom quark mass.
Comparing with the relaxation time at typical temperature $T\sim 0.4{\rm GeV}\sim 1/\tau_E$, the assumption $\tau_S\sim (2\text{-}3)\tau_E > \tau_E$ is not too far from reality. 
The situation is better for charmonium, whose mass is about three times lighter than bottomonium.
Furthermore, if we estimate by $\tau_S = 2\pi/E$ or by $\tau_E\sim 1/2\pi T$, the situation gets even better.

\subsubsection{Lindblad equation}
\label{sec:NRQCD_QQbar_Lindblad}
We apply the formula \eqref{eq:Lindblad_QBM} to non-relativistic heavy quark pair in the QGP.
The total Hamiltonian is given in Eq.~\eqref{eq:NRQCD_Hamiltonian}, which we quote here again,
\begin{align}
H_{\rm tot} = \left(\frac{p_Q^2}{2M} + \frac{p_{Q_c}^2}{2M}\right)\otimes I_E
+ I_S \otimes H_{q+A}
+ \int d^3 x \left[
\delta(\vec x - \vec x_Q) t^a_Q - \delta(\vec x - \vec x_{Q_c}) t^{a*}_{Q_c}
\right] \otimes gA_0^a(\vec x), \nonumber
\end{align}
and the operator correspondence is
\begin{subequations}
\begin{align}
V_S^{(i)} &\leftrightarrow \delta(\vec x - \vec x_Q) t^a_Q - \delta(\vec x - \vec x_{Q_c}) t^{a*}_{Q_c} \equiv V_S^a(\vec x), \\
\dot V_S^{(i)}=i[H_S, V_S^{(i)}] &\leftrightarrow 
\left(
\begin{aligned}
&\left[
-\frac{i}{2M}\nabla_x^2\delta(\vec x - \vec x_Q)
-\frac{1}{M}\vec\nabla_x\delta(\vec x - \vec x_Q)\cdot\vec p_Q
\right] t^a_Q \\
& -\left[
-\frac{i}{2M}\nabla_x^2\delta(\vec x - \vec x_{Q_c})
-\frac{1}{M}\vec\nabla_x\delta(\vec x - \vec x_{Q_c})\cdot\vec p_{Q_c}
\right] t^{a*}_{Q_c}
\end{aligned}
\right) \equiv \dot V_S^a(\vec x).
\end{align}
\end{subequations}
The coefficients are again quoted from Eqs.~\eqref{eq:gluon_corr} and \eqref{eq:gluon_corr_eta}
\begin{align}
\gamma_{ab}(\vec x-\vec y) &= T\frac{d}{d\omega}\sigma_{ab}(\omega, \vec x -\vec y)\Bigr|_{\omega=0}
\equiv \gamma(\vec x-\vec y)\delta_{ab}, \nonumber \\
S_{ab}(\vec x -\vec y) &= -\frac{1}{2}\int_{-\infty}^{\infty} \frac{d\omega}{2\pi}\frac{\sigma_{ab}(\omega, \vec x-\vec y)}{\omega}
\equiv S(\vec x - \vec y)\delta_{ab}, \nonumber \\
\sigma_{ab}(\omega, \vec x -\vec y) &\equiv
\int_{-\infty}^{\infty} dt e^{i\omega t} {\rm Tr}_E\left(\rho_E^{\rm th}\left[gA_0^a(t,\vec x), gA_0^b(0,\vec y)\right]\right)
\propto \delta_{ab}, \nonumber \\
\eta_{ab}(\vec x-\vec y)&\simeq -\frac{i}{4T}\gamma_{ab}(\vec x -\vec y) = -\frac{i}{4T}\gamma(\vec x - \vec y) \delta_{ab},\nonumber
\end{align}
whose explicit forms at the soft scale $|\vec x|\sim 1/gT$ are given in the Appendix \ref{app:Thermal_Corr_Gluons}.
Then the Lindblad operator reads
\begin{align}
\tilde V_S^{(i)} &\leftrightarrow V_S^a(\vec x) + \frac{i}{4T} \dot V_S^a(\vec x)  \nonumber\\
& \qquad =
\left(
\begin{aligned}
&\left[
\delta(\vec x - \vec x_Q)
+\frac{1}{8MT}\nabla_x^2\delta(\vec x - \vec x_Q)
-\frac{i}{4MT}\vec\nabla_x\delta(\vec x - \vec x_Q)\cdot\vec p_Q
\right] t^a_Q \\
& -\left[
\delta(\vec x - \vec x_{Q_c})
+\frac{1}{8MT}\nabla_x^2\delta(\vec x - \vec x_{Q_c})
-\frac{i}{4MT}\vec\nabla_x\delta(\vec x - \vec x_{Q_c})\cdot\vec p_{Q_c}
\right] t^{a*}_{Q_c}
\end{aligned}
\right) \equiv \tilde V_S^a(\vec x).
\end{align}
The correction to the Hamiltonian is
\begin{align}
\Delta H_S = \left[
-2S(\vec x_Q - \vec x_{Q_c})
-\frac{1}{8MT} \left\{\vec p_Q - \vec p_{Q_c}, \vec\nabla \gamma(\vec x_Q - \vec x_{Q_c}) \right\}
\right]t^a_Q t^{a*}_{Q_c},
\end{align}
where the first term gives the Coulomb interaction which is Debye screened at the long distance and the second term is time reversal odd.
Note that $t^b_Qt^{c*}_{Q_c}$ is not a matrix multiplication but a tensor product.

In the above derivation, we calculate $\dot V_S^{a}(\vec x)$ by the free Hamiltonian $(p_Q^2 + p_{Q_c}^2)/2M$.
We can improve this calculation by including the induced potential term in the Hamiltonian
\begin{align}
&\dot V_{S, \Delta}^{(i)} \equiv i \left[H_S + \Delta H_S\bigr|_{\partial^0}, V_S^{(i)}\right] \nonumber \\
&\leftrightarrow \dot V^a_{S, \Delta}(\vec x)\equiv \dot V_S^a(\vec x) -2iS(\vec x_Q-\vec x_{Q_c})\left[\delta(\vec x - \vec x_Q) - \delta(\vec x - \vec x_{Q_c})\right]
i f^{abc}t^b_Qt^{c*}_{Q_c},
\end{align}
and redefine the Lindblad operator as
\begin{align}
\tilde V_{S, \Delta}^a(\vec x) &\equiv V_S^a(\vec x) + \frac{i}{4T} \dot V_{S, \Delta}^a(\vec x)\nonumber \\
& = \left(
\begin{aligned}
&\left[\delta(\vec x - \vec x_Q)
+\frac{1}{8MT}\nabla_x^2\delta(\vec x - \vec x_Q)
-\frac{i}{4MT}\vec\nabla_x\delta(\vec x - \vec x_Q)\cdot\vec p_Q
\right] t^a_Q \\
& -\left[
\delta(\vec x - \vec x_{Q_c})
+\frac{1}{8MT}\nabla_x^2\delta(\vec x - \vec x_{Q_c})
-\frac{i}{4MT}\vec\nabla_x\delta(\vec x - \vec x_{Q_c})\cdot\vec p_{Q_c}
\right] t^{a*}_{Q_c} \\
& +\frac{S(\vec x_Q-\vec x_{Q_c})}{2T}\left[\delta(\vec x - \vec x_Q) - \delta(\vec x - \vec x_{Q_c})\right]
i f^{abc}t^b_Qt^{c*}_{Q_c}
\end{aligned}\right),
\end{align}
where $f^{abc}$ is the structure constant of the SU($N_c$) algebra $[t^a, t^b] = if^{abc}t^c$.
In the present example, $\Delta H_S$ is not affected by adopting the improved $\dot V_{S, \Delta}^a(\vec x)$ instead of $\dot V_S^a(\vec x)$ in $\Delta H_S\bigr|_{\partial^1}$.
Diagrammatically, this improvement corresponds to adding a cross-ladder exchange to the resummation kernel, which drives the time-evolution of the density matrix via the master equation.
By this improvement, we can take into account the potential energy difference between the heavy quark pair in the singlet and that in the octet.
Therefore, the equilibrium occupation depends on the color representations of the heavy quark pair because of this improvement.

Collecting these results, we obtain the Lindblad equation
\begin{subequations}
\label{eq:Lindblad_NRQCD_QQbar}
\begin{align}
\frac{d}{dt}\rho_S(t) &= -i\left[H_S + \Delta H_S, \rho_S\right]
+\int_{x,y} \gamma(\vec x-\vec y)\left[
\tilde V_{S, \Delta}^a(\vec y)\rho_S \tilde V_{S, \Delta}^{a\dagger}(\vec x)
-\frac{1}{2}\left\{
\tilde V_{S, \Delta}^{a\dagger}(\vec x)\tilde V_{S, \Delta}^a(\vec y), \rho_S
\right\}
\right] \nonumber \\
&= -i\left[H_S + \Delta H_S, \rho_S\right]
+\int_k \gamma(\vec k)\left[
\tilde V_{S, \Delta}^a(\vec k)\rho_S \tilde V_{S, \Delta}^{a\dagger}(\vec k)
-\frac{1}{2}\left\{
\tilde V_{S, \Delta}^{a\dagger}(\vec k)\tilde V_{S, \Delta}^a(\vec k), \rho_S
\right\}
\right],\\
\Delta H_S &= \left[
-2S(\vec x_Q - \vec x_{Q_c})
-\frac{1}{8MT} \left\{\vec p_Q - \vec p_{Q_c}, \vec\nabla \gamma(\vec x_Q - \vec x_{Q_c}) \right\}
\right]t^a_Q t^{a*}_{Q_c},\\
\tilde V_{S, \Delta}^{a}(\vec k) &= e^{i\vec k\cdot\vec x_Q/2}\left(1-\frac{\vec k\cdot\vec p_Q}{4MT}\right)e^{i\vec k\cdot\vec x_Q/2} t^a_Q
-e^{i\vec k\cdot\vec x_{Q_c}/2}\left(1-\frac{\vec k\cdot\vec p_{Q_c}}{4MT}\right)e^{i\vec k\cdot\vec x_{Q_c}/2} t^{a*}_{Q_c} \nonumber \\
&\quad +\frac{S(\vec x_Q - \vec x_{Q_c})}{2T} \left(e^{i\vec k\cdot \vec x_Q} - e^{i\vec k\cdot\vec x_{Q_c}}\right)i f^{abc}t^b_Qt^{c*}_{Q_c}.
\end{align}
\end{subequations}
The physical meaning of the Lindblad operator $\tilde V_{S, \Delta}^{a}(\vec k)$ can be understood similarly to the single heavy quark case: the first two terms describe a simultaneous scattering of $Q+g\to Q+g$ and $\bar Q+g\to\bar Q+g$ (and similar processes involving $q$ and $\bar q$) with momentum transfer $\vec k$.
Note that this simultaneous scattering induces quantum interference on the wave function.
To give physical meaning to the last term, singlet-octet basis for the color space is more convenient as explained below.

The color part of the Lindblad equation \eqref{eq:Lindblad_NRQCD_QQbar} can be expressed in the singlet-octet basis.
Switching from the tensor product basis ($|i\rangle_Q|j\rangle_{Q_c}$) to the singlet-octet basis ($|s\rangle, |a\rangle$) is done by the following unitary transformation
\begin{subequations}
\begin{align}
|s\rangle &= \frac{1}{\sqrt{N_c}}\delta_{ij}|i\rangle_Q|j\rangle_{Q_c}, \quad (i,j=1,2,\cdots, N_c), \\
|a\rangle &= \sqrt{2}(t^a_F)_{ij}|i\rangle_Q|j\rangle_{Q_c}, \quad (a=1,2,\cdots, N_c^2-1),
\end{align}
\end{subequations}
where $t^a_F$ is the fundamental representation of SU($N_c$) algebra with a conventional normalization ${\rm tr}(t^a_Ft^b_F)=\delta_{ab}/2$.
The color space operators in the singlet-octet basis are\footnote{
We provide a short derivation for $\langle s|f^{abc}t^b_Qt^{c*}_{Q_c}|d\rangle = -i\frac{1}{2}\sqrt{\frac{N_c}{2}}\delta_{ad}$ and $\langle d|f^{abc}t^b_Qt^{c*}_{Q_c}|e\rangle = 0$:
\begin{subequations}
\begin{align}
\langle s|f^{abc}t^b_Qt^{c*}_{Q_c}|d\rangle &= \sqrt{\frac{2}{N_c}}f^{abc}{\rm tr} (t_F^ct_F^bt_F^d) 
= -i\sqrt{\frac{2}{N_c}}{\rm tr}([t_F^a,t_F^b]t_F^bt_F^d)
= -i \frac{N_c}{2}\sqrt{\frac{1}{2N_c}}\delta_{ad}, \\
\langle d|f^{abc}t^b_Qt^{c*}_{Q_c}|e\rangle &=2f^{abc}{\rm tr}(t_F^bt_F^et_F^ct_F^d)
=-2i{\rm tr}(t_F^bt_F^e[t_F^a,t_F^b]t_F^d)
=0.
\end{align}
\end{subequations}
At the final steps for both, we use  $t_F^at_F^bt_F^a=-t_F^b/2N_c$ which follows from $(t_F^b)_{ij}(t_F^b)_{kl} = \frac{1}{2}(\delta_{il}\delta_{jk}-\frac{1}{N_c}\delta_{ij}\delta_{kl})$.
}
\begin{subequations}
\begin{align}
\langle s|t^a_Q|s\rangle &= \langle s|t^{a*}_{Q_c}|s\rangle = \langle s|f^{abc}t^b_Qt^{c*}_{Q_c}|s\rangle = \langle d|f^{abc}t^b_Qt^{c*}_{Q_c}|e\rangle = 0, \\
\langle s|t^a_Q|b\rangle &= \langle s|t^{a*}_{Q_c}|b\rangle = \frac{1}{\sqrt{2N_c}}\delta_{ab}, \quad
\langle s|f^{abc}t^b_Qt^{c*}_{Q_c}|d\rangle = -i\frac{1}{2}\sqrt{\frac{N_c}{2}}\delta_{ad},\\
\langle b|t^a_Q|c\rangle &= 2{\rm tr}(t_F^ct_F^bt_F^a) = \frac{d_{abc} - i f_{abc}}{2}, \quad
\langle b|t^{a*}_{Q_c}|c\rangle =2{\rm tr}(t_F^at_F^bt_F^c) = \frac{d_{abc} + i f_{abc}}{2},
\end{align}
\end{subequations}
where $d_{abc}\equiv 2{\rm tr}\left(\{t_F^a, t_F^b\} t_F^c\right)$ is a totally symmetric tensor.
In the singlet-octet basis, the Lindblad operator $\tilde V_{S, \Delta}^{a}(\vec k)$ is decomposed as
\begin{subequations}
\label{eq:Lindblad_NRQCD_QQbar_sa}
\begin{align}
\tilde V_{S, \Delta}^{a}(\vec k)
&= \tilde V_+(\vec k)\sqrt{\frac{1}{2N_c}}|a\rangle\langle s| + \tilde V_-(\vec k) \sqrt{\frac{1}{2N_c}}|s\rangle\langle a|
+ \tilde V_d(\vec k) \frac{1}{2}d^{abc}|b\rangle\langle c|
+ \tilde V_f(\vec k) \frac{1}{2}if^{abc}|b\rangle\langle c|, \\
\tilde V_{\pm}(\vec k) &= e^{i\vec k\cdot\vec x_Q/2}\left[
1-\frac{\vec k\cdot\vec p_Q}{4MT}
\mp \frac{N_c S(\vec x_Q - \vec x_{Q_c})}{4T}
\right]e^{i\vec k\cdot\vec x_Q/2} \nonumber \\
&\quad 
-e^{i\vec k\cdot\vec x_{Q_c}/2}\left[
1-\frac{\vec k\cdot\vec p_{Q_c}}{4MT}
\mp \frac{N_c S(\vec x_Q - \vec x_{Q_c})}{4T}
\right]e^{i\vec k\cdot\vec x_{Q_c}/2}, \\
\tilde V_{d/f}(\vec k) &= \pm e^{i\vec k\cdot\vec x_Q/2}\left(1-\frac{\vec k\cdot\vec p_Q}{4MT}\right)e^{i\vec k\cdot\vec x_Q/2}
-e^{i\vec k\cdot\vec x_{Q_c}/2}\left(1-\frac{\vec k\cdot\vec p_{Q_c}}{4MT}\right)e^{i\vec k\cdot\vec x_{Q_c}/2}.
\end{align}
\end{subequations}
In this basis, physical meaning of the last term in the Lindblad operators in \eqref{eq:Lindblad_NRQCD_QQbar} can be analyzed.
It only appears in $V_{\pm}(\vec k)$ that describes the transitions between the singlet and the octets.
Repeating the discussions on the detailed balance in Sec.~\ref{sec:NRQCD_HQ_Lindblad} leads to the following detailed balance in the color space
\begin{align}
\frac{\Gamma_{s\to a}}{\Gamma_{a\to s}}
\simeq \exp\left(-\frac{N_c S(\vec x_Q - \vec x_{Q_c})}{T}\right),
\end{align}
where $-N_c S(\vec x_Q - \vec x_{Q_c})<0$ is potential energy difference between the singlet and the octets.

In both basis, the internal color space has 9 dimensions.
Physically, we often do not need to know the detailed distribution in the octet space.
Therefore it is very convenient if we can derive a master equation for the projected density matrices
\begin{align}
\rho_{s}(t) \equiv \langle s| \rho_S(t)|s\rangle, \quad
\rho_{o}(t) \equiv \langle a| \rho_S(t)|a\rangle.
\end{align}
Indeed we can show that the off-diagonal parts $\langle s| \rho_S(t)|a\rangle$ and $\langle a| \rho_S(t)|s\rangle$ decouple from the diagonal parts because of the properties $f^{aac}=d^{aac}=f^{abc}d^{abd}=0$.
Therefore the master equation for the reduced density matrix with $2\times 2$ internal space  
\begin{align}
\label{eq:rho_so_diagonal}
\rho_S(t) = \begin{pmatrix}
\rho_s(t) & 0 \\
0 & \rho_o(t)
\end{pmatrix}
\end{align}
is obtained in the following Lindblad form with new Lindblad operators $\tilde C_n(\vec k)$
\begin{subequations}
\label{eq:Lindblad_NRQCD_QQbar_proj}
\begin{align}
\frac{d}{dt}\rho_S(t) &= -i\left[H_S + \Delta H_S, \rho_S\right]
+\int_k \gamma(\vec k)\sum_{n=+,-,d,f}\left[
\tilde C_n(\vec k)\rho_S \tilde C_n^{\dagger}(\vec k)
-\frac{1}{2}\left\{
\tilde C_n^{\dagger}(\vec k)\tilde C_n(\vec k), \rho_S
\right\}
\right],\\
\Delta H_S &= \left[
-2S(\vec x_Q - \vec x_{Q_c})
-\frac{1}{8MT} \left\{\vec p_Q - \vec p_{Q_c}, \vec\nabla \gamma(\vec x_Q - \vec x_{Q_c}) \right\}
\right]
\begin{pmatrix}
C_F & 0 \\
0 & -\frac{1}{2N_c}
\end{pmatrix}, \\
\tilde C_+(\vec k) &= \tilde V_+(\vec k) \sqrt{C_F}\begin{pmatrix} 0 & 0 \\ 1 & 0 \end{pmatrix}, \quad
\tilde C_-(\vec k) = \tilde V_-(\vec k) \sqrt{\frac{1}{2N_c}}\begin{pmatrix} 0 & 1 \\ 0 & 0 \end{pmatrix}, \\
\tilde C_d(\vec k) &= \tilde V_d(\vec k) \sqrt{\frac{N_c^2-4}{4N_c}}\begin{pmatrix} 0 & 0 \\ 0 & 1 \end{pmatrix}, \quad
\tilde C_f(\vec k) = \tilde V_f(\vec k) \sqrt{\frac{N_c}{4}}\begin{pmatrix} 0 & 0 \\ 0 & 1 \end{pmatrix},
\end{align}
\end{subequations}
where formulas $f^{abc}f^{abd}=N_c\delta_{cd}$ and $d^{abc}d^{abd} = \frac{N_c^2-4}{N_c}\delta_{cd}$ are used\footnote{
A short derivation of the latter is given here.
Using $\{t_F^a,t_F^b\} =\frac{\delta^{ab}}{N_c}  + d^{abc}t_F^c$ and $t_F^at_F^bt_F^a = -t_F^b/2N_c$, one gets
\begin{align}
\{\{t_F^a,t_F^b\},t_F^b\} =\frac{2}{N_c}t_F^a  + d^{abc}\{t_F^c, t_F^b\}
=\frac{2}{N_c}t_F^a  + d^{abc}d^{cbd}t_F^d
=\left(2C_F - \frac{1}{N_c}\right)t_F^a,
\end{align}
from which $d^{abc}d^{cbd}= \frac{N_c^2-4}{N_c}\delta_{ad}$ follows.
}.

The Lindblad equation for a quarkonium in the QGP was first derived in \cite{Akamatsu:2014qsa}.
In the original paper, it was obtained in the form of \eqref{eq:Lindblad_NRQCD_QQbar} without taking into account the color state dependence of the potential energy, i.e. the Lindblad operator was not $\tilde V^a_{S,\Delta}(\vec k)$ but $\tilde V^a_S(\vec k)$.
The Abelian limit of \eqref{eq:Lindblad_NRQCD_QQbar}, or to be specific \eqref{eq:Lindblad_QBM_2body}, was numerically simulated in 1-dimension for the first time in \cite{Miura:2019ssi} using the method of Quantum State Diffusion \cite{gisin1992quantum,percival1998quantum} and in \cite{Alund:2020ctu} by directly solving the master equation with a novel trace preserving algorithm.
The full non-Abelian version has recently been simulated in 1-dimension in \cite{Akamatsu:2021dot} using the Quantum State Diffusion method.
The equilibration to the Boltzmann distribution was confirmed within acceptable deviations.
It was also pointed out that the quantum dissipation is not negligible already at initial times if the initial quarkonium is a localized bound state such as the ground state, as is expected from the discussions in Sec.~\ref{sec:NRQCD_HQ}.

\subsubsection{Recoilless limit and the stochastic potential}
\label{sec:NRQCD_QQbar_Stochastic}
As discussed in Sec.~\ref{sec:NRQCD_HQ_DecDiss}, the Lindblad equation in the leading order gradient expansion, or in the recoilless limit, describes quantum decoherence.
Although its applicability is limited to shorter timescales than the heavy quark relaxation time, quantum decoherence is an essential feature of the early-time dynamics of quarkonium dissociation process in the QGP.
In this section, we give an intuitive picture of quantum decoherence by introducing a stochastic Hamiltonian which models the environmental thermal fluctuations as a noise term\footnote{
It is straightforward to apply the description here to the recoilless limit of the single heavy quark in Sec.~\ref{sec:NRQCD_HQ_DecDiss}.
In this case, $\Delta H_S=0$ and $V_S^a(\vec x)=\delta(\vec x - \vec x_Q) t^a_Q$ in Eq.~\eqref{eq:stochastic_Hamiltonian}.
}.

Let us start from the Lindblad equation in the tensor product basis
\begin{subequations}
\label{eq:Lindblad_NRQCD_QQbar_recoilless}
\begin{align}
\frac{d}{dt}\rho_S(t) &= -i\left[H_S + \Delta H_S, \rho_S\right]
+\int_{x,y} \gamma(\vec x-\vec y)\left[
V_{S}^a(\vec y)\rho_S V_{S}^{a}(\vec x)
-\frac{1}{2}\left\{
V_{S}^{a}(\vec x)V_{S}^a(\vec y), \rho_S
\right\}
\right],\\
\Delta H_S &= -2S(\vec x_Q - \vec x_{Q_c}) t^a_Q t^{a*}_{Q_c},\quad
V_S^a(\vec x) = \delta(\vec x - \vec x_Q) t^a_Q - \delta(\vec x - \vec x_{Q_c}) t^{a*}_{Q_c}.
\end{align}
\end{subequations}
This equation is equivalent to the evolution by a stochastic Hamiltonian in the system Hilbert space:
\begin{subequations}
\label{eq:stochastic_Hamiltonian}
\begin{align}
&H(t; \theta) \equiv H_S + \Delta H_S + \int_x \theta^a(\vec x, t)V_S^a(\vec x), \\
&E_{\theta}\left[\theta^a(\vec x, t)\theta^b(\vec y, t')\right] = \gamma(\vec x-\vec y)\delta^{ab}\delta(t-t'), \\
&|\psi(t+dt;\theta)\rangle = e^{-iH(t;\theta)dt}|\psi(t;\theta)\rangle,
\end{align}
\end{subequations}
where $E_{\theta}[\mathcal O]$ denotes the expectation value of $\mathcal O(\theta)$ over the stochastic variable $\theta$ and the stochastic evolution is discretized in the It$\hat {\rm o}$ scheme.
Note that this stochastic Hamiltonian is Hermitian because $\theta^a(\vec x, t)$ is real.
The equivalence is confirmed by deriving the time evolution of the system density matrix from
\begin{align}
\rho_S(t) \equiv E_{\theta}\left[|\psi(t;\theta)\rangle\langle\psi(t;\theta)|\right], \quad
\rho_S(t+dt) = E_{\theta}\left[e^{-iH(t;\theta)dt}\rho_S(t)e^{iH(t;\theta)dt}\right],
\end{align}
by noting that $\theta dt\sim dt^{1/2}$.
The Hermiticity of $H(t;\theta)$ is essential for this equivalence.
Comparing the stochastic Hamiltonian \eqref{eq:stochastic_Hamiltonian} with Eq.~\eqref{eq:NRQCD_Hamiltonian}, it is clear that the scalar potential $A_0^a(\vec x)$ is modeled by the noise field $\theta^a(\vec x,t)$.
By putting the screening potential and the noise term together, let us introduce a {\it stochastic potential}
\begin{align}
\label{eq:stochastic_potential}
U_{QQ_c} (\vec x_Q,\vec x_{Q_c};\theta)
\equiv -2S(\vec x_Q - \vec x_{Q_c})t^a_Q t^{a*}_{Q_c}  + \theta^a(\vec x_Q, t)t_Q^a - \theta^a(\vec x_{Q_c}, t) t_{Q_c}^{a*},
\end{align}
which summarizes the effects in the leading order gradient expansion.
The explicit analytic forms of the potential $S(\vec x)$ and the noise correlation function $\gamma(\vec x)$ at the soft scale $|\vec x|\sim 1/gT$ are given in \eqref{eq:stochastic_potential_HTL} in the Appendix \ref{app:Thermal_Corr_Gluons}, which we quote here
\begin{align}
S(\vec x) = \frac{g^2}{8\pi |\vec x|}e^{-m_D |\vec x|}, \quad
\gamma(\vec x) = g^2T \int \frac{d^3q}{(2\pi)^3}e^{i\vec q\cdot \vec x}
\frac{\pi m_D^2}{q(q^2 + m_D^2)^2},
\end{align}
where $m_D^2 = \frac{1}{3}g^2 T^2 \left(N_c + \frac{1}{2}N_f\right)$ for $N_c$ colors and $N_f$ light flavors.

The stochastic potential was first derived in an Abelian plasma \cite{Akamatsu:2011se} in which the noise field has no color index and was numerically simulated (after integrating out the center-of-mass motion) in one dimension \cite{Akamatsu:2011se, Kajimoto:2017rel} and in three dimensions \cite{Rothkopf:2013kya}.
The ${\rm SU}(N_c)$ version \eqref{eq:stochastic_potential} was first derived in \cite{Akamatsu:2014qsa} and numerically simulated in one dimension \cite{sharma2020quantum, Akamatsu:2021vsh}\footnote{
To be precise, Ref.~\cite{sharma2020quantum} takes the small dipole limit for the stochastic potential and solves in the angular momentum basis including only S and P waves.
So it may be called a $1_+$ dimensional simulation.
However, once the density matrix is projected onto angular momentum by $\rho_S^{(\ell)}(t,r_1,r_2) \equiv\sum_m\int_{\Omega_1,\Omega_2}Y_{\ell m}^*(\Omega_1)Y_{\ell m}(\Omega_2) \rho_S(t,r_1,\Omega_1,r_2,\Omega_2)$, the stochastic description is no longer available because the Lindblad operators are not Hermitian anymore.
}.
Conventional description for the quarkonium dissociation has mainly focused on the color screening in the QGP.
However, our analysis based on the open quantum system reveals that the decoherence is another important mechanism that contributes to the dissociation.
The importance of the decoherence can be estimated by the correlation length $\ell_{\rm corr}$ of the noise field.
If the size of a quarkonium $\ell_{\psi}$ is smaller than $\ell_{\rm corr}$, the noises for the heavy quark and antiquark cancel each other and quarkonium wave function remains almost intact\footnote{
In this case, quantum dissipation, which is neglected here, may be as important as quantum decoherence.
}.
If opposite $\ell_{\psi}\gtrsim \ell_{\rm corr}$, the wave function receives local random phase rotations and quickly mixes with the excited and unbound states.
Therefore, the fate of the quarkonium in the QGP is not only determined by the screening, but also by the decoherence.
In other words, there are two fundamental scales of the QGP concerning the quarkonium dissociation, namely the screening mass $m_D$ and the dynamical correlation length $\ell_{\rm corr}$.

Once we gain the intuitive picture by the stochastic Hamiltonian, several statements are made simpler to understand.
For example, since the stochastic Hamiltonian, or Hamiltonian dynamics in general, cannot treat non-potential forces, it is natural that the frictional force is not given by the leading order gradient expansion.
Another example is that the heavy quark momentum diffusion constant is given by $\kappa = -\frac{C_F\nabla^2\gamma(0)}{3}$.
Since the stochastic force is given by the gradient of the stochastic potential $\vec f(\vec x_Q,t) \equiv  -\vec\nabla\theta^a(\vec x_Q)t^a$ and the heavy quark wave function is localized in the classical limit, the averaged strength of the stochastic force is given by
\begin{align}
\kappa &= \frac{1}{3}\int dt E_{\theta}\left[\frac{1}{N_c}{\rm tr}\left(\vec f(\vec x_Q,t)\cdot \vec f(\vec x_Q,0)\right) \right]
= \frac{C_F}{3}\vec\nabla_x\cdot \vec\nabla_y\gamma(\vec x - \vec y)|_{\vec y\to\vec x},
\end{align}
where heavy quark color is assumed equally occupied\footnote{
Randomization of heavy quark color is achieved in a time scale of soft collision intervals $\sim 1/g^2 T$, so that this assumption is safely satisfied in the heavy quark equilibration time $\sim M/g^4 T^2$ \cite{Akamatsu:2015kaa}.
}.

Finally, let us make a comment on the unavailability of stochastic description in the projected singlet-octet basis.
The Lindblad equation in the leading order gradient expansion is
\begin{subequations}
\begin{align}
\frac{d}{dt}\rho_S(t) &= -i\left[H_S + \Delta H_S, \rho_S\right]
+\int_{x,y} \gamma(\vec x-\vec y)\sum_{n=+,-,d,f}\left[
C_n(\vec y)\rho_S C_n^{\dagger}(\vec x)
-\frac{1}{2}\left\{
C_n^{\dagger}(\vec x)C_n(\vec y), \rho_S
\right\}
\right],\\
\Delta H_S &= -2S(\vec x_Q - \vec x_{Q_c})
\begin{pmatrix}
C_F & 0 \\
0 & -\frac{1}{2N_c}
\end{pmatrix},\\
C_+(\vec x) &= \left[\delta(\vec x - \vec x_Q) - \delta(\vec x - \vec x_{Q_c})\right] \sqrt{C_F}\begin{pmatrix} 0 & 0 \\ 1 & 0 \end{pmatrix}, \\\
C_-(\vec x) &= \left[\delta(\vec x - \vec x_Q) - \delta(\vec x - \vec x_{Q_c})\right]  \sqrt{\frac{1}{2N_c}}\begin{pmatrix} 0 & 1 \\ 0 & 0 \end{pmatrix}, \\
C_d(\vec x) &= \left[\delta(\vec x - \vec x_Q) - \delta(\vec x - \vec x_{Q_c})\right]  \sqrt{\frac{N_c^2-4}{4N_c}}\begin{pmatrix} 0 & 0 \\ 0 & 1 \end{pmatrix}, \\
C_f(\vec x) &= \left[\delta(\vec x - \vec x_Q) + \delta(\vec x - \vec x_{Q_c})\right]  \sqrt{\frac{N_c}{4}}\begin{pmatrix} 0 & 0 \\ 0 & 1 \end{pmatrix}.
\end{align}
\end{subequations}
To construct a stochastic potential, one might start from introducing 4 independent noise fields $\theta_n(\vec x)$, each of which couples to $C_n(\vec x)$.
However, $C_{\pm}(\vec x)$ is not Hermitian and the resulting stochastic Hamiltonian would be non-Hermitian.
One might still try to make them Hermitian by some linear combination of $C_{\pm}(\vec x)$ as mentioned in Sec.~\ref{sec:Basics_Lindblad}, but it is impossible.
Physically, this is because the dimensions of the singlet and octet internal spaces are asymmetric and the excitation and de-excitation do not take place at the same rate.

\subsubsection{Static limit}
\label{sec:NRQCD_QQbar_Static}
Let us investigate a slightly different regime from the previous section.
Since quarkonium has singlet and octet sectors, it is interesting to study how the color configuration equilibrates.
For this purpose, the kinetic term complicates the situation because it mixes color internal spaces at different points.
Therefore, we take $M\to\infty$ limit and focus on the color space dynamics of a heavy quark pair placed at a fixed distance.

In the limit $M\to \infty$, the Lindblad equation in the projected singlet-octet basis is
\begin{subequations}
\label{eq:Lindblad_NRQCD_QQbar_static}
\begin{align}
\frac{d}{dt}\rho_S(t) &= -i\left[\Delta H_S, \rho_S\right]
+\int_{x,y} \gamma(\vec x-\vec y)\sum_{n=+,-,d,f}\left[
\tilde C_n(\vec y)\rho_S \tilde C_n^{\dagger}(\vec x)
-\frac{1}{2}\left\{
\tilde C_n^{\dagger}(\vec x)\tilde C_n(\vec y), \rho_S
\right\}
\right],\\
\Delta H_S &= -2S(\vec x_Q - \vec x_{Q_c})
\begin{pmatrix}
C_F & 0 \\
0 & -\frac{1}{2N_c}
\end{pmatrix}
\equiv \begin{pmatrix}
U_s(\vec x_Q - \vec x_{Q_c}) & 0 \\
0 & U_o(\vec x_Q - \vec x_{Q_c})
\end{pmatrix},\\
\tilde C_+(\vec x) &= \left[\delta(\vec x - \vec x_Q) - \delta(\vec x - \vec x_{Q_c})\right]
\left[1-\frac{N_c S(\vec x_Q-\vec x_{Q_c})}{4T}\right]\sqrt{C_F}\begin{pmatrix} 0 & 0 \\ 1 & 0 \end{pmatrix}, \\
\tilde C_-(\vec x) &= \left[\delta(\vec x - \vec x_Q) - \delta(\vec x - \vec x_{Q_c})\right]
\left[1+\frac{N_c S(\vec x_Q-\vec x_{Q_c})}{4T}\right]\sqrt{\frac{1}{2N_c}}\begin{pmatrix} 0 & 1 \\ 0 & 0 \end{pmatrix}, \\
\tilde C_d(\vec x) &= \left[\delta(\vec x - \vec x_Q) - \delta(\vec x - \vec x_{Q_c})\right]  \sqrt{\frac{N_c^2-4}{4N_c}}\begin{pmatrix} 0 & 0 \\ 0 & 1 \end{pmatrix}, \\
\tilde C_f(\vec x) &= \left[\delta(\vec x - \vec x_Q) + \delta(\vec x - \vec x_{Q_c})\right]  \sqrt{\frac{N_c}{4}}\begin{pmatrix} 0 & 0 \\ 0 & 1 \end{pmatrix}.
\end{align}
\end{subequations}
In this limit, the heavy quark pair is localized at $\vec r_Q$ and $\vec r_{Q_c}$ with $\vec r\equiv \vec r_Q - \vec r_{Q_c}$ and the density matrix takes the following form
\begin{align}
\rho_S(t) = 
\begin{pmatrix}
\rho_s(t) & 0 \\
0 & \rho_o(t) 
\end{pmatrix}, \quad
\rho_{s/o}(t) = n_{s/o}(t; \vec r) |\vec r_Q, \vec r_{Q_c}\rangle\langle \vec r_Q, \vec r_{Q_c}|.
\end{align}
Substituting this form, the Lindblad equation is greatly simplified to a coupled rate equations
\begin{subequations}
\begin{align}
\frac{dn_s}{dt} &= -2\left[\gamma(\vec 0)-\gamma(\vec r)\right]
\left(C_F\left[1-\frac{N_c S(\vec r)}{4T}\right]^2 n_s - \frac{1}{2N_c}\left[1+\frac{N_c S(\vec r)}{4T}\right]^2 n_o\right), \\
\frac{dn_o}{dt} &= -2\left[\gamma(\vec 0)-\gamma(\vec r)\right]
\left(\frac{1}{2N_c}\left[1+\frac{N_c S(\vec r)}{4T}\right]^2 n_o - C_F\left[1-\frac{N_c S(\vec r)}{4T}\right]^2 n_s\right).
\end{align}
\end{subequations}
In a relaxation time $\sim \left(N_c\left[\gamma(\vec 0) - \gamma(\vec r)\right]\right)^{-1}$, the color configuration reaches equilibrium
\begin{align}
\frac{n_s^{\rm eq}(\vec r)}{n_o^{\rm eq}(\vec r)}
&=\frac{1}{N_c^2-1}\left(\frac{1 + N_c S(\vec r)/4T}{1 - N_c S(\vec r)/4T}\right)^2
\simeq \frac{1}{N_c^2-1} \exp\left[\frac{N_c S(\vec r)}{T}\right],
\end{align}
which is very close to the Boltzmann distribution as long as the energy gap is not larger than the temperature; for example the error is less than $3\%$ for $N_cS(\vec r) = U_o(\vec r) - U_s(\vec r)=T$.

\subsubsection{Classical limit}
\label{sec:NRQCD_QQbar_CL}
In the previous two sections, the friction force is ignored because it vanishes in the recoilless and static limits.
Here, let us take the classical limit of Eq.~\eqref{eq:Lindblad_NRQCD_QQbar_proj} and derive corresponding classical dynamics.

First, let us first derive classical limit of quantum Brownian motion of two particles without internal degrees of freedom (Abelian case).
The Lindblad equation is
\begin{subequations}
\label{eq:Lindblad_QBM_2body}
\begin{align}
\frac{d}{dt}\rho_S(t) &= -i\left[\frac{p_1^2 + p_2^2 }{2M} + U(\vec x_1- \vec x_2) 
\pm \frac{1}{8MT} \left\{\vec p_1 - \vec p_2, \vec\nabla \gamma(\vec x_1 - \vec x_2) \right\}, \rho_S\right] \nonumber \\
& \quad +\int_k \gamma(\vec k)\left[
\tilde V_{S}(\vec k)\rho_S \tilde V_{S}^{\dagger}(\vec k)
-\frac{1}{2}\left\{
\tilde V_{S}^{\dagger}(\vec k)\tilde V_{S}(\vec k), \rho_S
\right\}
\right],\\
\tilde V_{S}(\vec k) &= e^{i\vec k\cdot\vec x_1/2}\left(1-\frac{\vec k\cdot\vec p_1}{4MT}\right)e^{i\vec k\cdot\vec x_1/2}
\pm e^{i\vec k\cdot\vec x_2/2}\left(1-\frac{\vec k\cdot\vec p_2}{4MT}\right)e^{i\vec k\cdot\vec x_2/2},
\end{align}
\end{subequations}
which is readily obtained from \eqref{eq:Lindblad_NRQCD_QQbar} by replacing $t^a\to 1$.
The signs $\pm$ in Eq.~\eqref{eq:Lindblad_QBM_2body} correspond to the cases where the two Brownian particles have the same charge ($+$) or the opposite charges ($-$).
We consider both cases because the Lindblad operators in Eq.~\eqref{eq:Lindblad_NRQCD_QQbar_proj} take both signs.
The derivation of the classical limit is straightforward but quite involved, so we summarize its procedure here.
It consists of the following 3 steps:
\begin{enumerate}
\item Write the Lindblad equation \eqref{eq:Lindblad_QBM_2body} in position space.
The density matrix is $\rho_S(t,\vec x_1,\vec x_2, \vec y_1, \vec y_2)$.
Note that $\vec x_{1,2}$ and $\vec y_{1,2}$ are just coordinates for particles 1 and 2 and not operators as in Eq.~\eqref{eq:Lindblad_QBM_2body}.
\item Put $\hbar$ where necessary ($\partial_t \to \hbar \partial_t$, $\nabla\to \hbar\nabla$, and $\gamma(\vec x)\to \gamma(\vec x)/\hbar$) and take $\hbar\to 0$ limit. 
It is simpler to
(i) substitute $\vec x_{1,2} = \vec r_{1,2} + \frac{1}{2}\vec s_{1,2}$ and $\vec y_{1,2} = \vec r_{1,2} - \frac{1}{2}\vec s_{1,2}$,
(ii) In the differential operators, Taylor expand the coefficient functions in terms of $s$, and
(iii) count $s\sim \hbar\nabla_p$ and $\nabla_s\sim p/\hbar$ and ignore the terms that vanish in $\hbar\to 0$.
\item Perform the Wigner transformation: 
\begin{align}
f(t,\vec r_1, \vec r_2, \vec p_1, \vec p_2) \equiv \int d^3 s_1 d^3 s_2  e^{-i\frac{\vec p_1\cdot\vec s_1 + \vec p_2\cdot\vec s_2}{\hbar}}\rho_S\left(t,\vec r_1+\frac{\vec s_1}{2},\vec r_2+\frac{\vec s_2}{2}, \vec r_1 - \frac{\vec s_1}{2}, \vec r_2 - \frac{\vec s_2}{2}\right).
\end{align}
\end{enumerate}
In this procedure, we use the fact that $\gamma(\vec k)$ is an even function.
The resulting kinetic equation is
\begin{align}
\label{eq:Kramers_2body}
\left[\begin{aligned}
&\frac{\partial}{\partial t} + \frac{\vec p_1 \cdot \vec\nabla_{r_1} + \vec p_2 \cdot \vec\nabla_{r_2}}{M}
-\vec\nabla U(\vec r_1-\vec r_2)\cdot \left(\vec \nabla_{p_1} - \vec\nabla_{p_2}\right) \\
&+\frac{1}{2}\partial_i\partial_j\gamma(\vec 0) \left(
\frac{\partial^2}{\partial p_{1i}\partial p_{1j}} + \frac{\partial}{\partial p_{1i}}\frac{p_{1j}}{MT}
+\frac{\partial^2}{\partial p_{2i}\partial p_{2j}} + \frac{\partial}{\partial p_{2i}}\frac{p_{2j}}{MT}
\right)\\
&\pm\frac{1}{2}\partial_i\partial_j\gamma(\vec r_1-\vec r_2) \left(
\frac{2\partial^2}{\partial p_{1i}\partial p_{2j}} + \frac{\partial}{\partial p_{1i}}\frac{p_{2j}}{MT}
+ \frac{\partial}{\partial p_{2i}}\frac{p_{1j}}{MT}
\right)
\end{aligned}\right]f(t,\vec r_1, \vec r_2, \vec p_1, \vec p_2) = 0.
\end{align}
Here $\partial_i\partial_j\gamma(\vec 0) = \nabla^2\gamma(\vec 0)\delta_{ij}/3$.
The static solution of this equation is the classical Boltzmann distribution $\propto \exp\left[-\frac{1}{T}\left(\frac{p_1^2+p_2^2}{2M}+U(\vec r)\right)\right]$ with $\vec r\equiv \vec r_1 - \vec r_2$.
The first two lines correspond to an intuitive picture of classical Brownian motion of interacting two particles.
However, in addition to the potential force, this kinetic equation contains an interesting coupling between the two particles in their relaxation dynamics.

To see the physical origin of the coupling, it is easier to analyze the corresponding Langevin equation.
\begin{subequations}
\label{eq:Langevin_2body}
\begin{align}
&\frac{d}{dt}\begin{pmatrix}\vec r_1 \\ \vec r_2\end{pmatrix} = \frac{1}{M}\begin{pmatrix}\vec p_1 \\ \vec p_2\end{pmatrix}, \\
&\frac{d}{dt}\begin{pmatrix}\vec p_1 \\ \vec p_2\end{pmatrix}_i 
= -\begin{pmatrix}\vec \nabla_{r_1} \\ \vec \nabla_{r_2}\end{pmatrix}_i U(\vec r)
+\frac{1}{2MT}\begin{pmatrix}
\partial_i\partial_j\gamma(\vec 0) & \pm\partial_i\partial_j\gamma(\vec r) \\
\pm\partial_i\partial_j\gamma(\vec r) & \partial_i\partial_j\gamma(\vec 0)
\end{pmatrix} 
\begin{pmatrix}\vec p_1 \\ \vec p_2 \end{pmatrix}_j + \begin{pmatrix}\vec \xi_1 \\ \vec \xi_2 \end{pmatrix}_i, \\
&E_{\xi}[\xi_{1i}(t)\xi_{1j}(t')] = E_{\xi}[\xi_{2i}(t)\xi_{2j}(t')] = -\partial_i\partial_j\gamma(\vec 0)\delta(t-t'), \\
&E_{\xi}[\xi_{1i}(t)\xi_{2j}(t')] = \mp \partial_i\partial_j\gamma(\vec r)\delta(t-t').
\end{align}
\end{subequations}
Recall now that in the stochastic potential picture, which is $U(\vec x_1 - \vec x_2) + \theta(\vec x_1, t) \pm \theta(\vec x_2, t)$ in this case, the random force is given by the gradient of the noise field ($-\vec\nabla\theta(\vec x_1,t)$ and $\mp\vec\nabla\theta(\vec x_2,t)$) as discussed in Sec.~\ref{sec:NRQCD_QQbar_Stochastic}.
Then the random forces of the two particles are correlated because the noise field has finite correlation length.
Note that the sign of the correlation can be naturally understood from the signs of noise fields in the stochastic potential.

The Langevin equation \eqref{eq:Langevin_2body} for a pair of heavy fermions in an Abelian plasma with correlated noise was first derived in \cite{Blaizot:2015hya}.
It was extended to more than two heavy fermions and simulated for tens of heavy fermion pairs in \cite{Blaizot:2015hya}.
Note that the equation \eqref{eq:Langevin_2body} is an example of the generalized Langevin equation, often used in the critical dynamics \cite{chaikin_lubensky_1995}.
Let $\phi_i$ be a set of slow variables.
They evolve according to the stochastic equation
\begin{align}
\label{eq:generalized_Langevin}
\dot\phi_i = \left\{\phi_i, F(\phi)\right\}_{\rm PB} - \frac{K_{ij}(\phi)}{2T}\frac{\partial F(\phi)}{\partial\phi_j} + \xi_i, \quad
E_{\xi}\left[\xi_i(t)\xi_j(t')\right] = K_{ij}(\phi)\delta(t-t'),
\end{align}
and their equilibrium distribution is $\propto \exp\left[-F(\phi)/T\right]$ with $F(\phi)$ being the free energy.
Here $\{A,B\}_{\rm PB}$ denotes the Poisson bracket and we assume that $\left\{\phi_i, \phi_j\right\}_{\rm PB}$ is independent of $\phi_k$.
In our case, $\phi = (\vec r_1, \vec r_2, \vec p_1, \vec p_2)$, $F(\phi)=(p_1^2+p_2^2)/2M+U(\vec r_1 - \vec r_2)$, and $K$ is finite only in the momentum sector
\begin{align}
K=\begin{pmatrix}
0 & 0 \\
0 & \kappa
\end{pmatrix},\quad
\kappa=
\begin{pmatrix}
-\partial_i\partial_j\gamma(\vec 0) & \mp\partial_i\partial_j\gamma(\vec r) \\
\mp\partial_i\partial_j\gamma(\vec r) & -\partial_i\partial_j\gamma(\vec 0)
\end{pmatrix} .
\end{align}

Now that we understand the classical limit of quantum Brownian motion for two particles without internal degrees of freedom, let us derive the same limit for particles with color degrees of freedom.
In Eq.~\eqref{eq:Lindblad_NRQCD_QQbar_proj}, the Lindblad operators $\tilde C_{\pm}(\vec k)$ describe the transitions between the singlet and the octet and $\tilde C_{d,f}(\vec k)$ describe the collisions among the octet channels, all with momentum transfer $\vec k$.
However, when taking the classical limit, we face with a strange situation: the color sector is divided into singlet and octet but the transitions between them have no classical counterpart.
To see the problem, let us rewrite the Lindblad equations \eqref{eq:Lindblad_NRQCD_QQbar_proj} for the singlet ($\rho_s$) and the octet ($\rho_o$) density matrices by explicitly separating the transition processes
\begin{subequations}
\label{eq:Lindblad_NRQCD_QQbar_proj_explicit}
\begin{align}
\frac{d}{dt}\rho_s(t) &= -i\left[\frac{p_Q^2 + p_{Q_c}^2 }{2M} + U_s(\vec x_Q- \vec x_{Q_c}) -\frac{C_F}{8MT} \left\{\vec p_Q - \vec p_{Q_c}, \vec\nabla \gamma(\vec x_Q - \vec x_{Q_c}) \right\}, \rho_s\right] \nonumber \\
&\quad +\int_k \gamma(\vec k)\left[
\frac{1}{2N_c}\tilde V_-(\vec k)\rho_o \tilde V_-^{\dagger}(\vec k)
-C_F\tilde V_+(\vec k)\rho_s\tilde V_+^{\dagger}(\vec k)
\right]\nonumber \\
&\quad + C_F\int_k \gamma(\vec k)\left[
\tilde V_+(\vec k)\rho_s \tilde V_+^{\dagger}(\vec k)
-\frac{1}{2}\left\{
\tilde V_+^{\dagger}(\vec k)\tilde V_+(\vec k), \rho_s
\right\}
\right], \\%
\label{eq:Lindblad_NRQCD_QQbar_proj_explicit_octet}
\frac{d}{dt}\rho_o(t) &= -i\left[\frac{p_Q^2 + p_{Q_c}^2 }{2M} + U_o(\vec x_Q- \vec x_{Q_c}) +\frac{1}{2N_c}\frac{1}{8MT} \left\{\vec p_Q - \vec p_{Q_c}, \vec\nabla \gamma(\vec x_Q - \vec x_{Q_c}) \right\}, \rho_o\right] \nonumber \\
&\quad +\int_k \gamma(\vec k)\left[
C_F\tilde V_+(\vec k)\rho_s \tilde V_+^{\dagger}(\vec k)
-\frac{1}{2N_c}\tilde V_-(\vec k)\rho_o\tilde V_-^{\dagger}(\vec k)
\right]\nonumber \\
&\quad + \frac{1}{2N_c}\int_k \gamma(\vec k)\left[
\tilde V_-(\vec k)\rho_o \tilde V_-^{\dagger}(\vec k)
-\frac{1}{2}\left\{
\tilde V_-^{\dagger}(\vec k)\tilde V_-(\vec k), \rho_o
\right\}
\right]\nonumber \\
&\quad +\frac{N_c^2-4}{4N_c}\int_k \gamma(\vec k)\left[
\tilde V_d(\vec k)\rho_o \tilde V_d^{\dagger}(\vec k) 
-\frac{1}{2}\left\{\tilde V_d^{\dagger}(\vec k)\tilde V_d(\vec k), \rho_o\right\} 
\right]\nonumber \\
&\quad +\frac{N_c}{4} \int_k \gamma(\vec k)\left[
\tilde V_f(\vec k)\rho_o \tilde V_f^{\dagger}(\vec k) 
-\frac{1}{2}\left\{\tilde V_f^{\dagger}(\vec k)\tilde V_f(\vec k), \rho_o\right\} 
\right],
\end{align}
\end{subequations}
with $\tilde V_{+,-,d,f}(\vec k)$ defined in Eq.~\eqref{eq:Lindblad_NRQCD_QQbar_sa} and $U_s(\vec r) \equiv -2C_F S(\vec r)$ and $U_o(\vec r) \equiv (1/N_c)S(\vec r)$.
Here we can see that the physical meaning of $\tilde C_{\pm}(\vec k)$ (or $\tilde V_{\pm}(\vec k)$) is split into transitions and collisions.
In the equations for singlet and octet, the second lines describe transitions between them while the third lines and below describe collisions without changing the color sector.
To give a classical interpretation, we must slightly modify the Lindblad operators for these processes:
\begin{subequations}
\begin{align}
\text{\rm Transitions: } \tilde V_{\pm}(\vec k) 
&\to e^{i\vec k\cdot\vec x_Q}\left[1 \mp \frac{N_c S(\vec x_Q - \vec x_{Q_c})}{4T}\right]
-e^{i\vec k\cdot\vec x_{Q_c}}\left[1 \mp \frac{N_c S(\vec x_Q - \vec x_{Q_c})}{4T}\right], \\
\text{\rm Collisions: } \tilde V_{\pm}(\vec k) 
&\to e^{i\vec k\cdot\vec x_Q/2}\left(1-\frac{\vec k\cdot\vec p_Q}{4MT}\right)e^{i\vec k\cdot\vec x_Q/2} 
-e^{i\vec k\cdot\vec x_{Q_c}/2}\left(1-\frac{\vec k\cdot\vec p_{Q_c}}{4MT}\right)e^{i\vec k\cdot\vec x_{Q_c}/2}.
\end{align}
\end{subequations}
The modification for ``Transitions" corresponds to ignoring the change of kinetic energy in the transitions (because it is accounted for by the ``Collisions") and that for ``Collisions" corresponds to discarding the change of the potential energy in the collisions (because it does not change)\footnote{
After this modification, $\tilde V_-(\vec k)$ for the ``Collisions" is identical to $\tilde V_d(\vec k)$ and the third and fourth lines of octet equations in Eq.~\eqref{eq:Lindblad_NRQCD_QQbar_proj_explicit} can be brought together with a new coefficient $(N_c^2-2)/4N_c$.
}.
The ``Transitions" are purely quantum processes $(\propto \hbar^{-2})$ and their formal classical limit is still ill-defined while the ``Collisions" have a classical counter part.
By recovering $\hbar$ and picking up the leading terms in $\hbar\to 0$ for each process, we arrive at the following Langevin equations for the singlet and octet:
\begin{subequations}
\label{eq:Langevin_QQbar_singlet}
\begin{align}
& \hspace{-5mm}\text{\rm Singlet :}\\
&\frac{d}{dt}\begin{pmatrix}\vec r_Q \\ \vec r_{Q_c}\end{pmatrix} = \frac{1}{M}\begin{pmatrix}\vec p_Q \\ \vec p_{Q_c}\end{pmatrix}, \\
&\frac{d}{dt}\begin{pmatrix}\vec p_Q \\ \vec p_{Q_c}\end{pmatrix}_i 
= -\begin{pmatrix}\vec \nabla_{r_Q} \\ \vec \nabla_{r_{Q_c}}\end{pmatrix}_i U_s(\vec r)
+\frac{C_F}{2MT}\begin{pmatrix}
\partial_i\partial_j\gamma(\vec 0) & -\partial_i\partial_j\gamma(\vec r) \\
-\partial_i\partial_j\gamma(\vec r) & \partial_i\partial_j\gamma(\vec 0)
\end{pmatrix} 
\begin{pmatrix}\vec p_Q \\ \vec p_{Q_c} \end{pmatrix}_j + \begin{pmatrix}\vec \xi_Q \\ \vec \xi_{Q_c} \end{pmatrix}_i, \\
&E_{\xi}[\xi_{Qi}(t)\xi_{Qj}(t')] = E_{\xi}[\xi_{Q_c i}(t)\xi_{Q_c j}(t')] = -C_F\partial_i\partial_j\gamma(\vec 0)\delta(t-t'), \\
&E_{\xi}[\xi_{Qi}(t)\xi_{Q_c j}(t')] = C_F \partial_i\partial_j\gamma(\vec r)\delta(t-t'), 
\end{align}
\end{subequations}
\begin{subequations}
\label{eq:Langevin_QQbar_octet}
\begin{align}
& \hspace{-5mm}\text{\rm Octet :} \\
&\frac{d}{dt}\begin{pmatrix}\vec r_Q \\ \vec r_{Q_c}\end{pmatrix} = \frac{1}{M}\begin{pmatrix}\vec p_Q \\ \vec p_{Q_c}\end{pmatrix}, \\
&\frac{d}{dt}\begin{pmatrix}\vec p_Q \\ \vec p_{Q_c}\end{pmatrix}_i 
= -\begin{pmatrix}\vec \nabla_{r_Q} \\ \vec \nabla_{r_{Q_c}}\end{pmatrix}_i U_o(\vec r)
+\frac{1}{2MT}\begin{pmatrix}
C_F\partial_i\partial_j\gamma(\vec 0) & \frac{1}{2N_c}\partial_i\partial_j\gamma(\vec r) \\
\frac{1}{2N_c}\partial_i\partial_j\gamma(\vec r) & C_F\partial_i\partial_j\gamma(\vec 0)
\end{pmatrix} 
\begin{pmatrix}\vec p_Q \\ \vec p_{Q_c} \end{pmatrix}_j + \begin{pmatrix}\vec \xi_Q \\ \vec \xi_{Q_c} \end{pmatrix}_i, \\
&E_{\xi}[\xi_{Qi}(t)\xi_{Qj}(t')] = E_{\xi}[\xi_{Q_c i}(t)\xi_{Q_c j}(t')] = -C_F\partial_i\partial_j\gamma(\vec 0)\delta(t-t'), \\
&E_{\xi}[\xi_{Qi}(t)\xi_{Q_c j}(t')] = -\frac{1}{2N_c} \partial_i\partial_j\gamma(\vec r)\delta(t-t'),
\end{align}
\end{subequations}
with $\vec r\equiv \vec r_Q - \vec r_{Q_c}$.
The color state changes with flipping probabilities ($\Gamma_{s\to o}\Delta t$ and $\Gamma_{o\to s}\Delta t$) for each time step ($t\to t + \Delta t$)
\begin{subequations}
\label{eq:color_flipping_rate}
\begin{align}
\Gamma_{s\to o}&=2C_F\left[1-\frac{U_o(\vec r)-U_s(\vec r)}{4T}\right]^2\frac{\gamma(\vec 0)-\gamma(\vec r)}{\hbar^2}, \\
\Gamma_{o\to s}&=\frac{1}{N_c}\left[1+\frac{U_o(\vec r) - U_s(\vec r)}{4T}\right]^2\frac{\gamma(\vec 0)-\gamma(\vec r)}{\hbar^2}.
\end{align}
\end{subequations}
Since the detailed balance between the color sectors holds approximately 
\begin{align}
\frac{\Gamma_{o\to s}}{\Gamma_{s\to o}}
=\frac{1}{N_c^2-1} \left(\frac{1+[U_o(\vec r)-U_s(\vec r)]/4T}{1-[U_o(\vec r)-U_s(\vec r)]/4T}\right)^2
\simeq \frac{1}{N_c^2-1} \exp\left[\frac{U_o(\vec r) - U_s(\vec r)}{T}\right],
\end{align}
and the phase space dynamics is again given by the generalized Langevin equation, the equilibrium distribution for the Langevin equation is $\propto \exp\left[-\frac{(p_Q^2+p_{Q_c}^2)/2M+U_s(\vec r)}{T}\right]$ for the singlet and $\propto (N_c^2-1)\cdot\exp\left[-\frac{(p_Q^2+p_{Q_c}^2)/2M+U_o(\vec r)}{T}\right]$ for the octet within the same approximation.
This approximation holds quite well when $U_o(\vec r)-U_s(\vec r) \lesssim T$ but breaks down for larger energy gap.
In particular, the rate for singlet-to-octet transition is divergent at $U_o(\vec r)-U_s(\vec r) = 4T$.
This is where the gradient expansion for the quantum Brownian motion fails completely and gives a reasonable scale for the ultraviolet cutoff for the wave functions of relative motion.
If one dares to simulate classically, the transition rate needs to be regulated when the heavy quark pair is too close, for example by
\begin{subequations}
\begin{align}
\Gamma_{s\to o}&\approx 2C_F\exp\left[-\frac{U_o(\vec r)-U_s(\vec r)}{2T}\right]\frac{\gamma(\vec 0)-\gamma(\vec r)}{\hbar^2}, \\
\Gamma_{o\to s}&\approx \frac{1}{N_c}\exp\left[\frac{U_o(\vec r) - U_s(\vec r)}{2T}\right]\frac{\gamma(\vec 0)-\gamma(\vec r)}{\hbar^2}.
\end{align}
\end{subequations}

This Langevin equation for a heavy quark pair in the quark-gluon plasma was first derived in \cite{Blaizot:2017ypk}.
In the original paper, an important factor $\left[1 \mp (U_o(\vec r) - U_s(\vec r))/4T\right]^2$ or $\exp[\mp(U_o(\vec r)-U_s(\vec r))/2T]$ in the flipping rate was missing, i.e. the color dependence of the potential energy is not taken into account in the relaxation process.

\subsubsection{Small dipole limit}
\label{sec:NRQCD_QQbar_Dipole}
The final limit we take is the small dipole limit.
We examine this limit in order to compare with the results in the Sec.~\ref{sec:pNRQCD} where the quarkonium Lindblad equation is studied in the potential NRQCD (pNRQCD) approach.

When the heavy quark pair is close to each other, the pair can be regarded as a localized object in the color singlet or octet states.
To get this limit, we employ the center-of-mass and the relative coordinates (for the operators)
\begin{align}
\vec X\equiv \frac{\vec x_Q + \vec x_{Q_c}}{2}, \quad
\vec x\equiv \vec x_Q - \vec x_{Q_c},\quad
\vec P\equiv \vec p_Q + \vec p_{Q_c}, \quad
\vec p\equiv \frac{\vec p_Q - \vec p_{Q_c}}{2}.
\end{align}
In these coordinates, the matrix structure of the Lindblad operators $\tilde C_n(\vec k)$ in Eq.~\eqref{eq:Lindblad_NRQCD_QQbar_proj} is unchanged while $\tilde V_n(\vec k)$ is
\begin{subequations}
\begin{align}
\tilde V_{\pm}(\vec k) &= 
e^{i\vec k\cdot\vec X}\left[1-\frac{\vec k}{4MT}\cdot\left(\frac{1}{2}\vec P + \vec p\right) \mp \frac{N_c S(\vec x)}{4T}\right]
e^{i\vec k\cdot\vec x/2} \nonumber \\
&\quad -e^{i\vec k\cdot\vec X}\left[1-\frac{\vec k}{4MT}\cdot\left(\frac{1}{2}\vec P - \vec p\right) \mp \frac{N_c S(\vec x)}{4T}\right]
e^{-i\vec k\cdot\vec x/2}, \\
\tilde V_{d/f}(\vec k) &= 
\pm e^{i\vec k\cdot\vec X}\left[1-\frac{\vec k}{4MT}\cdot\left(\frac{1}{2}\vec P + \vec p\right)\right]e^{i\vec k\cdot\vec x/2} 
-e^{i\vec k\cdot\vec X}\left[1-\frac{\vec k}{4MT}\cdot\left(\frac{1}{2}\vec P - \vec p\right)\right]e^{-i\vec k\cdot\vec x/2},
\end{align}
\end{subequations}
where an unitary operator $e^{i\vec k\cdot\vec X}$ is moved to the left.
Since we are interested in the relative motion of the heavy quark pair, we define a further reduced density matrix
\begin{align}
\rho^{(r)}_S \equiv {\rm Tr}_X\rho_S,
\end{align}
where ${\rm Tr}_X$ denotes tracing over the Hilbert space for $X$.
In general, the reduced density matrix is entangled between the Hilbert spaces of the center-of-mass motion and of the relative and internal dynamics.
For an entangled density matrix, the operation of partial trace ${\rm Tr}_X$ gives rise to such terms as ${\rm Tr}_X\left(\vec P \rho_S\right)$ from which $\rho^{(r)}_S$ does not factorize.
Therefore, for simplicity we assume the following factorized form of the reduced density matrix
\begin{align}
\rho_S = |\vec P\rangle\langle \vec P|\otimes \rho^{(r)}_S.
\end{align}
Since the pNRQCD calculation is performed for a quarkonium staying at rest, we take $\vec P=\vec 0$.
Then, the evolution of $\rho^{(r)}_S$ is obtained in the Lindblad form
\begin{subequations}
\label{eq:Lindblad_NRQCD_QQbar_relative}
\begin{align}
\frac{d}{dt}\rho^{(r)}_S(t) &= -i\left[H_S^{(r)} + \Delta H_S^{(r)}, \rho_S^{(r)}\right]
+\int_k \gamma(\vec k)\sum_{n=+,-,d,f}\left[
\tilde C^{(r)}_n(\vec k)\rho_S^{(r)} \tilde C_n^{(r)\dagger}(\vec k)
-\frac{1}{2}\left\{
\tilde C_n^{(r)\dagger}(\vec k)\tilde C_n^{(r)}(\vec k), \rho_S^{(r)}
\right\}
\right],\\
H_S^{(r)}&=\frac{p^2}{M} ,\quad
\Delta H_S^{(r)} = \left[
-2S(\vec x)
-\frac{1}{4MT} \left\{\vec p, \vec\nabla \gamma(\vec x) \right\}
\right]
\begin{pmatrix}
C_F & 0 \\
0 & -\frac{1}{2N_c}
\end{pmatrix}, \\
\tilde C^{(r)}_+(\vec k) &= \tilde V_+^{(r)}(\vec k) \sqrt{C_F}\begin{pmatrix} 0 & 0 \\ 1 & 0 \end{pmatrix}, \quad
\tilde C^{(r)}_-(\vec k) = \tilde V_-^{(r)}(\vec k) \sqrt{\frac{1}{2N_c}}\begin{pmatrix} 0 & 1 \\ 0 & 0 \end{pmatrix}, \\
\tilde C^{(r)}_d(\vec k) &= \tilde V_d^{(r)}(\vec k) \sqrt{\frac{N_c^2-4}{4N_c}}\begin{pmatrix} 0 & 0 \\ 0 & 1 \end{pmatrix}, \quad
\tilde C^{(r)}_f(\vec k) = \tilde V_f^{(r)}(\vec k) \sqrt{\frac{N_c}{4}}\begin{pmatrix} 0 & 0 \\ 0 & 1 \end{pmatrix},
\end{align}
\end{subequations}
with
\begin{subequations}
\begin{align}
\tilde V_{\pm}^{(r)}(\vec k) &= 
\left[1-\frac{\vec k\cdot\vec p}{4MT} \mp \frac{N_c S(\vec x)}{4T}\right]e^{i\vec k\cdot\vec x/2}
-\left[1+\frac{\vec k\cdot\vec p}{4MT} \mp \frac{N_c S(\vec x)}{4T}\right]e^{-i\vec k\cdot\vec x/2}, \\
\tilde V_{d/f}^{(r)}(\vec k) &= 
\pm \left[1-\frac{\vec k\cdot\vec p}{4MT}\right]e^{i\vec k\cdot\vec x/2} 
-\left[1+\frac{\vec k\cdot\vec p}{4MT}\right]e^{-i\vec k\cdot\vec x/2}.
\end{align}
\end{subequations}
Instead of taking $\vec P = \vec 0$, one may also be able to model $\vec P(t)$ by a solution of some classical model such as Langevin dynamics of a quarkonium.

Now we can take the small dipole limit as follows.
\begin{enumerate}
\item Take the small $\vec x$ limit for $\Delta H_S^{(r)}$ and $\tilde C_n^{(r)}(\vec k)$:
\begin{subequations}
\label{eq:Lindblad_NRQCD_QQbar_smallsize}
\begin{align}
\Delta H_S^{(r)} &= \left[
-2S(\vec x)
-\frac{\nabla^2 \gamma (\vec 0)}{12MT} \left\{\vec p, \vec x\right\} + \mathcal O(x^3)
\right]
\begin{pmatrix}
C_F & 0 \\
0 & -\frac{1}{2N_c}
\end{pmatrix}, \\
\tilde C_+^{(r)}(\vec k) &= \left[i\vec k\cdot \vec x\left(1 - \frac{N_c S(\vec x)}{4T}\right) -\frac{\vec k\cdot \vec p}{2MT} + \mathcal O(x^3)\right]
\sqrt{C_F}\begin{pmatrix} 0 & 0 \\ 1 & 0 \end{pmatrix}, \\
\tilde C_-^{(r)}(\vec k) &= \left[i\vec k\cdot \vec x\left(1 + \frac{N_c S(\vec x)}{4T}\right) -\frac{\vec k\cdot \vec p}{2MT} + \mathcal O(x^3)\right]
\sqrt{\frac{1}{2N_c}}\begin{pmatrix} 0 & 1 \\ 0 & 0 \end{pmatrix},\\
\tilde C_d^{(r)}(\vec k) &=  \left[i\vec k\cdot \vec x -\frac{\vec k\cdot \vec p}{2MT} + \mathcal O(x^3)\right]
\sqrt{\frac{N_c^2-4}{4N_c}}\begin{pmatrix} 0 & 0 \\ 0 & 1 \end{pmatrix}, \\
\tilde C_f^{(r)}(\vec k) &=\left[-2 + \frac{(\vec k\cdot\vec x)^2}{4} + \frac{i(\vec k\cdot\vec p)(\vec k\cdot\vec x)}{4MT} + \mathcal O(x^3)\right]
\sqrt{\frac{N_c}{4}}\begin{pmatrix} 0 & 0 \\ 0 & 1 \end{pmatrix}.
\end{align}
\end{subequations}
Here, we count $\vec p = \frac{M}{2}\dot {\vec x}\sim \mathcal O(x)$.
In this limit, $\tilde C_{+,-,d}^{(r)}(\vec k)$ implies the coupling between the quarkonium dipole and the color electric fields while $\tilde C_f^{(r)}(\vec k)$ reminds us of the adjoint gauge interaction of an octet quarkonium with finite size corrections.
For the latter, remember that $\tilde C_f^{(r)}(\vec k)$ comes from $\tilde V_f(\vec k) if^{abc}|b\rangle\langle c| = -\tilde V_f(\vec k)[t_A^a]_{bc}|b\rangle\langle c|$ in Eq.~\eqref{eq:Lindblad_NRQCD_QQbar_sa}, where $t_A^a$ denotes an SU($N_c$) generator in the adjoint representation.
\item Isolate the Coulomb singularity from $S(\vec x)$ and evaluate the remaining thermal correction to $\mathcal O(x^2)$:
\begin{subequations}
\label{eq:S_small_r}
\begin{align}
S(\vec x) &=S_{T=0}(\vec x) + S_{T\neq 0}(\vec x)=\frac{\alpha_s}{2 |\vec x|} + S_{T\neq 0}(\vec 0) + \frac{\nabla^2 S_{T\neq 0}(\vec 0)}{6} x^2 + \cdots,\\
\nabla^2 S_{T\neq 0}(\vec 0) &=\frac{1}{2(N_c^2-1)}{\rm Re}\left[i \int_0^{\infty} dt {\rm Tr}_E
\left(\rho_E^{\rm th} \left[g\vec\nabla A_0^a(t,\vec 0), g\vec \nabla A_0^a(0,\vec 0)\right]\right) \right]_{T\neq 0}.
\end{align}
\end{subequations}
We drop $S_{T\neq 0}(\vec 0)$ in the Hamiltonian $\Delta H_S^{(r)}$ because it is a constant. 
Note that $\nabla^2 S_{T\neq 0}(\vec 0)$ is a perturbative limit of the two point function of color electric fields.
\item Truncate the small-$\vec x$ expansion of the Lindblad equation at $\mathcal O(x^2)$.
At this order, the dissipator with $\tilde C_f^{(r)}(\vec k)$ only plays a role of Hamiltonian
\begin{align}
\label{eq:Hamiltonian_Cf_Lindblad}
\Delta H_{f}^{(r)} = -\frac{\nabla^2 \gamma(\vec 0)}{12MT} \frac{N_c}{4}\{\vec p, \vec x\}
\begin{pmatrix}0 & 0 \\ 0 & 1\end{pmatrix},
\end{align}
which we include in $\Delta H_S^{(r)}$.
\item Complete the integration over $\vec k$.
\end{enumerate}
With this procedure, the Lindblad equation is finally obtained as
\begin{subequations}
\label{eq:Lindblad_NRQCD_QQbar_dipole}
\begin{align}
\frac{d}{dt}\rho^{(r)}_S(t) &= -i\left[H_S^{(r)} + \Delta H_S^{(r)}, \rho_S^{(r)}\right]
-\frac{\nabla^2\gamma(\vec 0)}{3} \sum_{n=+,-,d}\left[
\tilde C_{ni}\rho_S^{(r)} \tilde C_{ni}^{\dagger}
-\frac{1}{2}\left\{
\tilde C_{ni}^{\dagger} \tilde C_{ni}, \rho_S^{(r)}
\right\}
\right],\\
\Delta H_S^{(r)} &= \left[
-\frac{\alpha_s}{|\vec x|}
-\frac{\nabla^2 S_{T\neq 0}(\vec 0)}{3}x^2
\right]
\begin{pmatrix}
C_F & 0 \\
0 & -\frac{1}{2N_c}
\end{pmatrix}
-\frac{\nabla^2 \gamma (\vec 0)}{12MT} \left\{\vec p, \vec x\right\}
\begin{pmatrix}
C_F & 0 \\
0 & \frac{N_c}{4}-\frac{1}{2N_c}
\end{pmatrix}, \\
\tilde C_{+i} &= \left[\vec x\left(1 -\frac{N_c\alpha_s}{8T |\vec x|} \right) + \frac{i\vec p}{2MT} \right]_i \sqrt{C_F}\begin{pmatrix} 0 & 0 \\ 1 & 0 \end{pmatrix}, \\
\tilde C_{-i} &= \left[\vec x\left(1 + \frac{N_c\alpha_s}{8T |\vec x|} \right) + \frac{i\vec p}{2MT} \right]_i  \sqrt{\frac{1}{2N_c}}\begin{pmatrix} 0 & 1 \\ 0 & 0 \end{pmatrix}, \\
\tilde C_{di} &= \left[\vec x + \frac{i\vec p}{2MT} \right]_i  \sqrt{\frac{N_c^2-4}{4N_c}}\begin{pmatrix} 0 & 0 \\ 0 & 1 \end{pmatrix}.
\end{align}
\end{subequations}
In this limit, the Lindblad equation is parametrized by two constants: the momentum diffusion constant $\kappa = -\frac{C_F\nabla^2\gamma(\vec 0)}{3}>0$ and the coefficient for dipole self-energy of the singlet $\lambda = -\frac{2C_F\nabla^2 S_{T\neq 0}(\vec 0)}{3}$, which we call thermal dipole self-energy constant in this paper\footnote{
Since the master equation is obtained by weak coupling expansion, $\lambda$ here should be the leading order perturbative one
$\lambda_{\rm LO} = -2\zeta(3)C_F\left(\frac{4}{3}N_c + N_f\right)\alpha_s^2 T^3$ \cite{Brambilla:2008cx}.
See Appendix \ref{app:Thermal_Corr_Gluons} for details.
}.

The form of the Lindblad equation \eqref{eq:Lindblad_NRQCD_QQbar_dipole} is identical to that of the pNRQCD result \eqref{eq:Lindblad_pNRQCD_QQbar_proj} in Sec.~\ref{sec:pNRQCD_strong} except for the quadratic term $\propto x^2$ of $\Delta H_S^{(r)}$ in the octet channel.
Not only are the Lindblad equations almost identical, but the two constants $-\frac{\nabla^2\gamma(\vec 0)}{3}>0$ and $-\frac{\nabla^2 S_{T\neq 0}(\vec 0)}{3} < 0$ of \eqref{eq:Lindblad_NRQCD_QQbar_dipole} are weak coupling limit of the corresponding constants $\gamma$ and $S$ of \eqref{eq:Lindblad_pNRQCD_QQbar_proj}.
The difference of these Lindblad equations comes from the gauge interaction of an octet quarkonium.
In NRQCD, the gauge interaction contains corrections due to the finite size of the quarkonium, while in pNRQCD the octet quarkonium interacts with the gauge field as a point-like particle in the next-to-leading order multipole expansion.
Up to $\mathcal O(x^2)$, the only effect of the finite size correction in the Lindblad equation  is the quadratic term in $\Delta H_S^{(r)}$.
This term is absent in the pNRQCD because the finite size correction is not included.

To verify more explicitly that the difference is due to the finite size correction, recall that $\Delta H_S$ is given by the Lindblad operators as in Eq.~\eqref{eq:Lindblad_QBM}.
The gauge interaction for an octet quarkonium with finite size correction is described by the Lindblad operator $\tilde C_f^{(r)}(\vec k)$ in Eq.~\eqref{eq:Lindblad_NRQCD_QQbar_smallsize}
\begin{align}
\tilde C_f^{(r)}(\vec k) 
=\left[\left(-2 + \frac{(\vec k\cdot\vec x)^2}{4}\right) + \frac{i}{4T}\frac{(\vec k\cdot\vec p)(\vec k\cdot\vec x)}{M} +
\mathcal O(x^3)\right]
\sqrt{\frac{N_c}{4}}\begin{pmatrix} 0 & 0 \\ 0 & 1 \end{pmatrix}
\equiv C_f^{(r)}(\vec k) + \frac{i}{4T}\dot C_f^{(r)}(\vec k),
\end{align}
with which $\Delta H_S^{(r)}$ is given by
\begin{align}
\Delta H_S^{(r)}\Bigr|_{\tilde C^{(r)}_f}&=\int_k S_{T\neq 0}(\vec k) C_f^{(r)\dagger}(\vec k)C_f^{(r)}(\vec k)
+\frac{1}{8T}\left[\gamma(\vec k)C_f^{(r)\dagger}(\vec k)\dot C_f^{(r)}(\vec k)
+h.c.\right] \nonumber\\
&= \left[4S_{T\neq 0}(\vec 0) + \frac{\nabla^2 S_{T\neq 0}(\vec 0)}{3}x^2 + \frac{\nabla^2\gamma(\vec 0)}{12MT}\left\{\vec p,\vec x\right\}
+\mathcal O(x^3)\right]
\frac{N_c}{4}\begin{pmatrix} 0 & 0 \\ 0 & 1 \end{pmatrix}.
\end{align}
The first term does not play any role for a density matrix of the form \eqref{eq:rho_so_diagonal} and the third term is exactly canceled by $\Delta H_{f}^{(r)}$ of Eq.~\eqref{eq:Hamiltonian_Cf_Lindblad} from the Lindblad form, leaving the second term as the only difference from the pNRQCD result.

Finally, as is clear from the order of the gradient expansion of each term, the recoilless limit for a small dipole corresponds to
\begin{subequations}
\begin{align}
\Delta H_S^{(r)} &= \left[
-\frac{\alpha_s}{|\vec x|}
-\frac{\nabla^2 S_{T\neq 0}(\vec 0)}{3}x^2
\right]
\begin{pmatrix}
C_F & 0 \\
0 & -\frac{1}{2N_c}
\end{pmatrix}, \\
C_{+i} &= x_i \sqrt{C_F}\begin{pmatrix} 0 & 0 \\ 1 & 0 \end{pmatrix}, \quad
C_{-i} = x_i  \sqrt{\frac{1}{2N_c}}\begin{pmatrix} 0 & 1 \\ 0 & 0 \end{pmatrix}, \quad
C_{di} = x_i  \sqrt{\frac{N_c^2-4}{4N_c}}\begin{pmatrix} 0 & 0 \\ 0 & 1 \end{pmatrix}.
\end{align}
\end{subequations}
This is again different from the recoilless limit of the pNRQCD results \eqref{eq:Lindblad_pNRQCD_QQbar_proj_recoilless} for the same reason as above.

\subsection{Concluding remarks of Section \ref{sec:NRQCD}}
\label{sec:NRQCD_conclusion}
In this section, the main result is the Lindblad equation \eqref{eq:Lindblad_NRQCD_QQbar_proj} for quantum Brownian motion of a quarkonium in a weakly coupled quark-gluon plasma, obtained by applying the formulas of section \ref{sec:Basics} to quarkonium, which is described by the non-relativistic QCD.
A density matrix in the projected singlet-octet basis
\begin{align}
\rho_S(t) = \begin{pmatrix}
\rho_s(t) & 0 \\
0 & \rho_o(t)
\end{pmatrix} \nonumber
\end{align}
evolves according to the following Lindblad equation:
\begin{align}
\frac{d}{dt}\rho_S(t) &= -i\left[H_S + \Delta H_S, \rho_S\right]
+\int_k \gamma(\vec k)\sum_{n=+,-,d,f}\left[
\tilde C_n(\vec k)\rho_S \tilde C_n^{\dagger}(\vec k)
-\frac{1}{2}\left\{
\tilde C_n^{\dagger}(\vec k)\tilde C_n(\vec k), \rho_S
\right\}
\right], \nonumber\\
\Delta H_S &= \left[
-2S(\vec x_Q - \vec x_{Q_c})
-\frac{1}{8MT} \left\{\vec p_Q - \vec p_{Q_c}, \vec\nabla \gamma(\vec x_Q - \vec x_{Q_c}) \right\}
\right]
\begin{pmatrix}
C_F & 0 \\
0 & -\frac{1}{2N_c}
\end{pmatrix}, \nonumber \\
\tilde C_+(\vec k) &= \tilde V_+(\vec k) \sqrt{C_F}\begin{pmatrix} 0 & 0 \\ 1 & 0 \end{pmatrix}, \quad
\tilde C_-(\vec k) = \tilde V_-(\vec k) \sqrt{\frac{1}{2N_c}}\begin{pmatrix} 0 & 1 \\ 0 & 0 \end{pmatrix}, \nonumber \\
\tilde C_d(\vec k) &= \tilde V_d(\vec k) \sqrt{\frac{N_c^2-4}{4N_c}}\begin{pmatrix} 0 & 0 \\ 0 & 1 \end{pmatrix}, \quad
\tilde C_f(\vec k) = \tilde V_f(\vec k) \sqrt{\frac{N_c}{4}}\begin{pmatrix} 0 & 0 \\ 0 & 1 \end{pmatrix}, \nonumber
\end{align}
where the operators $\tilde V_{+,-,d,f}(\vec k)$ are
\begin{align}
\tilde V_{\pm}(\vec k) &= e^{i\vec k\cdot\vec x_Q/2}\left[
1-\frac{\vec k\cdot\vec p_Q}{4MT}
\mp \frac{N_c S(\vec x_Q - \vec x_{Q_c})}{4T}
\right]e^{i\vec k\cdot\vec x_Q/2} 
-e^{i\vec k\cdot\vec x_{Q_c}/2}\left[
1-\frac{\vec k\cdot\vec p_{Q_c}}{4MT}
\mp \frac{N_c S(\vec x_Q - \vec x_{Q_c})}{4T}
\right]e^{i\vec k\cdot\vec x_{Q_c}/2}, \nonumber \\
\tilde V_{d/f}(\vec k) &= \pm e^{i\vec k\cdot\vec x_Q/2}\left(1-\frac{\vec k\cdot\vec p_Q}{4MT}\right)e^{i\vec k\cdot\vec x_Q/2}
-e^{i\vec k\cdot\vec x_{Q_c}/2}\left(1-\frac{\vec k\cdot\vec p_{Q_c}}{4MT}\right)e^{i\vec k\cdot\vec x_{Q_c}/2} \nonumber
\end{align}
and the coefficients $S(\vec x)$ and $\gamma(\vec x)$ are defined by a correlation function of gauge fields and are calculated perturbatively at $|\vec x|\sim 1/gT$ as (quoted from Appendix \ref{app:Thermal_Corr_Gluons})
\begin{align}
S(\vec x) = \frac{g^2}{8\pi |\vec x|}e^{-m_D |\vec x|}, \quad
\gamma(\vec x) = g^2T \int \frac{d^3q}{(2\pi)^3}e^{i\vec q\cdot \vec x}
\frac{\pi m_D^2}{q(q^2 + m_D^2)^2}. \nonumber
\end{align}
In the Table \ref{tab:simulation_NRQCD}, a list of numerical simulations of the Lindblad equation from NRQCD is shown.

\begin{itemize}
\item This Lindblad equation contains rich physics but also non-essential physics in some specific situations.
By taking various limits of the Lindblad equation (recoilless, static, classical, and small dipole limits), essential physics in each regime becomes apparent.
One should note however that the time evolution of quarkonium is not necessarily confined in one particular regime.
\begin{itemize}
\item For example, it was shown in \cite{Akamatsu:2021dot} that off-diagonal elements of the density matrix show nontrivial time dependence and the heavy quark pair distance grows in time.
\end{itemize}
\item To be strict, when the temperature is not high enough for perturbative analysis, this Lindblad equation is not applicable.
However, we believe that one can construct a model Lindblad equation by assuming some forms for $S(\vec x)$ and $\gamma(\vec x)$ above, as is done in the numerical simulations.
\item If the temperature is too low for the quantum Brownian regime, one would need to consider coupled Boltzmann equation, which will be explained in Sec.~\ref{sec:pNRQCD_deep}.
\item If the diagonalization in the heavy-ion collisions proceeds only within the applicability of dipole approximation, the general description above can be simplified to the master equation in the small dipole limit \eqref{eq:Lindblad_NRQCD_QQbar_dipole}.
In that case, non-perturbative results can be obtained in the framework of potential non-relativistic QCD, which is the main topic of the next section.
\end{itemize}

\begin{table}
\centering
\caption{Numerical simulations of Lindblad equation from NRQCD.
For ``color" U(1), it means simulations of the Lindblad equation \eqref{eq:Lindblad_QBM_2body} and its recoilless limit.}
\label{tab:simulation_NRQCD} 
\vspace{3mm}
\begin{tabular}{c|c|c|l} \hline
Dimension & Color & Gradient Expansion & \qquad\quad Numerical Method \\ \hline 
1D & U(1) & LO & Stochastic Potential \cite{Akamatsu:2011se, Kajimoto:2017rel}\\
3D & U(1) & LO & Stochastic Potential \cite{Rothkopf:2013kya}\\
1D & SU(3) & LO & Stochastic Potential \cite{sharma2020quantum, Akamatsu:2021vsh}\\
1D & U(1) & NLO & Quantum State Diffusion \cite{Akamatsu:2018xim, Miura:2019ssi} \\
1D & SU(3) & NLO & Quantum State Diffusion \cite{Akamatsu:2021dot}\\
1D & U(1) & NLO & Direct evolution \cite{Alund:2020ctu}\\ \hline
\end{tabular}
\end{table}

\newpage
\section{Lindblad equations from potential non-relativistic QCD (pNRQCD)}
\label{sec:pNRQCD}
In this section, I first give a brief introduction to an effective field theory for quarkonium: potential non-relativistic QCD (pNRQCD) in Sec.~\ref{sec:pNRQCD_Review}.
Then, I apply the general framework of Sec.~\ref{sec:Basics} to the derivation of the Lindblad equations from pNRQCD.
There are several interesting regimes for quarkonium in the quark-gluon plasma (QGP) with different hierarchies of scales: (i) $1/r \gg T\sim m_D\gg E$, (ii) $1/r \gg T\gg E \gg m_D$, and (iii) $1/r \gg T\sim E\gg m_D$, where $T$ is the temperature of QGP, $m_D$ is the Debye screening mass, and $r$ and $E$ are the radius and the binding energy of the quarkonium.
Each regime is discussed in Sec.~\ref{sec:pNRQCD_strong}, \ref{sec:pNRQCD_weak}, and \ref{sec:pNRQCD_deep}.
The regimes (i) and (ii) correspond to loosely bound quarkonium because $T\gg E$ while (iii) corresponds to deeply bound quarkonium because $T\sim E$.
Since the Debye mass is given by $m_D\sim gT$, QGP is strongly coupled in the regime (i) and is weakly coupled in the regimes (ii) and (iii).
In terms of the open quantum system classification, the regimes (i) and (ii) are described by the quantum Brownian motion.
The previous analyses in Sec.~\ref{sec:NRQCD_QQbar_Dipole} overlap with (the perturbative limit of) the regime (i).
The regime (iii) seems to be in the quantum optical limit, but is not classified into either of the two basic regimes of the open quantum system and thus the standard approximation schemes are not applicable.
As a result, there is no well-defined Lindblad equation in this regime.
I will explain that the classical approximation and the point particle approximation for bound state quarkonium are essential in obtaining classical kinetic equations often adopted in this regime.

\subsection{A short review of potential non-relativistic QCD (pNRQCD)}
\label{sec:pNRQCD_Review}
In Sec.~\ref{sec:NRQCD_Review}, an effective field theory for non-relativistic heavy quarks (NRQCD) was reviewed.
When a pair of heavy quark and antiquark form a non-relativistic bound state of size $r$, typical scales involved are $p\sim Mv\sim 1/r$ for heavy quark momentum and $E\sim Mv^2\ll Mv$ for heavy quark energy, where $v$ is the relative velocity of the heavy quark pair.
We can focus on this particular setting to construct an effective field theory for a quarkonium: potential NRQCD (pNRQCD).

The degrees of freedom in pNRQCD are quarkonium in the singlet $S(t, \vec R, \vec r)$ and in the octet $O^a(t, \vec R,\vec r)$, and the gluons $A_{\mu}^a(t, \vec R)$.
Here $\vec R$ and $\vec r$ denote the center-of-mass and relative coordinates for the quarkonium. 
The cutoff $\Lambda_1$ is chosen to satisfy $Mv^2\ll\Lambda_1\ll Mv$ except for the cutoff $\Lambda_2$ for the quarkonium relative momentum, i.e. the resolution of $(t, \vec R)$ is $\sim 1/\Lambda_1$ and that of $\vec r$ is $\sim 1/\Lambda_2$.
The latter cutoff $\Lambda_2$ satisfies $Mv \ll \Lambda_2 \ll M$, so that the quarkonium is described as an extended non-relativistic object, whose structure is, however, not resolved by the gluon fields.
We further assume that $\Lambda_{\rm QCD}, T \ll Mv$, which enables us to put cutoffs at $Mv^2, \Lambda_{\rm QCD}, T \ll \Lambda_1 \ll Mv \ll \Lambda_2 \ll M$.
In this case, perturbative matching between NRQCD and pNRQCD is possible.

With these field contents and scales, the quarkonium dipole of size $r\sim 1/Mv$ is a localized color source for the gluons and their coupling is approximated by the multipole expansion.
Therefore, the pNRQCD Lagrangian is expanded in terms of the inverse heavy quark mass $1/M$ and the dipole size $r$.
The non-analytic terms of $r$, such as the Coulomb potential $\propto 1/r$, only enter through the Wilson coefficients.
In the leading order in $1/M$ and the next-to-leading order in $r$, the pNRQCD Lagrangian is
\begin{align}
\mathcal L_{\rm pNRQCD} &= \mathcal L_{q+A} + \int d^3 r
{\rm Tr}\left[
{\rm S}^{\dagger}\left(i\partial_t - V_s(r) +\cdots\right){\rm S} + {\rm O}^{\dagger}\left(iD_t - V_o(r) +\cdots\right){\rm O}
\right]  \\
&\quad +V_A(r){\rm Tr}\left[
{\rm O}^{\dagger}\vec r\cdot g\vec E {\rm S} + {\rm S}^{\dagger}\vec r\cdot g\vec E {\rm O}
\right]
+\frac{V_B(r)}{2}{\rm Tr}\left[
{\rm O}^{\dagger}\vec r\cdot g\vec E {\rm O} + {\rm O}^{\dagger}{\rm O}\vec r\cdot g\vec E
\right]
+\mathcal O(1/M), \nonumber \\
{\rm S}(t, \vec R,\vec r)&\equiv \frac{S(t, \vec R,\vec r)}{\sqrt{N_c}}\bm 1, \quad
{\rm O}(t, \vec R,\vec r)\equiv \sqrt{2} O^a(t, \vec R,\vec r)t_F^a.
\end{align}
By matching with the NRQCD, the leading order Wilson coefficients are\footnote{
To avoid the abusive use of $``V"$, hereafter we explicitly write $V_s(r)$ and $V_o(r)$ as $ -\frac{C_F\alpha_s}{r}$ and $\frac{\alpha_s}{2N_c r}$.
}
\begin{align}
V_s(r) = -\frac{C_F\alpha_s}{r}, \quad
V_o(r) = \frac{\alpha_s}{2N_c r}, \quad
V_A(r) = V_B(r) = 1.
\end{align}
In the pNRQCD Lagrangian, the color ${\rm SU}(3)$ gauge symmetry applies only for the center-of-mass coordinates $(t, \vec R)$.
Since the octet quarkonium $O^a$ has an adjoint color, its gauge interaction is given by the covariant derivative $D_t O=\partial_t O+ ig [A^a_0(t, \vec R)t_F^a, O]$.
By field redefinitions
\begin{subequations}
\label{eq:field_redef}
\begin{align}
{\rm O}(t, \vec R,\vec r) &= \Omega(t, \vec R){\rm O}'(t, \vec R,\vec r)\Omega^{\dagger}(t, \vec R), \quad
\vec E(t, \vec R) = \Omega(t, \vec R) \vec E'(t, \vec R)\Omega^{\dagger}(t, \vec R), \\
\Omega(t, \vec R)&\equiv \mathcal P \exp\left[
-ig \int_{-\infty}^t ds A_0(s, \vec R)
\right],
\end{align}
\end{subequations}
this covariant derivative can be eliminated.
Hereafter, the redefined fields are denoted by ${\rm O}(t, \vec R,\vec r)$ and $\vec E(t, \vec R)$.
According to the Virial theorem $Mv^2\sim C_F\alpha_s/r$ for the bound states, the kinetic term for the relative motion $\nabla_r^2/M$ turns out necessary in the singlet and thus in the octet as well for consistency:
\begin{align}
\int d^3 r
{\rm Tr}\left[
{\rm S}^{\dagger}\left(i\partial_t +\frac{\nabla_r^2}{M} + \frac{C_F\alpha_s}{r} +\cdots\right){\rm S} 
+ {\rm O}^{\dagger}\left(i\partial_t +\frac{\nabla_r^2}{M}- \frac{\alpha_s}{2N_c r} +\cdots\right){\rm O}
\right].
\end{align}
At this order of expansion, the center-of-mass kinetic energy for the quarkonium is irrelevant.

In order to apply the formula in the Section~\ref{sec:Basics}, let us give a quantum mechanical Hamiltonian corresponding to the pNRQCD Lagrangian
\begin{align}
H=\frac{p^2}{M} - \frac{C_F\alpha_s}{r}|s\rangle\langle s| + \frac{\alpha_s}{2N_c r} |a\rangle\langle a|
-\vec r\cdot g\vec E^a(\vec R)\left[
\sqrt{\frac{1}{2N_c}}\left(|a\rangle\langle s| + |s\rangle\langle a|\right) + \frac{1}{2}d^{abc}|b\rangle\langle c|
\right],
\end{align}
where $\vec R$ is the center-of-mass position of a quarkonium.
To explicitly distinguish the system and environment operators as in Eq.~\eqref{eq:vonNeumann}, the total Hamiltonian is written as
\begin{align}
\label{eq:pNRQCD_Hamiltonian}
H_{\rm tot} &= \left(\frac{p^2}{M} - \frac{C_F\alpha_s}{r}|s\rangle\langle s| + \frac{\alpha_s}{2N_c r} |a\rangle\langle a|\right) \otimes I_E + I_S\otimes H_{q+A} \nonumber \\
&\quad - r_i\left[
\sqrt{\frac{1}{2N_c}}\left(|a\rangle\langle s| + |s\rangle\langle a|\right) + \frac{1}{2}d^{abc}|b\rangle\langle c|
\right]\otimes g E_i^a(\vec R).
\end{align}

Before closing this section, let us focus on the vacuum $T=0$ for a moment and discuss how heavy quark potential is defined in the pNRQCD.
As is clear from construction, the Coulomb potential originates from the gauge fields above the cutoff scale $\Lambda_1$, typically with $Mv$.
If the next relevant energy scale is $Mv^2\gtrsim \Lambda_{\rm QCD}$, the gauge fields in the pNRQCD interact with quarkonium {\it non-potentially}, i.e. their effects cannot be integrated into an instantaneous potential term.
In this case, the Wilson coefficients $V_{s,o}(r)$ are potentials for the quarkonium.
If the next scale is $\Lambda_{\rm QCD}\gg Mv^2$, the gauge fields do contribute to the potential\footnote{
A keen reader might notice that the condition for the potential effect is very similar to one of the criteria for the quantum Brownian motion $\tau_S\gg \tau_E$.
In this analogy, the condition for the non-potential effect corresponds to $\tau_S\lesssim \tau_E$, which implies the quantum optical regime if we assume $\tau_E\ll \tau_R$.
}, which means that $V_{s,o}(r)$ are not precisely potentials for the quarkonium.
In this case, one needs to integrate out the pNRQCD degrees of freedom with $\Lambda_{\rm QCD}$ and construct a new effective field theory pNRQCD' by matching them at a scale between $\Lambda_{\rm QCD}$ and $Mv^2$.
The Wilson coefficients of the latter properly define the heavy quark potentials.
For further details about pNRQCD, see the reviews \cite{brambilla2005effective, petrov2015effective, brambilla2000potential}.

\subsection{Quantum Brownian motion of a quarkonium with $1/r\gg T\sim m_D\gg E$}
\label{sec:pNRQCD_strong}
In this section, we derive the Lindblad equation for quantum Brownian motion of a quarkonium in the quark-gluon plasma (QGP).
The relation $T\sim m_D$ indicates that the QGP is strongly coupled.
In this regime, the system relaxation time is estimated by its kinetic equilibration time $\tau_R\sim M/T^2$, the system time scale is by the inverse of the binding energy $\tau_S=1/E$, and the environment correlation time is the duration of the collisions $\tau_E\sim 1/m_D$.
Since $M\gg T$, we can see that the scale hierarchy for the quantum Brownian motion ($\tau_R\gg \tau_E$ and $\tau_S\gg \tau_E$) is satisfied.

We apply the formula \eqref{eq:Lindblad_QBM} to a quarkonium in the QGP.
The formula relies on perturbative expansion with the system-environment interaction.
Even though the QGP is strongly coupled in this regime, the perturbative expansion is justified in the limit of small dipole.
The total Hamiltonian is given in Eq.~\eqref{eq:pNRQCD_Hamiltonian}, which we quote here again,
\begin{align}
H_{\rm tot} &= \left(\frac{p^2}{M} - \frac{C_F\alpha_s}{r}|s\rangle\langle s| + \frac{\alpha_s}{2N_c r} |a\rangle\langle a|\right) \otimes I_E + I_S\otimes H_{q+A} \nonumber \\
&\quad - r_i\left[
\sqrt{\frac{1}{2N_c}}\left(|a\rangle\langle s| + |s\rangle\langle a|\right) + \frac{1}{2}d^{abc}|b\rangle\langle c|
\right]\otimes g E_i^a(\vec R), \nonumber
\end{align}
and the operator correspondence is
\begin{subequations}
\label{eq:Lindblad_correspondence_pNRQCD}
\begin{align}
V_S^{(i)} &\leftrightarrow -r_i\left[
\sqrt{\frac{1}{2N_c}}\left(|a\rangle\langle s| + |s\rangle\langle a|\right) + \frac{1}{2}d^{abc}|b\rangle\langle c|
\right]  \equiv V_S^{ai}, \\
\dot V_S^{(i)}=i[H_S, V_S^{(i)}] &\leftrightarrow 
-\frac{2p_i}{M}
\left[
\sqrt{\frac{1}{2N_c}}\left(|a\rangle\langle s| + |s\rangle\langle a|\right) + \frac{1}{2}d^{abc}|b\rangle\langle c|
\right]
+ i\sqrt{\frac{N_c}{8}}\frac{\alpha_s r_i}{r} \left(
|s\rangle\langle a| - |a\rangle\langle s|
\right) \equiv \dot V_S^{ai}.
\end{align}
\end{subequations}
The coefficients are given by non-perturbative gauge invariant color electric correlators\footnote{
Recall that we have made the field redefinition $\vec E(t, \vec R) = \Omega(t, \vec R) \vec E'(t, \vec R)\Omega^{\dagger}(t, \vec R)$ in Eq.~\eqref{eq:field_redef}, so that the spectral density $\sigma_{ab,ij}(\omega)$ for $\vec E'(t, \vec R)$ and thus all the other coefficients are gauge invariant.
Note that $S$ contains vacuum contributions which need to be subtracted.
}
\begin{subequations}
\label{eq:Lindblad_coefficients_pNRQCD}
\begin{align}
\gamma_{ab,ij}&= T\frac{d}{d\omega}\sigma_{ab,ij}(\omega)\Bigr|_{\omega=0}
\equiv \gamma\delta_{ab}\delta_{ij}, \\
S_{ab,ij}&= -\frac{1}{2}\int_{-\infty}^{\infty} \frac{d\omega}{2\pi}\frac{\sigma_{ab,ij}(\omega)}{\omega}
\equiv S\delta_{ab}\delta_{ij}, \\
\sigma_{ab,ij}(\omega) &\equiv
\int_{-\infty}^{\infty} dt e^{i\omega t} {\rm Tr}_E\left(\rho_E^{\rm th}\left[gE_i^a(t,\vec R), gE_j^b(0,\vec R)\right]\right)
\propto \delta_{ab}\delta_{ij}, \\
\eta_{ab,ij}&\simeq -\frac{i}{4T}\gamma_{ab,ij} = -\frac{i}{4T}\gamma\delta_{ab}\delta_{ij}.
\end{align}
\end{subequations}
Here, the real part of $\eta_{ab,ij}$ can be neglected\footnote{
Using \eqref{eq:master_QBM}, \eqref{eq:Lindblad_correspondence_pNRQCD}, and \eqref{eq:Lindblad_coefficients_pNRQCD}, singlet complex potential in $M\to\infty$ limit is
$-\frac{C_F\alpha_s}{r} + C_FSr^2 - i\frac{C_F\gamma}{2}r^2\left[1- \frac{N_c\alpha_s}{4Tr}\right]$
in the first order gradient expansion.
The absence of linear term $\propto r$ in the real part is due to the approximation ${\rm Re}\eta_{ab,ij}\approx 0$. 
}
because of the time scale hierarchy $\tau_E/\tau_S\ll 1$ (see the discussion at the end of Sec.~\ref{sec:Basics_Approx_QBM}).
Then the Lindblad operator reads
\begin{align}
\tilde V_S^{(i)} &\leftrightarrow V_S^{ai} + \frac{i}{4T} \dot V_S^{ai}
= \left(\begin{aligned}
&-\left(r_i+\frac{ip_i}{2MT} - \frac{N_c}{8T}\frac{\alpha_s r_i}{r}\right)\sqrt{\frac{1}{2N_c}}|a\rangle\langle s| \\
&-\left(r_i+\frac{ip_i}{2MT} + \frac{N_c}{8T}\frac{\alpha_s r_i}{r}\right)\sqrt{\frac{1}{2N_c}}|s\rangle\langle a|\\
&-\left(r_i+\frac{ip_i}{2MT} \right)\frac{1}{2}d^{abc}|b\rangle\langle c|
\end{aligned}\right)
\equiv \tilde V_S^{ai} ,
\end{align}
and the correction to the Hamiltonian is
\begin{align}
\Delta H_S = \left(S r^2 + \frac{\gamma}{4MT}\{\vec p,\vec r\}\right)\left[
C_F|s\rangle\langle s| + \frac{N_c^2-2}{4N_c}|a\rangle\langle a|
\right].
\end{align}
With these operators, we obtain the Lindblad equation
\begin{align}
\label{eq:Lindblad_pNRQCD_QQbar}
\frac{d}{dt}\rho_S(t) &= -i\left[H_S + \Delta H_S, \rho_S\right]
+ \gamma\left[
\tilde V_S^{ai}\rho_S \tilde V_S^{ai\dagger}
-\frac{1}{2}\left\{
\tilde V_S^{ai\dagger}\tilde V_S^{ai}, \rho_S
\right\}
\right].
\end{align}

As in Sec.~\ref{sec:NRQCD_QQbar_Lindblad}, we can express the Lindblad equation in the projected singlet-octet basis.
Let us first define
\begin{subequations}
\label{eq:Lindblad_pNRQCD_QQbar_proj_V}
\begin{align}
\tilde V_S^{ai}
&= \tilde V_{+i} \sqrt{\frac{1}{2N_c}} |a\rangle\langle s| + \tilde V_{-i}\sqrt{\frac{1}{2N_c}} |s\rangle\langle a| +  \tilde V_{di}\frac{1}{2}d^{abc}|b\rangle\langle c|, \\
\tilde V_{\pm i} &\equiv -\left(r_i+\frac{ip_i}{2MT} \mp \frac{N_c}{8T}\frac{\alpha_s r_i}{r}\right), \quad
\tilde V_{di} \equiv -\left(r_i+\frac{ip_i}{2MT} \right).
\end{align}
\end{subequations}
The master equation for the diagonal parts of the density matrix
\begin{align}
\rho_S(t) = \begin{pmatrix} \rho_s(t) & 0 \\ 0 & \rho_o(t) \end{pmatrix}, \quad
\rho_s(t) \equiv \langle s | \rho_S(t) |s\rangle, \quad
\rho_o(t) \equiv \langle a | \rho_S(t) |a\rangle
\end{align}
is again written in the Lindblad form
\begin{subequations}
\label{eq:Lindblad_pNRQCD_QQbar_proj}
\begin{align}
\frac{d}{dt}\rho_S(t) &= -i\left[H_S + \Delta H_S, \rho_S\right]
+\gamma \sum_{n=+,-,d}\left[
\tilde C_{ni}\rho_S \tilde C_{ni}^{\dagger}
-\frac{1}{2}\left\{
\tilde C_{ni}^{\dagger}\tilde C_{ni}, \rho_S
\right\}
\right],\\
\Delta H_S &= \left(S r^2 + \frac{\gamma}{4MT}\{\vec p,\vec r\}\right)
\begin{pmatrix} C_F & 0 \\ 0 & \frac{N_c^2-2}{4N_c}\end{pmatrix},
\\
\tilde C_{+i} &= \tilde V_{+i} \sqrt{C_F}\begin{pmatrix} 0 & 0 \\ 1 & 0 \end{pmatrix}, \quad
\tilde C_{-i} = \tilde V_{-i} \sqrt{\frac{1}{2N_c}}\begin{pmatrix} 0 & 1 \\ 0 & 0 \end{pmatrix}, \quad
\tilde C_{d i} = \tilde V_{d i} \sqrt{\frac{N_c^2-4}{4N_c}}\begin{pmatrix} 0 & 0 \\ 0 & 1 \end{pmatrix}.
\end{align}
\end{subequations}

The Lindblad equation \eqref{eq:Lindblad_pNRQCD_QQbar_proj} is parametrized by two non-perturbative coefficients $\gamma$ and $S$.
The former is proportional to the heavy quark momentum diffusion constant $\kappa = C_F\gamma$ and was calculated non-perturbatively by lattice QCD simulations for a pure gluon plasma with $N_c=3$
\begin{align}
1.8 \lesssim \frac{\kappa}{T^3} \lesssim 3.4
\end{align}
at around $T=1.5 T_c$ where $T_c$ is the deconfinement transition temperature \cite{francis2015nonperturbative}.
It was also calculated in a wider temperature range $T_c < T < 2 T_c$ with similar values $\kappa/T^3 \sim 1-4$ \cite{banerjee2012heavy}.
Lattice QCD simulations at higher temperatures $1.1 T_c < T < 10^4T_c$ \cite{Brambilla:2020siz} found that $\kappa/T^3$ decreases with temperature and is consistent with the next-to-leading order perturbative result \cite{CaronHuot:2007gq,CaronHuot:2008uh} at the highest temperature $T=10^4 T_c$.
The latter coefficient $S$ is proportional to the thermal dipole self-energy constant $\lambda = 2C_F S$ but has not been calculated non-perturbatively although it is computable by lattice QCD simulations without analytic continuation to Minkowski space \cite{eller2019thermal}.
Using the fact that the thermal dipole self energy $\lambda r^2/2$ shifts the in-medium quarkonium mass, a nonperturbative estimate of $\lambda$ was given in \cite{Brambilla:2019tpt}\footnote{
The thermal dipole self-energy constant $\lambda$ corresponds to $\gamma$ in \cite{eller2019thermal} and \cite{Brambilla:2019tpt}; readers should not get confused.
}
\begin{align}
-3.8\lesssim \frac{\lambda}{T^3} \lesssim -0.7
\end{align}
by examining the data of lattice QCD simulations \cite{kim2018quarkonium} for mass shifts of $J/\psi$ at $T=251{\rm MeV}$ and $\Upsilon (1S)$ at $T=251, 407 {\rm MeV}$.

Let us see the relation to the small dipole limit of the Lindblad equation from NRQCD in Sec.~\ref{sec:NRQCD_QQbar_Dipole}.
In the weak coupling and in the long-time limit, only the potential force part is relevant in the color electric fields, so that the spectral density can be approximated by
\begin{align}
\sigma_{ab,ij}(\omega)&\simeq
\int_{-\infty}^{\infty} dt e^{i\omega t} {\rm Tr}_E\left(\rho_E^{\rm th}\left[g\partial_i A_0^a(t,\vec R), g\partial_j A_0^b(0,\vec R)\right]\right) \nonumber \\
&= -\frac{\partial}{\partial R_i}\frac{\partial}{\partial R_j}\int_{-\infty}^{\infty} dt e^{i\omega t} {\rm Tr}_E\left(\rho_E^{\rm th}\left[gA_0^a(t,\vec R), gA_0^b(0,\vec R')\right]\right)\Bigr|_{R'\to R},
\end{align}
and thus
\begin{align}
\gamma \simeq -\frac{\nabla^2 \gamma(\vec 0)}{3}, \quad
S \simeq -\frac{\nabla^2 S_{T\neq 0}(\vec 0)}{3},
\end{align}
where the left hand sides are the coefficients of the Lindblad equation from pNRQCD while the right hand sides are those from NRQCD.
By substituting these relations, we can reproduce the NRQCD result Eq.~\eqref{eq:Lindblad_NRQCD_QQbar_dipole} except for the quadratic term $\propto r^2$ in the Hamiltonian in the octet channel.
As explained in Sec.~\ref{sec:NRQCD_QQbar_Dipole}, the reason for the discrepancy is that the gauge interaction of the octet quarkonium is point-like (and is gauged away) in pNRQCD while that in NRQCD takes account of the finite size of the dipole.

The recoilless limit of this Lindblad equation can be obtained by keeping the leading terms in the gradient expansion.
The resulting Lindblad equation is
\begin{subequations}
\label{eq:Lindblad_pNRQCD_QQbar_proj_recoilless}
\begin{align}
\frac{d}{dt}\rho_S(t) &= -i\left[H_S + \Delta H_S, \rho_S\right]
+\gamma \sum_{n=+,-,d}\left[
C_{ni}\rho_S C_{ni}^{\dagger}
-\frac{1}{2}\left\{
C_{ni}^{\dagger}C_{ni}, \rho_S
\right\}
\right],\\
\Delta H_S &= S r^2
\begin{pmatrix} C_F & 0 \\ 0 & \frac{N_c^2-2}{4N_c}\end{pmatrix},
\\
C_{+i} &= r_i \sqrt{C_F}\begin{pmatrix} 0 & 0 \\ 1 & 0 \end{pmatrix}, \quad
C_{-i} = r_i \sqrt{\frac{1}{2N_c}}\begin{pmatrix} 0 & 1 \\ 0 & 0 \end{pmatrix}, \quad
C_{d i} = r_i \sqrt{\frac{N_c^2-4}{4N_c}}\begin{pmatrix} 0 & 0 \\ 0 & 1 \end{pmatrix},
\end{align}
\end{subequations}
which was first derived in \cite{Brambilla:2016wgg, Brambilla:2017zei} and numerically solved in \cite{Brambilla:2016wgg, Brambilla:2017zei, Brambilla:2020qwo, Brambilla:2021wkt}.
In the numerical simulations, the Lindblad equation is written in the angular momentum basis  (see Appendix \ref{app:Lindblad_pNRQCD_LM}) and Refs.~\cite{Brambilla:2016wgg, Brambilla:2017zei} directly solved it including only $S$ and $P$ waves while Refs.~\cite{Brambilla:2020qwo, Brambilla:2021wkt} solved using the Quantum Jump Method \cite{Plenio:1997ep} without the angular momentum cutoff.
This Lindblad equation in the projected singlet-octet basis does not have an equivalent representation by the stochastic potential for the same reason given in Sec.~\ref{sec:NRQCD_QQbar_Stochastic}.
The stochastic potential can be obtained from the Lindblad equation in the full color space Eq.~\eqref{eq:Lindblad_pNRQCD_QQbar} in the recoilless limit $\tilde V_S^{ai}\to V_S^{ai} = V_S^{ai\dagger}$:
\begin{subequations}
\label{eq:stochastic_potential_dipole}
\begin{align}
&U_{QQ_c}(\theta)
=\left[
C_F\left(-\frac{\alpha_s}{r} + S r^2\right)|s\rangle\langle s| + 
\left(\frac{\alpha_s}{2N_c r}+\frac{N_c^2-2}{4N_c}S r^2\right)|a\rangle\langle a|
\right] \nonumber \\
& \hspace{2cm} + \theta^{ai}r_i\left[
\sqrt{\frac{1}{2N_c}}\left(|a\rangle\langle s| + |s\rangle\langle a|\right) + \frac{1}{2}d^{abc}|b\rangle\langle c|
\right],\\
& E_{\theta}[\theta^{ai}(t)\theta^{bj}(t')] = \gamma\delta_{ij}\delta^{ab}\delta(t-t').
\end{align}
\end{subequations}
As emphasized in Secs.~\ref{sec:NRQCD_HQ_DecDiss} and \ref{sec:NRQCD_QQbar_Stochastic}, it should be reminded that the master equation in the recoilless limit is applicable only when the decoherence is dominant over the dissipation.

\subsection{Quantum Brownian motion of a quarkonium with $1/r\gg T\gg E \gg m_D$}
\label{sec:pNRQCD_weak}
In this section, we derive the Lindblad equation for quantum Brownian motion of a quarkonium in the quark-gluon plasma (QGP).
The relation $T\gg m_D$ indicates that the QGP is weakly coupled.
In this regime, the system relaxation time is estimated by its kinetic equilibration time $\tau_R\sim M/g^4T^2$, the system time scale is by the inverse of the binding energy $\tau_S=1/E$.
The environment correlation time for the collisional processes is the duration of the collision, that is $\tau^{(\rm soft)}_E\sim 1/m_D$ for soft collisions and $\tau^{(\rm hard)}_E\sim 1/T$ for hard collisions, and that for the absorptions and emissions of a thermal gluon is the duration of these processes $\tau^{(g)}_E\sim 1/T$.
In this regime, only the hard collisions and gluon absorptions and emissions satisfy the scale hierarchy of quantum Brownian motion $\tau^{(\rm hard)}_E, \tau^{(g)}_E \ll \tau_S$.
In \cite{brambilla2010heavy, brambilla2011thermal, brambilla2013thermal}, it was shown that the thermal width of a quarkonium is dominated by {\it gluo-dissociation}, a process in which a singlet quarkonium is excited to an octet unbound state by absorbing a thermal gluon.
Therefore we here analyze the quantum Brownian motion driven by the gluon absorptions and emissions.

In the Coulomb gauge, the correlation functions are perturbatively calculated from the spectral density of thermal gluons $\sigma_{ab,ij}(\omega) = \frac{g^2}{3\pi}\omega^3 \delta_{ij}\delta^{ab}$ (see Appendix \ref{app:Thermal_Corr_Gluons}):
\begin{subequations}
\begin{align}
\gamma_{ab,ij}(\omega) &= \frac{g^2}{3\pi}\omega^3 (1 + n_B(\omega)) \delta_{ij}\delta^{ab}, \\
S_{ab,ij}(\omega) &= -\frac{g^2}{3\pi}\mathcal P \int_0^{\infty}\frac{d\omega'}{2\pi}\left\{
\left[1 + n_B(\omega')\right]\frac{\omega'^3}{\omega'-\omega}+ n_B(\omega')\frac{\omega'^3}{-\omega'-\omega} \right\}, \\
&= S_{ab,ij}(\omega)\Bigr|_{T=0}-\frac{g^2}{3\pi}\mathcal P \int_0^{\infty}\frac{d\omega'}{2\pi}\omega'^3 n_B(\omega')\left(
\frac{1}{\omega'-\omega}-\frac{1}{\omega'+\omega} \right).
\end{align}
\end{subequations}
Here $S_{ab,ij}(\omega)\Bigr|_{T=0}$ isolates ultraviolet divergence which is renormalized in the vacuum.
In the limit of vanishing $\omega$, we can approximate the correlation functions as
\begin{align}
\gamma_{ab,ij}(\omega)\simeq \frac{g^2T}{3\pi}\omega^2 \delta_{ij}\delta^{ab} + \mathcal O(\omega^3), \quad
S_{ab,ij}(\omega)\Bigr|_{T\neq 0} \simeq -\frac{g^2T^2}{18}\omega \delta_{ij}\delta^{ab} + \ldots
\end{align}
Since $\gamma_{ab,ij}(0)=0$, we need to use the formula \eqref{eq:Lindblad_exQBM}, instead of \eqref{eq:Lindblad_QBM}, in this regime.
The coefficients in the formula correspond to
\begin{align}
\gamma_{ab,ij}^{(2)} = \frac{2g^2 T}{3\pi}\delta_{ij}\delta^{ab}, \quad
S_{ab,ij}^{(0)}=0, \quad
S_{ab,ij}^{(1)}=-\frac{g^2T^2}{18}\delta_{ij}\delta^{ab},
\end{align}
and $S_{ab,ij}^{(2)}$ is unavailable because it is infrared divergent which might require resummation of higher order perturbative expansions.
The correction to the Hamiltonian is thus
\begin{align}
\Delta H_S &= \frac{i}{2}S_{ab,ij}^{(1)}\left[V_S^{ai}, \dot V_S^{bj}\right]\nonumber \\
&= \frac{g^2T^2}{36}\left[
\left(N_c\alpha_s r + \frac{6}{M}\right)C_F|s\rangle\langle s|
+\left(-\frac{\alpha_s r}{2}  + \frac{3(N_c^2-2)}{2N_cM}\right)|a\rangle\langle a|
\right].
\end{align}
By keeping the lowest non-vanishing order in the gradient expansion in the $\Delta H_S$ and the Lindblad operators, we obtain the Lindblad equation
\begin{subequations}
\begin{align}
\label{eq:Lindblad_pNRQCD_QQbar2}
\frac{d}{dt}\rho_S(t) &= -i\left[H_S + \Delta H_S, \rho_S\right]
+ \frac{g^2 T}{3\pi} \left[
\dot V_S^{ai}\rho_S \dot V_S^{ai}
-\frac{1}{2}\left\{
\dot V_S^{ai}\dot V_S^{ai}, \rho_S
\right\}
\right], \\
\dot V_S^{ai} &=-\frac{2p_i}{M}
\left[
\sqrt{\frac{1}{2N_c}}\left(|a\rangle\langle s| + |s\rangle\langle a|\right) + \frac{1}{2}d^{abc}|b\rangle\langle c|
\right]
+ i\sqrt{\frac{N_c}{8}}\frac{\alpha_s r_i}{r} \left(
|s\rangle\langle a| - |a\rangle\langle s|
\right),
\end{align}
\end{subequations}
which is in the projected singlet-octet basis
\begin{subequations}
\label{eq:Lindblad_pNRQCD_QQbar2_proj}
\begin{align}
\frac{d}{dt}\rho_S(t) &= -i\left[H_S + \Delta H_S, \rho_S\right]
+\frac{g^2T}{3\pi} \sum_{n=+,-,d}\left[
\dot C_{ni}\rho_S \dot C_{ni}^{\dagger}
-\frac{1}{2}\left\{
\dot C_{ni}^{\dagger} \dot C_{ni}, \rho_S
\right\}
\right],\\
\Delta H_S &= \frac{g^2T^2}{36}
\begin{pmatrix} \left(N_c\alpha_s r + \frac{6}{M}\right)C_F & 0 \\
0 & -\frac{\alpha_s r}{2}  + \frac{3(N_c^2-2)}{2N_cM}\end{pmatrix},
\\
\dot C_{+i} &= \left(-\frac{2p_i}{M} - i \frac{N_c}{2}\frac{\alpha_s r_i}{r}\right) \sqrt{C_F}
\begin{pmatrix} 0 & 0 \\ 1 & 0 \end{pmatrix}, \\
\dot C_{-i} &=  \left(-\frac{2p_i}{M} + i \frac{N_c}{2}\frac{\alpha_s r_i}{r}\right) \sqrt{\frac{1}{2N_c}}
\begin{pmatrix} 0 & 1 \\ 0 & 0 \end{pmatrix}, \quad
\dot C_{d i} = -\frac{2p_i}{M} \sqrt{\frac{N_c^2-4}{4N_c}}\begin{pmatrix} 0 & 0 \\ 0 & 1 \end{pmatrix}.
\end{align}
\end{subequations}
This Lindblad equation for a quarkonium in the weakly-coupled quark-gluon plasma was first derived in \cite{Brambilla:2017zei}.
The long-time behavior of this Lindblad equation has not been fully analyzed yet and one may need to continue the gradient expansion to higher order to describe the relaxation of quarkonium in this regime.

\subsection{Classical approximation for a quarkonium with $1/r \gg T\sim E\gg m_D$}
\label{sec:pNRQCD_deep}
In this section, let us focus on a quarkonium whose binding energy $E$ is as large as the temperature $T$.
In this regime, the dominant process of quarkonium dissociation is the gluo-dissociation \cite{brambilla2011thermal}.
Relevant time scales are $\tau_R\sim M/g^4T^2$ for heavy quark kinetic equilibration time, $\tau_S\sim 1/E\sim 1/T$ for the bound state time scale, and $\tau^{(g)}_E\sim 1/T$ for the duration of the gluon absorption/emission processes.
Clearly, this regime is not the quantum Brownian motion because $\tau_S\sim \tau^{(g)}_E$.
One would easily expect that it is in the quantum optical limit because $\tau_R\gg \tau_S$ holds.
However, by detailed examination of the gluon absorption and emission process, one can find that the rotating wave approximation
\begin{align}
\sum_{\omega, \omega'} e^{i(\omega'-\omega)t}\simeq \sum_{\omega, \omega'}\delta_{\omega,\omega'},
\end{align}
where $\omega$ is the energy of a gluon, is not applicable.
The rotating wave approximation takes advantage of the gapped spectrum for $\omega$, which is a stricter condition than the existence of the energy gap.
For gluo-dissociation processes, $\omega$ is gapped from zero by the binding energy $E$, but it takes continuous values because a singlet quarkonium is excited to an octet unbound state in the continuum.

The purpose of the analysis here is to examine the result of \cite{Yao:2017fuc,Yao:2018nmy}, in which the authors claim that coupled Boltzmann equations for quarkonium and unbound heavy quark pair are derived on the basis of the open quantum system approach, which is however not applicable.
Since the result is physically natural, I will try to discuss how one can nevertheless get the Boltzmann equations by overcoming the inapplicability of any of the open quantum system techniques.

We start from adding the center-of-mass energy of the heavy quark pair to the Hamiltonian 
\begin{align}
\label{eq:pNRQCD_Hamiltonian+cm}
H_{\rm tot} &= \left(\frac{P^2}{4M} + \frac{p^2}{M} - \frac{C_F\alpha_s}{r}|s\rangle\langle s| + \frac{\alpha_s}{2N_c r} |a\rangle\langle a|\right) \otimes I_E + I_S\otimes H_{q+A} \nonumber \\
&\quad - g\int_x r_i  \delta(\vec x -\vec R)\left[
\sqrt{\frac{1}{2N_c}}\left(|a\rangle\langle s| + |s\rangle\langle a|\right) + \frac{1}{2}d^{abc}|b\rangle\langle c|
\right]
 \otimes  E_i^a(\vec x).
\end{align}
Here $\vec P$ and $\vec R$ are operators for center-of-mass momentum and coordinate and $\vec p$ and $\vec r$ are those for relative momentum and coordinate.
Since the classical kinetic equation is based on the conservation laws for energy and momentum, it is desirable that the dynamics of center-of-mass motion is included in the description.
From this Hamiltonian, the operator correspondence to Sec.~\ref{sec:Basics} is found by
\begin{align}
V_S^{(i)} \leftrightarrow - r_i  \delta(\vec x -\vec R)\left[
\sqrt{\frac{1}{2N_c}}\left(|a\rangle\langle s| + |s\rangle\langle a|\right) + \frac{1}{2}d^{abc}|b\rangle\langle c|
\right] \equiv V_S^{ai}(\vec x).
\end{align}

Although we know that the formula \eqref{eq:Lindblad_QOL} for the quantum optical limit is not applicable on the physical grounds, let us anyway apply it here and see what happens.
To define the transition operators \eqref{eq:transitions} $V_S^{(i)}(\omega)\equiv\sum_{\epsilon,\epsilon'}\delta_{\epsilon+\omega, \epsilon'}\Pi(\epsilon)V_S^{(i)}\Pi(\epsilon')$ in the present case, we first need to prepare eigenstates of the Hamiltonian.
We adopt the large $N_c$ limit to simplify the discussions as below.
In the singlet sector, we only consider bound states.
This is because only $1/(N_c^2-1)\ll1$ portion of the unbound states are singlet. 
In the octet sector, the repulsive potential is neglected and the heavy quark pair in the octet does not interact with each other.
This is justified because the large $N_c$ limit is taken with $N_c g^2 \simeq 8\pi C_F \alpha_s$ fixed and thus $\alpha_s/2N_c \to 0$ in the limit.
With these simplifying assumptions, the eigenstates are
\begin{subequations}
\begin{align}
H_S |\vec k, n, s\rangle &= \left(\frac{k^2}{4M} + E_n\right)|\vec k, n, s\rangle \equiv E_{k,n}|\vec k, n, s\rangle, \\
H_S |\vec P, \vec p, a\rangle 
&=\left(\frac{P^2}{4M} + \frac{p^2}{M} \right)|\vec P, \vec p, a\rangle
\equiv E_{P, p}|\vec P, \vec p, a\rangle,
\end{align}
\end{subequations}
where $\vec k$ and $E_n < 0$ are the center-of-mass momentum and the bound state energy of the quarkonium and $\vec P$ and $\vec p$ are center-of-mass and relative momenta of the heavy quark pair.
For simplicity, we further assume the bound states are not degenerate.
The extension to the degenerate case must be relatively straightforward.
The definition of the transition operator \eqref{eq:transitions} is extended to the case when the transition energy $\omega$ is continuous.
Since we are interested in how to describe the singlet-octet transitions, let us omit from $V_S^{ai}(\vec x)$ the octet-octet scattering processes which is in the regime of quantum Brownian motion.
Transition operator from the singlet to the octet ($\sim |a\rangle\langle s|$) is restricted to negative frequency and is given by\footnote{
Summation over repeated indices is assumed for the adjoint color $a$ and the vector component $i$.
However, we explicitly write $\sum_n$ to sum the bound state labels.
}
\begin{align}
V_S^{ai}(\vec x, \omega<0) &=\frac{2\pi}{\mathcal T}\sum_n \int_{P, p, k}
\delta(\omega - E_{k,n} + E_{P, p}) 
|\vec P, \vec p, b\rangle \langle \vec P, \vec p, b|
V_S^{ai}(\vec x)|\vec k, n, s\rangle\langle \vec k, n ,s|\nonumber \\
&=-\frac{2\pi}{\mathcal T}\sum_n\int_{P, p, k}\delta(\omega - E_{k,n} + E_{P, p})
\frac{\langle \vec p|r_i|n\rangle e^{i(\vec k - \vec P)\cdot\vec x}}{\sqrt{2N_c}}  |\vec P, \vec p, a\rangle\langle \vec k, n, s| \nonumber \\
&=V_S^{ai\dagger}(\vec x, -\omega).
\end{align}
Here $\mathcal T$ is a range of time integration and is introduced to extend $\sum_{\epsilon,\epsilon'}$ and $\delta_{\epsilon+\omega, \epsilon'}$ when energy spectrum is continuous.
The matrix element $\langle \vec P, \vec p, b|V_S^{ai}(\vec x)|\vec k, n, s\rangle$ is calculated in each subspace of the heavy quark pair.
The transition operator from the octet to the singlet ($\sim |s\rangle\langle a|$) with positive frequency is also given by
\begin{align}
V_S^{ai}(\vec x, \omega>0) &=-\frac{2\pi}{\mathcal T}\sum_n\int_{P, p, k}\delta(\omega + E_{k,n} - E_{P, p})
\frac{\langle n|r_i|\vec p\rangle e^{-i(\vec k - \vec P)\cdot\vec x}}{\sqrt{2N_c}}  | \vec k, n, s\rangle\langle \vec P, \vec p, a|.
\end{align}

\subsubsection{Singlet-to-octet transition by gluon absorption}
\label{sec:pNRQCD_deep_so}
Here we analyze the transition process from the singlet to the octet.
The transition term in the Lindblad equation again needs an extension to the case with a continuous energy spectrum
\begin{align}
\sum_{\omega<0}\sum_{i,j}
\gamma_{ij}(\omega)V_S^{(j)}(\omega) \rho_S(t)V_S^{(i)\dagger}(\omega)
\leftrightarrow
\mathcal T\int_{\omega,x,y} \theta(-\omega)\gamma_{ab, ij}(\omega, \vec x, \vec y)
V_S^{bj}(\vec y, \omega)\rho_S(t) V_S^{ai\dagger}(\vec x, \omega),
\end{align}
where $\int_{\omega}\equiv\int\frac{d\omega}{2\pi}$.
After a straightforward calculation with substitution of $\gamma_{ab,ij}(\omega, \vec q)$ for thermal gluons in the Appendix \ref{app:Thermal_Corr_Gluons}, the time evolution of the octet density matrix due to gluon absorption is given by
\begin{align}
\frac{\partial}{\partial t} \langle \vec P, \vec p, a| \rho_S|\vec P', \vec p^{\ \prime}, a'\rangle
\Bigr|_{s\to o}
&=\frac{(2\pi)^8}{\mathcal T}\delta_{aa'} \frac{g^2}{2N_c}\sum_{n,n'} \int_{q,k,k'}  [\text{8 $\delta$-functions}] \nonumber \\
&\qquad \times n_B(q) \langle \vec k, n, s|\rho_S|\vec k', n', s\rangle 
\cdot q^2\delta_T^{ij}(\hat q) \langle \vec p |r_i|n\rangle\langle n'|r_j|\vec p^{\ \prime}\rangle,
\end{align}
where the 8 delta functions impose energy and momentum conservation in the forward and backward amplitudes:
\begin{align}
\label{eq:8-delta}
\delta(q + E_{k,n} - E_{P, p})
\delta(\vec q + \vec k - \vec P)
\delta(q + E_{k',n'} - E_{P', p'})
\delta(\vec q + \vec k' - \vec P').
\end{align}
Here we introduce the real gluon momentum $\vec q$ and a shorthand notation for its integration $\int_q\equiv\int\frac{d^3 q}{(2\pi^3)2q}$ with covariant normalization.
Unless one of the energy delta functions is trivially satisfied $\delta(0) = \mathcal T/2\pi$, the transition rate contains a peculiar dependence on $\mathcal T$.
This indicates that the formula for the quantum optical limit is not applicable.

To save this situation, we need to make two additional approximations:
(i) classical approximation by Wigner transformation, and
(ii) local approximation for bound states in the singlet-to-octet transitions.
In the Wigner transformation, we assume that the density matrix is diagonal in the singlet and color neutral in the octet, that is $\rho_S\sim |n,s\rangle\langle n,s| + |a\rangle\langle a|$, and define the phase space distribution function for a singlet bound state $n$ and unbound pair in the octet as
\begin{subequations}
\begin{align}
f_n(t,\vec R,\vec k) &\equiv \int_{\Delta k}
\left\langle \vec k_+, n,s\Big|\rho_S(t)\Big|\vec k_-, n,s\right\rangle
e^{i\Delta\vec k\cdot\vec R},\\
f_o(t,\vec R,\vec r, \vec P, \vec p) &\equiv \int_{\Delta P,\Delta p}
\left\langle \vec P_+, \vec p_+, a\Big|\rho_S(t)\Big|\vec P_-, \vec p_-, a\right\rangle
e^{i\Delta\vec P\cdot\vec R + i\Delta\vec p\cdot\vec r}, \\
\vec k_{\pm} &\equiv \vec k \pm \frac{1}{2}\Delta \vec k, \quad
\vec P_{\pm}\equiv \vec P\pm \frac{1}{2}\Delta\vec P, \quad
\vec p_{\pm} \equiv \vec p \pm \frac{1}{2}\Delta \vec p.
\end{align}
\end{subequations}
By the Wigner transformation, the transition term in the Lindblad equation turns into a collision term of the kinetic equation for the octet\footnote{
To get this form, we approximate one of the energy delta functions as
\begin{align}
\delta\left(\frac{2\vec p\cdot\Delta\vec p}{M} + \frac{\vec q\cdot\Delta\vec k}{2M}\right)
\approx\delta\left(\frac{2\vec p\cdot\Delta\vec p}{M}\right),
\end{align}
using the estimates $q\sim T$ for thermal gluons, $p\sim \Delta p\sim Mv$ for bound states, and $\Delta k\sim \sqrt{MT}$ for total momentum of a quarkonium in equilibrium, and the assumption $Mv^2\sim T$ in this regime.
The same approximation is made in the derivation of Eq.~\eqref{eq:wign_singlet_os}.
}
\begin{align}
\label{eq:kinetic_octet_so}
\frac{\partial}{\partial t}f_o(t,\vec R,\vec r, \vec P, \vec p)\Bigr|_{s\to o}
&=\frac{(2\pi)^5}{\mathcal T}C_Fg^2\sum_n\int_{q,k} [\text{4 $\delta$-functions}] \cdot n_B(q) f_n(t,\vec R,\vec k) \nonumber \\
&\quad \times q^2\delta_T^{ij}(\hat q)\int_{\Delta p}e^{i\Delta\vec p\cdot\vec r}
\langle \vec p_+|r_i|n\rangle\langle n|r_j|\vec p_-\rangle
\delta\left(\frac{2\vec p\cdot\Delta\vec p}{M}\right),
\end{align}
and the $\mathcal T$ does not yet disappear from the equation.
The energy-momentum conservation is satisfied by the 4 delta functions
\begin{align}
\label{eq:4-delta}
\delta(q + E_{k,n} - E_{P, p})\delta(\vec q + \vec k - \vec P).
\end{align}
From the uncertainty principle, the matrix elements $\langle \vec p_+|r_i|n\rangle$ and $\langle n|r_j|\vec p_-\rangle$ are simultaneously finite for $\Delta p\sim Mv$.
Then the right hand side of \eqref{eq:kinetic_octet_so} is finite only in a small domain of the bound state size $r\lesssim 1/Mv$ as is naturally expected.
This scale is too microscopic for the kinetic description to resolve and thus the singlet-to-octet transition can be approximated as a local process.
Therefore, the kinetic equation is approximated by
\begin{align}
\label{eq:local_octetkin_so}
\frac{\partial}{\partial t}f_o(t,\vec R,\vec r, \vec P, \vec p)\Bigr|_{s\to o}
&=\delta(\vec r) C_{s\to o}, \quad
C_{s\to o}=\int_r\text{[r.h.s. of Eq.~\eqref{eq:kinetic_octet_so}]}.
\end{align}
The integration over $\vec r$ yields a delta function $\delta(\Delta \vec p)$ which makes one of the energy delta functions trivial
\begin{align}
\delta\left(\frac{2\vec p\cdot\Delta\vec p}{M}\right)\delta(\Delta \vec p)
= \delta(0)\delta(\Delta \vec p) = \frac{\mathcal T}{2\pi}\delta(\Delta \vec p).
\end{align}
Finally, $\mathcal T$ dependence disappears and we get
\begin{subequations}
\label{eq:coll_octet_so}
\begin{align}
\frac{\partial}{\partial t}f_o(t,\vec R,\vec r, \vec P, \vec p)\Bigr|_{s\to o}
&=\delta(\vec r) (2\pi)^4 \sum_n\int_{q,k} [\text{4 $\delta$-functions}] 
\cdot n_B(q) f_n(t,\vec R,\vec k)\cdot |\mathcal M_{s\to o}(\vec q, n; \vec p)|^2,\\
|\mathcal M_{s\to o}(\vec q, n; \vec p)|^2&\equiv C_Fg^2q^2\delta_T^{ij}(\hat q)
\langle n|r_i|\vec p\rangle\langle \vec p|r_j|n\rangle.
\end{align}
\end{subequations}

The damping term in the Lindblad equation is also written for continuous energy spectrum as
\begin{align}
\frac{1}{2}\sum_{\omega<0}\sum_{i,j}
\gamma_{ij}(\omega)\left\{V_S^{(i)\dagger}(\omega)V_S^{(j)}(\omega), \rho_S(t)\right\}\leftrightarrow
\frac{1}{2}\mathcal T\int_{\omega,x,y} \theta(-\omega)\gamma_{ab, ij}(\omega, \vec x, \vec y)
\left\{V_S^{ai\dagger}(\vec x, \omega)V_S^{bj}(\vec y, \omega), \rho_S(t)\right\},
\end{align}
from which the decay of the singlet density matrix is obtained
\begin{subequations}
\begin{align}
&\frac{\partial}{\partial t}\langle \vec k, n, s|\rho_S(t)|\vec k', n', s\rangle \Bigr|_{s\to o}
= -\Gamma_{nn'}(\vec k, \vec k') \langle \vec k, n, s|\rho_S(t)|\vec k', n', s\rangle , \\
&\Gamma_{nn'}(\vec k, \vec k')=(2\pi)^4C_F g^2 \int_{q, P, p} n_B(q) \cdot q^2\delta_T^{ij}(\hat q) \nonumber\\
&\qquad\qquad\quad \times
\frac{1}{2}\left[\begin{aligned}
&\langle n|r_i|\vec p\rangle\langle \vec p|r_j|n\rangle
\delta(q + E_{k,n} - E_{P, p})\delta(\vec q + \vec k - \vec P) \\
&+ \langle n'|r_i|\vec p\rangle\langle \vec p|r_j|n'\rangle
\delta(q + E_{k',n'} - E_{P', p'})\delta(\vec q + \vec k' - \vec P')
\end{aligned}
\right].
\end{align}
\end{subequations}
In this case, $\mathcal T$ dependence has already disappeared from the equation because one of the two energy delta functions is trivially satisfied for discrete bound state levels.
Assuming the same structure of the density matrix as above, the Wigner transform of the damping term turns into a collision term of the kinetic equation for the singlet
\begin{align}
\label{eq:coll_singlet_so}
\frac{\partial}{\partial t}f_n(t,\vec R, \vec k)\Bigr|_{s\to o}
=-(2\pi)^4 \int_{q,P,p}[\text{4 $\delta$-functions}]\cdot n_B(q)f_n(t,\vec R,\vec k)
\cdot |\mathcal M_{s\to o}(\vec q, n; \vec p)|^2,
\end{align}
where the energy-momentum conservation is imposed by the 4 delta functions \eqref{eq:4-delta}.

After the peculiar $\mathcal T$ dependence disappears, singlet-to-octet transition is described as the collision terms \eqref{eq:coll_octet_so} and \eqref{eq:coll_singlet_so} in the kinetic equation.
Two particle spatial distribution of the heavy quark pair
\begin{align}
\label{eq:2particle_distribution}
N(t,\vec R,\vec r) \equiv \int_{P,p} f_o(t,\vec R,\vec r,\vec P,\vec p) + \delta(\vec r)\sum_n\int_k f_n(t,\vec R,\vec k)
\end{align}
is locally conserved in the singlet-to-octet transition process
\begin{align}
\frac{\partial}{\partial t}N(t,\vec R,\vec r)\Bigr|_{s\to o} = 0.
\end{align}
The difference from the classical description in Sec.~\ref{sec:NRQCD_QQbar_CL} is that here the singlet heavy quark pair is described as a point-like molecule while in Sec.~\ref{sec:NRQCD_QQbar_CL} the phase space trajectory of the relative motion in the singlet is resolved.

\subsubsection{Octet-to-singlet transition by gluon emission}
\label{sec:pNRQCD_deep_os}
Here we analyze the transition process from the octet to the singlet, the inverse process of the previous analysis.
The transition term in the Lindblad equation for the continuous energy spectrum is
\begin{align}
\sum_{\omega>0}\sum_{i,j}
\gamma_{ij}(\omega)V_S^{(j)}(\omega) \rho_S(t)V_S^{(i)\dagger}(\omega)
\leftrightarrow
\mathcal T\int_{\omega,x,y} \theta(\omega)\gamma_{ab, ij}(\omega, \vec x, \vec y)
V_S^{bj}(\vec y, \omega)\rho_S(t) V_S^{ai\dagger}(\vec x, \omega),
\end{align}
from which we obtain
\begin{align}
\frac{\partial}{\partial t} \langle \vec k, n, s| \rho_S|\vec k', n', s\rangle
\Bigr|_{o\to s}
&=\frac{(2\pi)^8}{\mathcal T} \frac{g^2}{2N_c} \int_{q,P,p,P',p'}  [\text{8 $\delta$-functions}] \nonumber \\
&\qquad \times (1+n_B(q)) \langle \vec P, \vec p, a|\rho_S|\vec P', \vec p^{\ \prime}, a\rangle 
\cdot q^2\delta_T^{ij}(\hat q) \langle n|r_i|\vec p\rangle \langle \vec p^{\ \prime} |r_j|n'\rangle,
\end{align}
where the 8 delta functions \eqref{eq:8-delta} impose energy and momentum conservation in the forward and backward amplitudes.
The Wigner transform of this equation yields
\begin{align}
\label{eq:wign_singlet_os}
\frac{\partial}{\partial t} f_n(t,\vec R,\vec k)\Bigr|_{o\to s}
&=\frac{(2\pi)^5}{\mathcal T}\frac{g^2}{2N_c}\int_{q,P,p}[\text{4 $\delta$-functions}]\cdot (1+n_B(q)) \cdot q^2\delta^{ij}_T(\hat q) \nonumber \\
&\qquad\times\int_{\Delta P, \Delta p} e^{i\Delta \vec P\cdot\vec R}
\langle \vec P_+, \vec p_+, a|\rho_S|\vec P_-, \vec p_-, a\rangle
\langle n|r_i|\vec p_+\rangle \langle \vec p_- |r_j|n\rangle
\delta\left(\frac{2\vec p\cdot\Delta\vec p}{M}\right),
\end{align}
with the $\mathcal T$ dependence still present.
The heavy quark pair in the octet is unbound and its wave function in the relative coordinate is expected to be extended, at least much larger than the bound state size.
We therefore assume that octet density matrix $\langle \vec P_+, \vec p_+, a|\rho_S|\vec P_-, \vec p_-, a\rangle$ is localized at $\Delta \vec p\sim\vec 0$.
Then, we can approximate 
\begin{align}
\label{eq:rho_octet_p_approx}
\langle \vec P_+, \vec p_+, a|\rho_S|\vec P_-, \vec p_-, a\rangle
\approx (2\pi)^3\delta(\Delta \vec p)\int_{\Delta p'} \langle \vec P_+, \vec p_+^{\ \prime}, a|\rho_S|\vec P_-, \vec p_-^{\ \prime}, a\rangle,\quad
\vec p_{\pm}^{\ \prime} \equiv \vec p \pm \frac{1}{2}\Delta\vec p^{\ \prime}
\end{align}
to get a collision term for the singlet:
\begin{subequations}
\label{eq:coll_singlet_os}
\begin{align}
\frac{\partial}{\partial t} f_n(t,\vec R,\vec k)\Bigr|_{o\to s}
&=(2\pi)^4 \int_{q,P,p}[\text{4 $\delta$-functions}]\cdot (1+n_B(q))f_o(t,\vec R,\vec 0,\vec P,\vec p)
\cdot |\mathcal M_{o\to s}(\vec p; \vec q, n)|^2, \\
|\mathcal M_{o\to s}(\vec p; \vec q, n)|^2&\equiv\frac{g^2}{2N_c}q^2\delta^{ij}_T(\hat q) 
\langle \vec p |r_i|n\rangle\langle n|r_j|\vec p\rangle,
\end{align}
\end{subequations}
where the energy-momentum conservation is imposed by the 4 delta functions \eqref{eq:4-delta}.
The $f_o(t,\vec R,\vec r = \vec 0,\vec P,\vec p)$ term in the right hand side of Eq.~\eqref{eq:coll_singlet_os} shows that the approximation made here for the octet density matrix \eqref{eq:rho_octet_p_approx}  is essentially the local approximation for singlet bound states, the same as in the singlet-to-octet transitions.

The damping term in the Lindblad equation for the continuous energy spectrum is 
\begin{align}
\frac{1}{2}\sum_{\omega>0}\sum_{i,j}
\gamma_{ij}(\omega)\left\{V_S^{(i)\dagger}(\omega)V_S^{(j)}(\omega), \rho_S(t)\right\}\leftrightarrow
\frac{1}{2}\mathcal T\int_{\omega,x,y} \theta(\omega)\gamma_{ab, ij}(\omega, \vec x, \vec y)
\left\{V_S^{ai\dagger}(\vec x, \omega)V_S^{bj}(\vec y, \omega), \rho_S(t)\right\},
\end{align}
from which the decay of the octet density matrix is obtained as
\begin{align}
&\frac{\partial}{\partial t}\langle \vec P, \vec p, a|\rho_S(t)|\vec P', \vec p^{\ \prime}, a'\rangle \Bigr|_{o\to s}
= \frac{(2\pi)^5}{\mathcal T}\sum_n\int_{q, k, p''}
\frac{g^2}{2N_c}(1+n_B(q))q^2 \\
&\qquad\times \frac{1}{2}
\left[
\begin{aligned}
&\delta(q + E_{k,n} - E_{P, p})\delta(\vec q + \vec k - \vec P)\delta\left(\frac{p^2}{M} - \frac{p''^2}{M}\right)
\langle p|r_i|n\rangle\langle n|r_j|p''\rangle \delta_T^{ij}(\hat q) \langle \vec P, \vec p^{\ \prime\prime}, a|\rho_S(t)|\vec P', \vec p^{\ \prime}, a'\rangle\\
&+\delta(q + E_{k,n} - E_{P', p'})\delta(\vec q + \vec k - \vec P')\delta\left(\frac{p'^2}{M} - \frac{p''^2}{M}\right)
\langle p''|r_i|n\rangle\langle n|r_j|p'\rangle \delta_T^{ij}(\hat q) \langle \vec P, \vec p, a|\rho_S(t)|\vec P', \vec p^{\ \prime\prime}, a'\rangle
\end{aligned}
\right]. \nonumber
\end{align}
Unlike the damping term in the singlet-to-octet transition, this equation depends explicitly on $\mathcal T$.
To get rid of $\mathcal T$ dependence, the energy delta functions $\delta\left(\frac{p^2}{M} - \frac{p''^2}{M}\right)$ and $\delta\left(\frac{p'^2}{M} - \frac{p''^2}{M}\right)$ need to be trivially satisfied.
Wigner transform alone does not remove $\mathcal T$ from the equation.
In addition, we must approximate the octet density matrix by Eq.~\eqref{eq:rho_octet_p_approx}, and assume that the octet-to-singlet process is local in the octet kinetic equation ($s\to o$ replaced with $o\to s$ in Eq.~\eqref{eq:local_octetkin_so}).
Then, we can eliminate $\mathcal T$ dependence and get a collision term for the octet:
\begin{align}
\label{eq:coll_octet_os}
\frac{\partial}{\partial t} f_o(t,\vec R, \vec r, \vec P, \vec p)\Bigr|_{o\to s}
&=-\delta(\vec r)(2\pi)^4 \sum_n\int_{q,k}[\text{4 $\delta$-functions}]
\cdot (1+n_B(q))f_o(t,\vec R,\vec 0,\vec P,\vec p)
\cdot |\mathcal M_{o\to s}(\vec p; \vec q, n)|^2,
\end{align}
where the energy-momentum conservation is imposed by the 4 delta functions \eqref{eq:4-delta}.
Again, we observe that the local approximation for singlet bound states as well as classical approximation is essential to get the collision term without explicit $\mathcal T$ dependence.

Once the $\mathcal T$ dependence disappears in the collision terms \eqref{eq:coll_singlet_os} and \eqref{eq:coll_octet_os}, we can check that the two particle distribution $N(t,\vec R,\vec r)$ defined in Eq.~\eqref{eq:2particle_distribution} is locally conserved in the octet-to-singlet transitions:
\begin{align}
\frac{\partial}{\partial t} N(t,\vec R,\vec r)\Bigr|_{o\to s} = 0.
\end{align}

\subsubsection{Coupled Boltzmann equations for singlet and octet}
\label{sec:pNRQCD_deep_Boltzmann}
With the results in Sec.~\ref{sec:pNRQCD_deep_so} and \ref{sec:pNRQCD_deep_os}, we can summarize the classical kinetic descriptions for the heavy quark pair by the following coupled Boltzmann equations for singlet and octet distribution functions:
\begin{subequations}
\begin{align}
&\left[\frac{\partial}{\partial t} + \frac{\vec k}{2M}\cdot\vec\nabla_R\right] f_n(t,\vec R,\vec k) \nonumber \\
& \qquad =(2\pi)^4 \int_{q,P,p}[\text{4 $\delta$-functions}] 
\left[\begin{aligned}
&(1+n_B(q))f_o(t,\vec R,\vec 0,\vec P,\vec p)
\cdot |\mathcal M_{o\to s}(\vec p; \vec q, n)|^2\\
&-n_B(q)f_n(t,\vec R,\vec k) \cdot |\mathcal M_{s\to o}(\vec q, n; \vec p)|^2
\end{aligned}\right], \\
&\left[\frac{\partial}{\partial t} + \frac{\vec P}{2M}\cdot\vec\nabla_R + \frac{2\vec p}{M}\cdot\vec\nabla_r\right] 
f_o(t,\vec R,\vec r, \vec P, \vec p) \nonumber \\
& \qquad =\delta(\vec r) (2\pi)^4 \sum_n\int_{q,k} [\text{4 $\delta$-functions}]
\left[\begin{aligned}
&n_B(q) f_n(t,\vec R,\vec k)\cdot |\mathcal M_{s\to o}(\vec q, n; \vec p)|^2 \\
&-(1+n_B(q))f_o(t,\vec R,\vec 0,\vec P,\vec p)\cdot |\mathcal M_{o\to s}(\vec p; \vec q, n)|^2
\end{aligned}\right].
\end{align}
\end{subequations}
The two particle distribution for the octet $f_o(t,\vec R,\vec r, \vec P, \vec p)$ contains correlation between the heavy quark pair.
It is a probability distribution to find the heavy quark pair in the octet with which single heavy quark distribution functions are obtained as
\begin{subequations}
\begin{align}
f_{Q}(t,\vec r_Q, \vec p_Q) &= \int_{r_{Q_c}, p_{Q_c}}
f_o\left(t, \frac{\vec r_Q + \vec r_{Q_c}}{2} ,\vec r_Q - \vec r_{Q_c},
\vec p_Q + \vec p_{Q_c}, \frac{\vec p_Q - \vec p_{Q_c}}{2} \right), \\
f_{Q_c}(t,\vec r_{Q_c}, \vec p_{Q_c}) &= \int_{r_Q, p_Q}
f_o\left(t, \frac{\vec r_Q + \vec r_{Q_c}}{2} ,\vec r_Q - \vec r_{Q_c},
\vec p_Q + \vec p_{Q_c}, \frac{\vec p_Q - \vec p_{Q_c}}{2} \right).
\end{align}
\end{subequations}
They are normalized to
\begin{align}
\int_{R, r, P, p}f_o(t,\vec R,\vec r, \vec P, \vec p)
= \int_{r_Q, p_Q}f_{Q}(t,\vec r_Q, \vec p_Q) 
= \int_{r_{Q_c}, p_{Q_c}}f_{Q_c}(t,\vec r_{Q_c}, \vec p_{Q_c})
= N_o(t),
\end{align}
with $0\leq N_o(t)\leq 1$.
When we make the molecular chaos assumption, $f_o(t,\vec R,\vec r, \vec P, \vec p)$ is factorized into
\begin{align}
f_o(t,\vec R,\vec r, \vec P, \vec p) = \frac{1}{N_o(t)}f_{Q}(t,\vec r_Q, \vec p_Q)f_{Q_c}(t,\vec r_{Q_c}, \vec p_{Q_c}).
\end{align}
The factor $1/N_o(t)$ takes account of the correlation that if there is an unbound heavy quark, there is always an unbound heavy antiquark (in the octet).
This factor is important especially when $N_o(t)$ is small, for example at the initial stage of quarkonium dissociation, and correctly normalizes the recombination probability\footnote{
If there are always several unbound heavy quarks $N_{Q}, N_{Q_c} >1$, normalizations of the distribution functions are schematically
\begin{align}
\int_Q f_Q = N_Q, \quad \int_{Q_c} f_{Q_c} = N_{Q_c}, \quad \int_{Q,Q_c}f_o = N_Q N_{Q_c}, \quad
\frac{1}{N_{Q_c}}\int_{Q_c} f_o = f_Q, \quad \frac{1}{N_Q}\int_{Q} f_o = f_{Q_c},
\end{align}
and the molecular chaos assumption means $f_o\approx f_Q f_{Q_c}$.
As a result, there is no $1/N_o$ factors in the Boltzmann equations \eqref{eq:Boltzmann_coll} in this case.
}.
The Boltzmann equations for single particle distribution functions are
\begin{subequations}
\label{eq:Boltzmann_coll}
\begin{align}
&\left[\frac{\partial}{\partial t}+\frac{\vec k}{M}\cdot\nabla_R \right]f_n(t,\vec R,\vec k)
=(2\pi)^4 \int_{q,p_Q,p_{Q_c}}[\text{4 $\delta$-functions}] \\
& \qquad\qquad\qquad\qquad\times
\left[\begin{aligned}
&\frac{1}{N_o(t)}f_{Q}(t,\vec R, \vec p_Q)f_{Q_c}(t,\vec R, \vec p_{Q_c}) (1+n_B(q))
\cdot \Bigl|\mathcal M_{o\to s}\left(\frac{\vec p_Q - \vec p_{Q_c}}{2}; \vec q, n\right)\Bigr|^2\\
&-n_B(q)f_n(t,\vec R,\vec k) \cdot \Bigl|\mathcal M_{s\to o}\left(\vec q, n; \frac{\vec p_Q - \vec p_{Q_c}}{2}\right)\Bigr|^2
\end{aligned}\right],\nonumber \\
\label{eq:Boltzmann_coll_octet}
&\left[\frac{\partial}{\partial t}+\frac{\vec p_Q}{M}\cdot \vec\nabla_{r_Q}\right]f_{Q}(t,\vec r_Q, \vec p_Q) 
=(2\pi)^4 \sum_n\int_{q,k,p_{Q_c}} [\text{4 $\delta$-functions}] \\
& \qquad \qquad\qquad\qquad\times 
\left[\begin{aligned}
&n_B(q) f_n(t,\vec r_Q,\vec k)\cdot \Bigl|\mathcal M_{s\to o}\left(\vec q, n; \frac{\vec p_Q - \vec p_{Q_c}}{2}\right)\Bigr|^2 \\
&-\frac{1}{N_o(t)}f_{Q}(t,\vec r_Q, \vec p_Q)f_{Q_c}(t,\vec r_{Q}, \vec p_{Q_c})(1+n_B(q))\cdot
\Bigl|\mathcal M_{o\to s}\left(\frac{\vec p_Q - \vec p_{Q_c}}{2}; \vec q, n\right)\Bigr|^2
\end{aligned}\right],\nonumber
\end{align}
\end{subequations}
and similarly for $f_{Q_c}(t,\vec r_{Q_c}, \vec p_{Q_c})$.
The product of 4 delta functions \eqref{eq:4-delta} is rewritten as
\begin{align}
\delta\left(q + E_{k,n} - \frac{p_Q^2}{2M} - \frac{p_{Q_c}^2}{2M}\right)
\delta(\vec q + \vec k - \vec p_Q - \vec p_{Q_c}),
\end{align}
and imposes the conservation of energy and momentum in the collisions.

The octet-to-octet collisions, which we so far have ignored, are in the regime of quantum Brownian motion and the results of Sec.~\ref{sec:NRQCD_QQbar_CL} can be used here.
The Lindblad operators describing the octet-to-octet collisions are $\tilde V_d(k)$ and $\tilde V_f(k)$ in Eq.~\eqref{eq:Lindblad_NRQCD_QQbar_proj_explicit_octet}, which yield the noise correlation in the Langevin equation for the octet.
\begin{subequations}
\begin{align}
 &E_{\xi}[\xi_{Qi}(t)\xi_{Qj}(t')] = E_{\xi}[\xi_{Q_c i}(t)\xi_{Q_c j}(t')] = -\frac{N_c^2-2}{2N_c}\partial_i\partial_j\gamma(\vec 0)\delta(t-t'), \\
&E_{\xi}[\xi_{Qi}(t)\xi_{Q_c j}(t')] = -\frac{1}{N_c} \partial_i\partial_j\gamma(\vec r)\delta(t-t').
\end{align}
\end{subequations}
This is different from Eq.~\eqref{eq:Langevin_QQbar_octet} by the contribution of $\tilde V_-(k)$, which describes the octet-to-singlet process and is already taken into account in the collision term in Sec.~\ref{sec:pNRQCD_deep_os}.
In the large $N_c$ limit, the noise correlation between the heavy quark pair vanishes and the Boltzmann equation for the octet \eqref{eq:Boltzmann_coll_octet} is modified to the Kramers equation with a collision term
\begin{align}
\label{eq:Kramers_coll_octet}
\left[
\frac{\partial}{\partial t}
+\frac{\vec p_Q}{M}\cdot\vec \nabla_{r_Q} - \frac{\kappa_{\infty}}{2MT}\frac{\partial}{\partial \vec p_Q}
\cdot\left(
\vec p_Q + MT\frac{\partial}{\partial \vec p_Q}
\right)
\right]f_Q(t,\vec r_Q, \vec p_Q)
= \text{[r.h.s. of  Eq.~\eqref{eq:Boltzmann_coll_octet}]},
\end{align}
where the momentum diffusion constant is $\kappa_{\infty} = -\lim_{N_c\to\infty}\frac{N_c^2-2}{2N_c}\frac{\nabla^2\gamma(\vec 0)}{3}>0$.
The coupled Boltzmann equations \eqref{eq:Boltzmann_coll} and \eqref{eq:Kramers_coll_octet} reproduce those in \cite{Yao:2017fuc} except that the factor $1/N_o(t)$ is absent, which is fine when $N_o>1$, and the octet-to-octet collisions are simplified by the relaxation time approximation.

\subsection{Concluding remarks of Section \ref{sec:pNRQCD}}
\label{sec:pNRQCD_conclusion}
In this section, we restrict ourselves to the situation where heavy quark pair is close to each other.
By this restriction, the description becomes far simpler.
For instance, in the regime $1/r\gg T\sim m_D\gg E$ (Sec.~\ref{sec:pNRQCD_strong}), the Lindblad equation for a density matrix in the projected singlet-octet basis is obtained in Eq.~\eqref{eq:Lindblad_pNRQCD_QQbar_proj} as
\begin{align}
\frac{d}{dt}\rho_S(t) &= -i\left[H_S + \Delta H_S, \rho_S\right]
+\gamma \sum_{n=+,-,d}\left[
\tilde C_{ni}\rho_S \tilde C_{ni}^{\dagger}
-\frac{1}{2}\left\{
\tilde C_{ni}^{\dagger}\tilde C_{ni}, \rho_S
\right\}
\right], \nonumber\\
\Delta H_S &= \left(S r^2 + \frac{\gamma}{4MT}\{\vec p,\vec r\}\right)
\begin{pmatrix} C_F & 0 \\ 0 & \frac{N_c^2-2}{4N_c}\end{pmatrix},
\nonumber\\
\tilde C_{+i} &= \tilde V_{+i} \sqrt{C_F}\begin{pmatrix} 0 & 0 \\ 1 & 0 \end{pmatrix}, \quad
\tilde C_{-i} = \tilde V_{-i} \sqrt{\frac{1}{2N_c}}\begin{pmatrix} 0 & 1 \\ 0 & 0 \end{pmatrix}, \quad
\tilde C_{d i} = \tilde V_{d i} \sqrt{\frac{N_c^2-4}{4N_c}}\begin{pmatrix} 0 & 0 \\ 0 & 1 \end{pmatrix}, \nonumber
\end{align}
where the operators $ \tilde V_{ni} \ (n=+,i,d)$ are
\begin{align}
\tilde V_{\pm i} &\equiv -\left(r_i+\frac{ip_i}{2MT} \mp \frac{N_c}{8T}\frac{\alpha_s r_i}{r}\right), \quad
\tilde V_{di} \equiv -\left(r_i+\frac{ip_i}{2MT} \right) \nonumber
\end{align}
and constants $S$ and $\gamma$ are defined non-perturbatively in terms of gauge invariant gluon correlators.
Table~\ref{tab:simulation_pNRQCD} lists numerical simulations of the Lindblad equation from pNRQCD.
\begin{itemize}
\item A nontrivial question is whether equilibration of the quarkonium system is achieved within this regime.
In equilibrium, heavy quark pair is separately distributed and thus the dipole approximation must break down at some point even if the system starts from a small dipole.
Therefore, one must know to what extent this Lindblad equation can describe the equilibration process.
\item At lower temperature, one of the conditions for the quantum Brownian motion ($T\gg E$) becomes inapplicable and one needs to match and switch to an alternative description by coupled Boltzmann equation \eqref{eq:Boltzmann_coll}.
Although the Boltzmann equation is derived assuming the coupling $g$ is small, it is expected to capture the essence of physical process even when extrapolated to a larger value of $g$.
\end{itemize}

\begin{table}
\centering
\caption{Numerical simulations of Lindblad equation from pNRQCD.
The dimension $1_+$D means that it takes partial account of the angular directions by S and P waves, in addition to the radial direction.}
\label{tab:simulation_pNRQCD} 
\vspace{3mm}
\begin{tabular}{c|c|c|l} \hline
Dimension & Color & Gradient Expansion & \qquad\quad Numerical Method \\ \hline 
1$_{+}$D & SU(3) & LO & Direct evolution for S and P waves \cite{Brambilla:2016wgg, Brambilla:2017zei}\\
3D & SU(3) & LO & Quantum Jump \cite{Brambilla:2020qwo,Brambilla:2021wkt} \\ \hline
\end{tabular}
\end{table}

\newpage
\section{Summary}
\label{sec:summary}
In this review, I re-examined all of the open quantum system approaches to quarkonium dynamics in the quark-gluon plasma (QGP) in a systematic way.
I first introduced in Sec.~\ref{sec:Basics} the regimes and the approximation methods for an open quantum system weakly coupled to its environment.
I showed useful formulas for Lindblad equations in two regimes: the quantum optical regime (Eq.~\eqref{eq:Lindblad_QOL}) and the quantum Brownian motion (Eqs.~\eqref{eq:Lindblad_QBM} and \eqref{eq:Lindblad_exQBM}).
The formulas are applied to quarkonium system in the QGP, when the coupling constant is small in Sec.~\ref{sec:NRQCD} and when the quarkonium can be treated as a small color dipole in Sec.~\ref{sec:pNRQCD}.
Note that the coupling constant is not necessarily small in the latter.
In most cases, the dynamics of quarkonium is the quantum Brownian motion.

Two most important results are the Lindblad equations \eqref{eq:Lindblad_NRQCD_QQbar_proj} and \eqref{eq:Lindblad_pNRQCD_QQbar_proj}, which are applicable when the QGP is weakly coupled and when the quarkonium is a small color dipole, respectively.
These equations are slightly different from the original results:
\eqref{eq:Lindblad_NRQCD_QQbar_proj} supplements the Lindblad operators of \cite{Akamatsu:2014qsa} with new terms that derive from the color dependent heavy quark potential, and
\eqref{eq:Lindblad_pNRQCD_QQbar_proj} adds dissipative terms to the Lindblad operators of \cite{Brambilla:2016wgg, Brambilla:2017zei}.
When quarkonium is a small color dipole in a weakly coupled QGP, these Lindblad equations are both applicable and indeed agree with each other except for a minor technical difference.
This suggests that the Lindblad equation for quarkonium in a strongly coupled QGP can be generally inferred from \eqref{eq:Lindblad_NRQCD_QQbar_proj} in such a way that the small color dipole limit corresponds to \eqref{eq:Lindblad_pNRQCD_QQbar_proj}.
Derivation of the Lindblad equations at the next-to-leading order in the weak coupling expansion or at the next-to-next-to-leading order in the multipole expansion will be a first step toward such construction.

Time evolution of quarkonium in the QGP is diagonalization process of density matrix, during which quantum dissipation becomes essential and quarkonium may cease to be a small color dipole.
Because of the high numerical cost of simulating Lindblad equations in 3 dimensions, two simplifications, if available, are crucial.
One must ask whether or not (i) quantum dissipation is negligible and (ii) quarkonium stays a small color dipole, during the lifetime of QGP fireballs.
Depending on the answers, one may choose to simplify as follows:
\begin{itemize}
\item[] (i) Yes (ii) Yes -- Lindblad equation \eqref{eq:Lindblad_pNRQCD_QQbar_proj_recoilless} or Schr\"odinger equation with stochastic potential \eqref{eq:stochastic_potential_dipole}
\item[] (i) Yes (ii) No -- Lindblad equation \eqref{eq:Lindblad_NRQCD_QQbar_recoilless} or Schr\"odinger equation with stochastic potential \eqref{eq:stochastic_potential}
\item[] (i) No (ii) Yes -- Lindblad equation \eqref{eq:Lindblad_pNRQCD_QQbar_proj}
\item[] (i) No (ii) No -- Lindblad equation \eqref{eq:Lindblad_NRQCD_QQbar_proj}
\end{itemize}
In the first two cases, stochastic evolution of 3-dimensional wave function is enough, which lowers the numerical cost dramatically.
In the third case, the number of Lindblad operators is small and thus the numerical cost is lower than the fourth case.
In the first and third cases, one can exploit the selection rules for dipole transitions and solve in the angular momentum basis (see Appendix \ref{app:Lindblad_pNRQCD_LM}) using the Quantum Jump method, which also lowers the numerical cost.
To answer these questions, however, one needs to solve the Lindblad equation \eqref{eq:Lindblad_NRQCD_QQbar_proj} in a realistic set up for heavy-ion collisions.
Therefore, a test calculation of the Lindblad equation \eqref{eq:Lindblad_NRQCD_QQbar_proj} even in 1 dimension to compare with the other simplified descriptions will have a very important impact on the future phenomenological applications of the open quantum system approaches, which is left for a future work.

\newpage
\begin{center}
{\bf Acknowledgements}
\end{center}
I am grateful to Alexander Rothkopf, Shiori Kajimoto, Takahiro Miura, and Masayuki Asakawa for collaborations, which brought me the valuable opportunity to write this review paper.
I also thank Takahiro Miura for carefully reading the manuscript.
Finally I would like to thank my family for patience and support during my writing.
This work is supported by JSPS KAKENHI Grant Number JP18K13538.

\newpage
\appendix
\section{Properties of thermal correlation functions $\Gamma_{ij}(\omega)$}
\label{app:Thermal_Corr}
In this Appendix, we summarize the properties of environment correlation functions when the environment is a thermal bath $\rho_E(0)=\rho_E^{\rm th}$ with temperature $T=1/\beta$.
In this case, the environment correlation function $\Gamma_{ij}(\omega)$ and their decompositions into Hermitian ($\gamma_{ij}$) and anti-Hermitian ($iS_{ij}$) parts are defined as in Eq.~\eqref{eq:env_corr}:
\begin{subequations}
\begin{align}
\Gamma_{ij}(\omega)
&\equiv \int_0^{\infty} ds e^{i\omega s}{\rm Tr}_E\left(\rho_E^{\rm th}V_E^{(i)}(s)V_E^{(j)}(0)\right)
\equiv \frac{1}{2}\gamma_{ij}(\omega) + iS_{ij}(\omega),\\
\gamma_{ij}^*(\omega)&=\gamma_{ji}(\omega), \quad
S_{ij}^*(\omega)=S_{ji}(\omega).
\end{align}
\end{subequations}

\subsection{General properties}
\label{app:Thermal_Corr_General}
Here we list several general properties of the correlation functions.
First, the Hermitian matrix $\gamma_{ij}(\omega)$ satisfies
\begin{align}
\label{eq:KMS}
\gamma_{ij}(\omega)
&=\int_{-\infty}^{\infty}dt e^{i\omega t}{\rm Tr}_E
\left(\rho_E^{\rm th} V_E^{(i)}(t)V_E^{(j)}(0)\right)\nonumber \\
&=\int_{-\infty}^{\infty}dt e^{i\omega t}{\rm Tr}_E
\left(\rho_E^{\rm th} V_E^{(j)}(0)V_E^{(i)}(t+i\beta)\right)
=e^{\beta\omega}\gamma_{ji}(-\omega),
\end{align}
the Kubo-Martin-Schwinger (KMS) relation, which is used to prove that the system density matrix $\propto e^{-\beta H_S}$ is a steady state solution in the quantum optical regime.
The first line of \eqref{eq:KMS} shows that the spectrum of $\gamma_{ij}(\omega)$ contains enough information to express $S_{ij}(\omega)$.
The explicit form is
\begin{align}
S_{ij}(\omega)
&=\frac{1}{2i}\left[\int_0^{\infty}dt - \int_{-\infty}^0 dt\right]e^{i\omega t}{\rm Tr}_E
\left(\rho_E^{\rm th} V_E^{(i)}(t)V_E^{(j)}(0)\right)\nonumber \\
&=\frac{1}{2}\int_{-\infty}^{\infty}\frac{d\omega'}{2\pi}\gamma_{ij}(\omega')\left(
\frac{1}{\omega-\omega'+i\epsilon} + \frac{1}{\omega-\omega'-i\epsilon}
\right)
=-\mathcal P\int_{-\infty}^{\infty}\frac{d\omega'}{2\pi}\frac{\gamma_{ij}(\omega') }{\omega'-\omega},
\end{align}
where $\mathcal P$ denotes the Cauchy principal value.
A straight forward calculation derives a relation between $\gamma_{ij}(\omega)$ and the spectral function $\sigma_{ij}(\omega)$ of environment operators:
\begin{subequations}
\begin{align}
\sigma_{ij}(\omega)&\equiv
\int_{-\infty}^{\infty}dt e^{i\omega t}{\rm Tr}_E
\left(\rho_E^{\rm th} \left[V_E^{(i)}(t), V_E^{(j)}(0)\right]\right)
=\left(\gamma_{ij}(\omega)-\gamma_{ji}(-\omega)\right)
=(1-e^{-\beta\omega})\gamma_{ij}(\omega),
\end{align}
and that between $S_{ij}(\omega)$ and $\sigma_{ij}(\omega)$ as well
\begin{align}
S_{ij}(\omega)
&=-\mathcal P\int_{-\infty}^{\infty}\frac{d\omega'}{2\pi}\frac{1}{\omega'-\omega}
\frac{\sigma_{ij}(\omega') }{1-e^{-\beta\omega'}}.
\end{align}
\end{subequations}
For completeness, we also list the relations between the spectral function $\sigma_{ij}(\omega)$ and various Green functions of the environment operators, namely the retarded/advanced ($G^{R/A}_{ij}$) and symmetrized ($G^S_{ij}$) Green functions:
\begin{subequations}
\begin{align}
G_{ij}^R(\omega) &\equiv i\int_0^{\infty} dt e^{i\omega t}{\rm Tr}_E
\left(\rho_E^{\rm th} \left[V_E^{(i)}(t), V_E^{(j)}(0)\right]\right)
=\int\frac{d\omega'}{2\pi} \frac{\sigma_{ij}(\omega')}{\omega' - \omega - i\epsilon},\\
G_{ij}^A(\omega) &\equiv -i\int_{-\infty}^{0} dt e^{i\omega t}{\rm Tr}_E
\left(\rho_E^{\rm th} \left[V_E^{(i)}(t), V_E^{(j)}(0)\right]\right)
=\int\frac{d\omega'}{2\pi} \frac{\sigma_{ij}(\omega')}{\omega' - \omega + i\epsilon},\\
G_{ij}^S(\omega) &\equiv \int_{-\infty}^{\infty} dt e^{i\omega t}{\rm Tr}_E
\left(\rho_E^{\rm th} \left\{V_E^{(i)}(t), V_E^{(j)}(0)\right\}\right)
=\gamma_{ij}(\omega) + \gamma_{ji}(-\omega)
=\coth\left(\frac{\beta\omega}{2}\right)\sigma_{ij}(\omega).
\end{align}
\end{subequations}

When the environment operators $V_E^{(i)}(t)$ and $V_E^{(j)}(t)$ are both Hermitian and have the same sign under time-reversal transformation, which is always the case in the examples in the main text, the spectral function $\sigma_{ij}(\omega)$ is shown to be a real and odd function of $\omega$ and thus is a symmetric matrix $\sigma_{ij}(\omega)=\sigma_{ji}(\omega)$.
In this case, $\gamma_{ij}(\omega)$ and $S_{ij}(\omega)$ are also real symmetric matrices and expressed by
\begin{subequations}
\begin{align}
\gamma_{ij}(\omega)&=\left[1+n_B(\omega)\right]\sigma_{ij}(\omega)
=n_B(-\omega)\sigma_{ij}(-\omega),\\
S_{ij}(\omega)
&=-\mathcal P\int_0^{\infty}\frac{d\omega'}{2\pi}
\left\{
\left[1+n_B(\omega')\right]\frac{\sigma_{ij}(\omega')}{\omega'-\omega}
+n_B(\omega')\frac{\sigma_{ij}(\omega')}{-\omega'-\omega}
\right\}.
\end{align}
\end{subequations}
In this form, it is clear that the eigenvalues of $\gamma_{ij}(\omega)$ is the rate of energy transfer $|\omega|$ between the environment and the system; when $\omega$ is positive/negative, the environment absorbs/emits the energy.
We can also interpret $S_{ij}(\omega)$ as virtual state contributions to the forward amplitude when the transferred energy is off-shell $\omega\neq \omega'$.
The first/second term corresponds to $E\to E\pm\omega'\to E$ with $\omega'>0$.

\subsection{Small $\omega$ behaviors}
\label{app:Thermal_Corr_Small_Omega}
When the transferred energy is small $\omega\sim 0$, the real and odd spectral density $\sigma_{ij}(\omega)$ can be expanded by
\begin{align}
\sigma_{ij}(\omega)\simeq 
\frac{d\sigma_{ij}}{d\omega}\Bigr|_0\omega + \frac{1}{3!}\frac{d^3\sigma_{ij}}{d\omega^3}\Bigr|_0\omega^3 + \cdots .
\end{align}
The lower order expansion of $\gamma_{ij}(\omega)$ is then obtained as
\begin{align}
\gamma_{ij}(\omega) 
\simeq \gamma_{ij}^{(0)} + \gamma_{ij}^{(1)}\omega + \cdots
\simeq T\frac{d\sigma_{ij}}{d\omega}\Bigr|_0\left(
1 + \frac{\omega}{2T}
\right) + \mathcal O(\omega^2).
\end{align}
Note that the leading and the next-to-leading order terms of $\gamma_{ij}(\omega)$ are characterized by a common transport coefficient:
\begin{align}
\frac{\gamma_{ij}^{(0)}}{T} = 2\gamma_{ij}^{(1)} = \frac{d\sigma_{ij}}{d\omega}\Bigr|_0
=\frac{d}{d\omega}\left[2{\rm Im} G_{ij}^R(\omega)\right]\Bigr|_{0}.
\end{align}
Similarly, $S_{ij}(\omega)$ can also be expanded by
\begin{align}
S_{ij}(\omega) \simeq S_{ij}^{(0)} + S_{ij}^{(1)}\omega + \cdots
\end{align}
with the coefficients being
\begin{subequations}
\begin{align}
S_{ij}^{(0)} &= -\mathcal P\int_{-\infty}^{\infty}\frac{d\omega}{2\pi}\frac{1}{\omega}
\frac{\sigma_{ij}(\omega) }{1-e^{-\beta\omega}} 
=-\frac{1}{2} \int_{-\infty}^{\infty}\frac{d\omega}{2\pi}\frac{\sigma_{ij}(\omega)}{\omega}
=-\frac{1}{2}{\rm Re} \left[G_{ij}^R(0)\right], \\
S_{ij}^{(1)} &= -\mathcal P\int_{-\infty}^{\infty}\frac{d\omega}{2\pi}\frac{1}{\omega}
\frac{d}{d\omega}\left[
\frac{\sigma_{ij}(\omega) }{1-e^{-\beta\omega}}\right]
=-\frac{1}{2}\int_{-\infty}^{\infty}\frac{d\omega}{2\pi}\frac{1}{\omega}
\frac{d}{d\omega}\left[
\coth\left(\frac{\beta\omega}{2}\right)\sigma_{ij}(\omega) \right] \nonumber\\
&= -\frac{1}{2}\int_{-\infty}^{\infty}\frac{d\omega}{2\pi}\frac{1}{\omega}
\frac{d}{d\omega}G^S_{ij}(\omega).
\end{align}
\end{subequations}

In the main text, we also encounter an exceptional case $d\sigma_{ij}/d\omega|_0 = 0$, that is,
\begin{align}
\sigma_{ij}(\omega)\simeq 
\frac{1}{3!}\frac{d^3\sigma_{ij}}{d\omega^3}\Bigr|_0\omega^3 + \frac{1}{5!}\frac{d^5\sigma_{ij}}{d\omega^5}\Bigr|_0\omega^5 + \cdots .
\end{align}
In this case, the expansion of $\gamma_{ij}(\omega)$ starts from the second order:
\begin{align}
\gamma_{ij}(\omega) 
\simeq \frac{1}{2!}\gamma_{ij}^{(2)}\omega^2 + \frac{1}{3!}\gamma_{ij}^{(3)}\omega^3 + \cdots
\simeq \frac{T}{3!}\frac{d^3\sigma_{ij}}{d\omega^3}\Bigr|_0 \omega^2\left(
1 + \frac{\omega}{2T}
\right) + \mathcal O(\omega^4).
\end{align}
From the small $\omega$ behavior of $\sigma_{ij}(\omega)$, the spectral integration for $S_{ij}^{(1)}$ can be defined without employing Cauchy principal value.

\newpage
\section{Thermal correlation functions of gluons}
\label{app:Thermal_Corr_Gluons}
In this Appendix, we quote several analytic results for thermal correlation functions of gluons in the leading order perturbation theory.

\subsection{$\gamma(\vec x)$ and $S(\vec x)$ in Section \ref{sec:NRQCD}}
\label{app:Thermal_Corr_Gluons_gammaS}
The retarded Green function
\begin{align}
G_{ab, \mu\nu}^{R}(Q) &\equiv i\int_0^{\infty}dt \int d^3 x e^{i\omega t - i\vec q\cdot\vec x} 
{\rm Tr}_E
\left(\rho_E^{\rm th} \left[A^a_{\mu}(t,\vec x), A^b_{\nu}(0, \vec 0)\right]\right)
\equiv G_{\mu\nu}^R(Q)\delta_{ab}, \quad
Q^{\mu} \equiv (\omega, \vec q)
\end{align}
is calculated in the Hard-Thermal Loop (HTL) approximation \cite{le2000thermal}.
In the Coulomb gauge and covariant gauge with parameter $\xi$, the retarded function is parametrized by
\begin{subequations}
\label{eq:retarded_corr_proj}
\begin{align}
&\text{(Coulomb gauge)}\qquad
G_{\mu\nu}^R(Q)
=\frac{-(P_T)_{\mu\nu}}{Q^2 - \Pi_T} + \frac{Q^2}{q^2}\frac{-\delta_{\mu 0}\delta_{\nu0}}{Q^2 - \Pi_L} ,\\
&\text{(Covariant gauge)}\qquad
G_{\mu\nu}^R(Q)
=\frac{-(P_T)_{\mu\nu}}{Q^2 - \Pi_T} + \frac{-(P_L)_{\mu\nu}}{Q^2 - \Pi_L} 
+\xi \frac{Q_{\mu}Q_{\nu}}{Q^4},
\end{align}
\end{subequations}
where $P_T$ and $P_L$ denote transverse and longitudinal projectors:
\begin{align}
(P_T)_{ij} = \delta_{ij} - \frac{q_i q_j}{q^2}, \quad
(P_T)_{00} = (P_T)_{0j} = (P_T)_{i0} = 0, \quad
(P_L)_{\mu\nu} + (P_T)_{\mu\nu} = -g_{\mu\nu} + \frac{Q_{\mu}Q_{\nu}}{Q^2}.
\end{align}
The self energy is obtained perturbatively in the HTL approximation for soft external momenta $Q\sim gT$
\begin{subequations}
\begin{align}
\Pi_L(Q) &= -m_D^2\frac{Q^2}{q^2}(1-F(\omega/q)), \quad
\Pi_T(Q) = \frac{m_D^2}{2}\left[1+\frac{Q^2}{q^2}(1-F(\omega/q)) \right], \\
F(x)&\equiv \frac{x}{2}\left[
\ln\left|\frac{x+1}{x-1}\right| - i\pi\theta(1-|x|)
\right], \quad
m_D^2 = \frac{1}{3}g^2 T^2 \left(N_c + \frac{1}{2}N_f\right),
\end{align}
\end{subequations}
where $N_c$ and $N_f$ denote the numbers of colors and massless quark flavors.
For $Q\sim m_D\sim gT$, the resummation of $\Pi_{T,L}$ is essential to obtain the leading order correlation functions.

In the section \ref{sec:NRQCD}, we use the following correlation functions in the Coulomb gauge
\begin{align}
\gamma(\vec x) = 2g^2T\frac{\partial}{\partial \omega}{\rm Im} G_{00}^R(\omega, \vec x)\Bigr|_{\omega=0}, \quad
S(\vec x) = -\frac{g^2}{2}{\rm Re}G_{00}^R(\omega, \vec x)\Bigr|_{\omega=0}.
\end{align}
The longitudinal self energy and the retarded Green function with small frequency are
\begin{subequations}
\begin{align}
\Pi_L(\omega < q, \vec q) &= -m_D^2\frac{Q^2}{q^2}\left(
1 - \frac{\omega}{2q} \ln \frac{1+\omega/q}{1-\omega/q} + i\pi\frac{\omega}{2q}
\right)
\simeq m_D^2 \left(1 + i\pi \frac{\omega}{2q}\right) + \mathcal O(\omega^2),\\
G_{00}^R(\omega, \vec x)
&\simeq \int \frac{d^3q}{(2\pi)^3}e^{i\vec q\cdot \vec x}
\frac{-1}{q^2 + m_D^2 + i\pi m_D^2\omega/2q} + \mathcal O(\omega^2),
\end{align}
\end{subequations}
and we get
\begin{align}
\label{eq:stochastic_potential_HTL}
\gamma(\vec x) = g^2T \int \frac{d^3q}{(2\pi)^3}e^{i\vec q\cdot \vec x}
\frac{\pi m_D^2}{q(q^2 + m_D^2)^2}, \quad
S(\vec x) = \frac{g^2}{8\pi |\vec x|}e^{-m_D |\vec x|}.
\end{align}
The same result is obtained in the covariant gauge in which $G_{00}^R(\omega, \vec x)$ is different from that in the Coulomb gauge only by $\mathcal O(\omega^2)$.
The expression \eqref{eq:stochastic_potential_HTL} is applicable for $|\vec x|\gtrsim 1/m_D$ where the HTL approximation works.

\subsection{$\kappa$ and $\lambda$ in Section \ref{sec:NRQCD}}
\label{app:Thermal_Corr_Gluons_Constants}
It is noteworthy to emphasize that the coefficients $\nabla^2\gamma(\vec 0)$ and $\nabla^2 S_{T\neq 0}(\vec 0)$ of the Lindblad equation in the small dipole limit (see Eq.~\eqref{eq:Lindblad_NRQCD_QQbar_dipole} in Sec.~\ref{sec:NRQCD_QQbar_Dipole}) cannot be calculated from \eqref{eq:stochastic_potential_HTL}.
As one would soon recognize, heavy quark momentum diffusion constant $\kappa = -\frac{C_F\nabla^2\gamma(\vec 0)}{3}$ is divergent from ultraviolet contributions.
Physically, this divergence originates from the fact that the relevant scattering processes for heavy quark momentum diffusion involve hard as well as soft momentum transfers.
The former is rare but significant, while the latter is frequent but individually less significant.
The HTL approximation for the gluon self energy is justified only when the exchanged momentum is soft $Q\sim m_D$.
By correcting the gluon self energy for hard momentum transfer, $\kappa$ is obtained to be finite \cite{moore2005much}
\begin{align}
\kappa_{\rm LO} = \frac{C_F g^4 T^3}{18\pi}\left[
N_c\left(\ln\frac{2T}{m_D} + \xi \right) + \frac{N_f}{2}\left(\ln\frac{4T}{m_D} + \xi \right)
\right], \quad
\xi = \frac{1}{2} - \gamma_E + \frac{\zeta'(2)}{\zeta(2)}\simeq -0.64718,
\end{align}
where $\gamma_E\simeq 0.5772$ is the Euler-Mascheroni constant and $\zeta(s)$ is the Riemann zeta function.
Non-perturbative definition of $\kappa (=C_F\gamma)$ is given by \cite{Casalderrey-Solana:2006fio, CaronHuot:2009uh} in terms of a gauge invariant correlator of the color electric fields \eqref{eq:Lindblad_coefficients_pNRQCD}.
The weak coupling expansion of $\kappa$ is known at present to the next-to-leading order \cite{CaronHuot:2007gq,CaronHuot:2008uh}
\begin{align}
\kappa_{\rm NLO} = \frac{C_F g^4 T^3}{18\pi}\left[
N_c\left(\ln\frac{2T}{m_D} + \xi \right) + \frac{N_f}{2}\left(\ln\frac{4T}{m_D} + \xi \right)
+\frac{N_cm_D}{T}C
\right], \quad
C 
\simeq 2.3302,
\end{align}
whose convergence is not very good.
For the same reason as above, the short distance behavior of $S(\vec x)$ in \eqref{eq:stochastic_potential_HTL} is not accurate, either.
Correct result for the thermal dipole self energy constant $\lambda=-\frac{2C_F\nabla^2 S_{T\neq 0}(\vec 0)}{3}$ of Eq.~\eqref{eq:S_small_r} can be extracted from \cite{Brambilla:2008cx} as shown in \cite{Brambilla:2016wgg, Brambilla:2017zei, Brambilla:2019tpt}
\begin{align}
\lambda_{\rm LO} = -2\zeta(3)C_F\left(\frac{4}{3}N_c + N_f\right)\alpha_s^2 T^3.
\end{align}
Non-perturbative definition of $\lambda(=2C_FS)$ is given in \cite{Brambilla:2016wgg, Brambilla:2017zei, Brambilla:2019tpt} in terms of a gauge invariant correlator of the color electric fields \eqref{eq:Lindblad_coefficients_pNRQCD} with subtraction of the vacuum contribution.
The weak coupling expansion of $\lambda$ is known up to the next-to-leading order \cite{eller2019thermal}
\begin{align}
\lambda_{\rm NLO} = -2\zeta(3)C_F\left(\frac{4}{3}N_c + N_f\right)\alpha_s^2 T^3 + \frac{\alpha_sC_Fm_D^3}{3},
\end{align}
which again can be extracted from \cite{Brambilla:2008cx}.

\subsection{$\gamma_{ab,ij}(\omega,\vec q)$ in Sections \ref{sec:pNRQCD_weak} and \ref{sec:pNRQCD_deep}}
\label{app:Thermal_Corr_Gluons_Transverse}
In the section \ref{sec:pNRQCD_weak}, we use a spectral density of transverse gluons
\begin{align}
\sigma_{ab,ij}(\omega) &= 2g^2\omega^2\delta^{ab} {\rm Im} \int \frac{d^3 q}{(2\pi)^3} 
(P_T)_{ik}(P_T)_{jl} G^R_{kl}(\omega, \vec q)\nonumber \\
&= 2g^2\omega^2\delta^{ab}{\rm Im} \int \frac{d^3 q}{(2\pi)^3}
\frac{-(P_T)_{ij}}{Q^2 - \Pi_T} 
= -\frac{4g^2}{3}\omega^2\delta^{ab}\delta_{ij}{\rm Im}\int \frac{d^3 q}{(2\pi)^3}
\frac{1}{Q^2 - \Pi_T}.
\end{align}
The self energy for the thermal gluons $Q\sim T$ is sub-leading and can be neglected.
The spectral density for the thermal gluons is thus
\begin{align}
\sigma_{ab,ij}(\omega) &= -\frac{4g^2}{3}\omega^2\delta_{ij}\delta^{ab}{\rm Im}\int \frac{d^3 q}{(2\pi)^3}
\frac{1}{(\omega+i\epsilon)^2 - q^2}\nonumber \\
&= -\frac{2g^2}{3}\omega\delta_{ij}\delta^{ab}{\rm Im}\int_0^{\infty} \frac{q^2 dq}{2\pi^2}
\left(\frac{1}{\omega-q + i\epsilon} + \frac{1}{\omega + q + i\epsilon}\right)
=\frac{g^2}{3\pi}\omega^3\delta_{ij}\delta^{ab},
\end{align}
from which $\gamma_{ab,ij}(\omega, \vec q)$ used in the section \ref{sec:pNRQCD_deep} is readily obtained as
\begin{align}
\gamma_{ab,ij}(\omega, \vec q)
=\pi g^2\delta^{ab}\delta_T^{ij}(\hat q) q \left[
(1+n_B(q))\delta(\omega - q) + n_B(q)\delta(\omega + q)
\right], \quad
\delta_T^{ij}(\hat q)\equiv (P_T)_{ij}.
\end{align}

\newpage
\section{Lindblad equation with angular momentum projection}
\label{app:Lindblad_pNRQCD_LM}
In this Appendix, we rewrite the Lindblad equation \eqref{eq:Lindblad_pNRQCD_QQbar_proj} for the density matrix projected onto angular momentum.
The result is partially obtained in \cite{Brambilla:2020qwo, Akamatsu:2021vsh} at the leading order of the gradient expansion, but let us give it here including the next-to-leading terms for the sake of possible future numerical implementation.
Let us first quote Eq.~\eqref{eq:Lindblad_pNRQCD_QQbar_proj}.
For a density matrix in the projected singlet-octet basis
\begin{align}
\rho_S(t) = \begin{pmatrix}
\rho_s(t) & 0 \\
0 & \rho_o(t)
\end{pmatrix}, \nonumber
\end{align}
the Lindblad equation is
\begin{align}
\frac{d}{dt}\rho_S(t) &= -i\left[H_S + \Delta H_S, \rho_S\right]
+\gamma \sum_{i=x,y,z}\sum_{n=+,-,d}\left[
\tilde C_{ni}\rho_S \tilde C_{ni}^{\dagger}
-\frac{1}{2}\left\{
\tilde C_{ni}^{\dagger}\tilde C_{ni}, \rho_S
\right\}
\right], \nonumber \\
\Delta H_S &= \left(S r^2 + \frac{\gamma}{4MT}\sum_{i=x,y,z}\{p_i, r_i\}\right)
\begin{pmatrix} C_F & 0 \\ 0 & \frac{N_c^2-2}{4N_c}\end{pmatrix}, \nonumber
\\
\tilde C_{+i} &= \tilde V_{+i} \sqrt{C_F}\begin{pmatrix} 0 & 0 \\ 1 & 0 \end{pmatrix}, \quad
\tilde C_{-i} = \tilde V_{-i} \sqrt{\frac{1}{2N_c}}\begin{pmatrix} 0 & 1 \\ 0 & 0 \end{pmatrix}, \quad
\tilde C_{d i} = \tilde V_{d i} \sqrt{\frac{N_c^2-4}{4N_c}}\begin{pmatrix} 0 & 0 \\ 0 & 1 \end{pmatrix}, \nonumber
\end{align}
with the operators $ \tilde V_{ni} \ (n=+,i,d)$ being
\begin{align}
\tilde V_{\pm i} &\equiv -\left(r_i+\frac{ip_i}{2MT} \mp \frac{N_c}{8T}\frac{\alpha_s r_i}{r}\right), \quad
\tilde V_{di} \equiv -\left(r_i+\frac{ip_i}{2MT} \right). \nonumber
\end{align}
To avoid confusion, summation over repeated indices $i$ is explicitly written with $\sum_{i=x,y,z}$.

We define the density matrices projected onto angular momentum
\begin{align}
\rho_{s/o}^{(\ell)}(t,r_1,r_2)\equiv \sum_{m=-\ell}^{\ell}\int_{\Omega_1,\Omega_2}
Y_{\ell m}^*(\Omega_1)Y_{\ell m}(\Omega_2)\rho_{s/o}(t,r_1,\Omega_1, r_2,\Omega_2),
\end{align}
where $\int_{\Omega}\equiv\int d(\cos\theta) d\phi$ in the spherical coordinates and $Y_{\ell m}(\Omega)$ is the spherical harmonics.
First, we show that the evolution of $\rho_{s/o}^{(\ell)}(t,r_1,r_2) \ (\ell=0,1,2,\cdots)$ is coupled but closed.
Since $H_S$, $\Delta H_S$, and $\sum_{i}\tilde V_{ni}^{\dagger}\tilde V_{ni}$ are all scalar operators, it amounts to show that
\begin{align}
\mathcal W_{\ell_1 m_1;\ell_2 m_2}^{(\ell)[n]}(r_1,r_2)&\equiv \sum_{m}\sum_{i}\int_{\Omega_1,\Omega_2}
Y_{\ell m}^*(\Omega_1)Y_{\ell m}(\Omega_2)\tilde V_{ni}(r_1,\Omega_1)\tilde V_{ni}^*(r_2,\Omega_2)
Y_{\ell_1 m_1}(\Omega_1)Y_{\ell_2 m_2}^*(\Omega_2)
\end{align}
is proportional to $\delta_{\ell_1\ell_2}\delta_{m_1m_2}$ with a $m_{1,2}$-independent coefficient.
Here $\tilde V_{ni}^*(r_2,\Omega_2)$ is complex conjugate, not Hermitian conjugate, of $\tilde V_{ni}(r_2,\Omega_2)$.
From the vector operator $\tilde V_{ni}(r_1,\Omega_1)=(\tilde V_{nx}, \tilde V_{ny}, \tilde V_{nz})$, we can make a rank-1 tensor operator $\tilde V_{nq}(r_1,\Omega_1)=(\tilde V_{n+}, \tilde V_{n0}, \tilde V_{n-})$ by
\begin{align}
\tilde V_{n+} = -\frac{\tilde V_{nx}+i\tilde V_{ny}}{\sqrt{2}}, \quad
\tilde V_{n0} = \tilde V_{nz}, \quad
\tilde V_{n-} = \frac{\tilde V_{nx}-i\tilde V_{ny}}{\sqrt{2}},
\end{align}
which leads to $\sum_i\tilde V_{ni}(r_1,\Omega_1)\tilde V_{ni}^*(r_2,\Omega_2) = \sum_q\tilde V_{nq}(r_1,\Omega_1)\tilde V_{nq}^*(r_2,\Omega_2)$.
The Wigner-Eckart theorem dictates that
\begin{align}
\int_{\Omega}Y_{\ell m}^*(\Omega)\tilde V_{nq}(r,\Omega)
Y_{\ell'm'}(\Omega)
=C_{1\ell'}(\ell m;qm') \left[\tilde V_{n}(r)\right]_{\ell\ell'},
\end{align}
where $\left[\tilde V_{n}(r)\right]_{\ell\ell'}$ is a reduced matrix element, more precisely a reduced ``operator", and $C_{j_1j_2}(jm;m_1m_2)$ is the Clebsch-Gordan coefficient.
We then get
\begin{align}
\mathcal W_{\ell_1 m_1;\ell_2 m_2}^{(\ell)[n]}(r_1,r_2)
=\sum_{m,q} C_{1\ell_1}(\ell m;qm_1)C_{1\ell_2}(\ell m;qm_2)
\left[\tilde V_{n}(r_1)\right]_{\ell\ell_1}\left[\tilde V_{n}(r_2)\right]^*_{\ell\ell_2}.
\end{align}
Using the symmetry and orthogonality of the Clebsch-Gordan coefficients
\begin{subequations}
\begin{align}
&C_{j_1j_2}(jm;m_1m_2)
=(-1)^{j_1-m_1}\sqrt{\frac{2j+1}{2j_2+1}}C_{jj_1}(j_2m_2;m,-m_1),\\
&\sum_{m_1,m_2}C_{j_1j_2}(jm;m_1m_2)C_{j_1j_2}(j'm';m_1m_2) = \delta_{jj'}\delta_{mm'},
\end{align}
\end{subequations}
their sum is explicitly calculated
\begin{align}
\sum_{m,q} C_{1\ell_1}(\ell m;qm_1)C_{1\ell_2}(\ell m;qm_2)
&=\frac{2\ell+1}{\sqrt{(2\ell_1+1)(2\ell_2+1)}}\sum_{m,q}C_{\ell 1}(\ell_1 m_1;m,-q)C_{\ell 1}(\ell_2 m_2;m,-q) \nonumber\\
&=\frac{2\ell+1}{2\ell_1+1}\delta_{\ell_1\ell_2}\delta_{m_1m_2} \quad (|\ell-1|\leq\ell_1 \leq\ell+1)
\end{align}
and finally we obtain for $\ell\geq 1$ (for $\ell=0$ only $\ell_1=\ell+1=1$ is allowed)
\begin{align}
\mathcal W_{\ell_1 m_1;\ell_2 m_2}^{(\ell)[n]}(r_1,r_2)
=\frac{2\ell+1}{2\ell_1+1}\delta_{\ell_1\ell_2}\delta_{m_1m_2}
\left[\tilde V_{n}(r_1)\right]_{\ell\ell_1}\left[\tilde V_{n}(r_2)\right]^*_{\ell\ell_1} \quad (\ell_1=\ell\pm 1).
\end{align}
Here $\left[\tilde V_{n}(r)\right]_{\ell\ell}=0$ because $\tilde V_{nq}(r)$ has odd parity.
The action of $\mathcal W_{\ell_1 m_1;\ell_2 m_2}^{(\ell)[n]}(r_1,r_2)$ is
\begin{align}
\label{eq:transition_angular_basis}
\frac{\partial}{\partial t}\rho^{(\ell)}_{s/o}(t,r_1,r_2)
&\ni\sum_{\ell_1,m_1}\sum_{\ell_2,m_2}\mathcal W_{\ell_1 m_1;\ell_2 m_2}^{(\ell)[n]}(r_1,r_2)
\int_{\Omega_1,\Omega_2}Y_{\ell_1m_1}^*(\Omega_1)Y_{\ell_2m_2}(\Omega_2)
\rho_{s/o}(t, r_1,\Omega_1,r_2,\Omega_2) \nonumber \\
&=\sum_{\ell' = \ell\pm 1}
\frac{2\ell+1}{2\ell'+1}\left[\tilde V_{n}(r_1)\right]_{\ell\ell'}\left[\tilde V_{n}(r_2)\right]^*_{\ell\ell'}
\rho^{(\ell')}_{s/o}(t,r_1,r_2) \nonumber \\
&=\sum_{\ell' = \ell\pm 1}
\frac{2\ell+1}{2\ell'+1}\left[\tilde V_{n}(r_1)\right]_{\ell\ell'}
\rho^{(\ell')}_{s/o}(t,r_1,r_2)\left[\tilde V_{n}(r_2)\right]^{\dagger}_{\ell\ell'},
\end{align}
showing that the evolution is coupled but closed in $\rho^{(\ell)}_{s/o}(t,r_1,r_2)$.
Since the number of angular momentum states increases from $2\ell'+1$ to $2\ell+1$ in the process $\ell'\to\ell$ , the reduced operator $\left[\tilde V_{n}(r)\right]_{\ell\ell'}$ is interpreted as an averaged all-to-one transition operator from $\ell'$ to $\ell$.

Next, let us explicitly derive the Lindblad operators in the angular momentum basis, which correctly reproduce the time evolution of
\begin{align}
\rho_{s/o}(t,r_1,r_2) \equiv \sum_{\ell=0}^{\infty}\rho^{(\ell)}_{s/o}(t,r_1,r_2)|\ell\rangle\langle\ell|.
\end{align}
It is clear from Eq.~\eqref{eq:transition_angular_basis} that the Lindblad operators need to be split into raising and lowering parts in the angular momentum ladder
\begin{subequations}
\begin{align}
\left[\tilde V_{n\uparrow}(r)\right] &\equiv \sum_{\ell=1}^{\infty}\sqrt{\frac{2\ell+1}{2\ell-1}}
\left[\tilde V_{n}(r)\right]_{\ell,\ell-1}|\ell\rangle\langle\ell-1 |, \\
\left[\tilde V_{n\downarrow}(r)\right] &\equiv \sum_{\ell=0}^{\infty}\sqrt{\frac{2\ell+1}{2\ell+3}}
\left[\tilde V_{n}(r)\right]_{\ell,\ell+1}|\ell\rangle\langle\ell+1|,
\end{align} 
\end{subequations}
otherwise the density matrix gains off-diagonal components in the angular momentum.
Here, $|\ell\rangle$ is just a formal basis to express infinite dimensional matrices.
It might be technical, but one should keep in mind that
\begin{align}
\left[\tilde V_{n}(r)\right]_{\ell\ell'}^{\dagger} \neq \left[\tilde V_{n}^{\dagger}(r)\right]_{\ell'\ell}.
\end{align}
This is because $\left[\tilde V_{n}(r)\right]_{\ell\ell'}$ is a reduced operator and a simple operation for a true basis $|i\rangle\otimes|a\rangle$
\begin{align}
\langle a|\mathcal O|b\rangle^{\dagger} = \langle b|\mathcal O^{\dagger}|a\rangle
\end{align}
does not hold.
In this way, the Lindblad operator in the angular momentum basis is obtained by replacing $\tilde V_{ni}$ with $\left[\tilde V_{n\uparrow}(r)\right]$ and $\left[\tilde V_{n\downarrow}(r)\right]$ for $n=+,-,d$.

Finally, we calculate the operator $\left[\tilde V_{n}(r)\right]_{\ell\ell'}$.
We only need to calculate the left hand side of the relation
\begin{align}
\int_{\Omega}Y_{\ell 0}^*(\Omega)\tilde V_{n0}(r,\Omega)
Y_{\ell'0}(\Omega)
&=C_{1\ell'}(\ell 0;00) \left[\tilde V_{n}(r)\right]_{\ell\ell'}
=\frac{-\sqrt{\ell'}\delta_{\ell+1,\ell'} + \sqrt{\ell}\delta_{\ell-1,\ell'}}{\sqrt{2\ell'+1}}\left[\tilde V_{n}(r)\right]_{\ell\ell'},
\end{align}
where $\tilde V_{n0}(r,\Omega)$ contains operators such as
\begin{align}
z=r\cos\theta, \quad
p_z = -i\left(\cos\theta\frac{\partial}{\partial r} + \frac{\sin^2\theta}{r}\frac{\partial}{\partial \cos\theta}\right),\quad
e_z=\cos\theta.
\end{align}
A straightforward calculation shows
\begin{subequations}
\begin{align}
\int_{\Omega}Y_{\ell 0}^*(\Omega)\cos\theta Y_{\ell'0}(\Omega)
&=\frac{\ell'\delta_{\ell+1,\ell'} + \ell \delta_{\ell-1,\ell'}}{\sqrt{(2\ell+1)(2\ell'+1)}}, \\
\int_{\Omega}Y_{\ell 0}^*(\Omega)\sin^2\theta\frac{\partial}{\partial\cos\theta}Y_{\ell'0}(\Omega)
&=\frac{\ell'(\ell'+1)}{\sqrt{(2\ell+1)(2\ell'+1)}}(\delta_{\ell+1,\ell'} - \delta_{\ell-1,\ell'})
\end{align}
\end{subequations}
and we get
\begin{subequations}
\begin{align}
\left[r\right]_{\ell,\ell+1} &= r\left[e_r\right]_{\ell,\ell+1} = -r\sqrt{\frac{\ell+1}{2\ell+1}}, \quad
\left[r\right]_{\ell,\ell-1} = r\left[e_r\right]_{\ell,\ell-1} = r\sqrt{\frac{\ell}{2\ell+1}}, \quad \\
\left[p\right]_{\ell,\ell+1} &= i\sqrt{\frac{\ell+1}{2\ell+1}}\left(\frac{\partial}{\partial r} + \frac{\ell+2}{r}\right), \quad
\left[p\right]_{\ell,\ell-1} = -i\sqrt{\frac{\ell}{2\ell+1}}\left(\frac{\partial}{\partial r} - \frac{\ell-1}{r}\right).
\end{align}
\end{subequations}
To conclude, the orbital parts of the Lindblad operators are
\begin{subequations}
\begin{align}
\left[\tilde V_{\pm\uparrow}(r)\right] &= -\sum_{\ell=0}^{\infty}\sqrt{\frac{\ell+1}{2\ell+1}}
\left[r+\frac{1}{2MT}\left(\frac{\partial}{\partial r} - \frac{\ell}{r}\right) \mp \frac{N_c\alpha_s }{8T}\right]
|\ell+1\rangle\langle\ell|, \\
\left[\tilde V_{\pm\downarrow}(r)\right] &= \sum_{\ell=1}^{\infty}\sqrt{\frac{\ell}{2\ell+1}}
\left[r+\frac{1}{2MT}\left(\frac{\partial}{\partial r} + \frac{\ell+1}{r}\right) \mp \frac{N_c\alpha_s }{8T}\right]
|\ell-1\rangle\langle\ell|, \\
\left[\tilde V_{d\uparrow}(r)\right] &= -\sum_{\ell=0}^{\infty}\sqrt{\frac{\ell+1}{2\ell+1}}
\left[r+\frac{1}{2MT}\left(\frac{\partial}{\partial r} - \frac{\ell}{r}\right)\right]
|\ell+1\rangle\langle\ell|, \\
\left[\tilde V_{d\downarrow}(r)\right] &= \sum_{\ell=1}^{\infty}\sqrt{\frac{\ell}{2\ell+1}}
\left[r+\frac{1}{2MT}\left(\frac{\partial}{\partial r} + \frac{\ell+1}{r}\right)\right]
|\ell-1\rangle\langle\ell|
\end{align}
\end{subequations}
and the full Lindblad operators are constructed by making a direct product with appropriate color space operators.

\newpage
\bibliography{opensys_review.bib}

\end{document}